\documentclass[10pt,a4paper]{article}%{elsart}%amsmath,amssymb]%{revtex4}
\usepackage{amssymb,amsmath} 
 \usepackage{amsthm}
\usepackage{mathrsfs}
\usepackage{graphicx}
\usepackage{fullpage} 
\usepackage{color}
\usepackage{bm}
\usepackage[all]{xy} 
\usepackage{makeidx}
\makeindex
\begin{document} 
%%%%%%%%%%%%%%%%
\newtheorem{Def}{Definition}[section]
\newtheorem{Thm}{Theorem}[section]
\newtheorem{Proposition}{Proposition}[section]
\newtheorem{Lemma}{Lemma}[section]
\theoremstyle{definition}
\newtheorem*{Proof}{Proof}%[section]
\newtheorem{Postulate}{Postulate}[section]
\newtheorem{Corollary}{Corollary}[section]
\newtheorem{Example}{Example}[section]
\newtheorem{Remark}{Remark}[section]
\newcommand{\beq}{\begin{equation}}
\newcommand{\beqa}{\begin{eqnarray}}
\newcommand{\eeq}{\end{equation}}
\newcommand{\eeqa}{\end{eqnarray}}
\newcommand{\non}{\nonumber}
\newcommand{\fr}[1]{(\ref{#1})}
\newcommand{\abs}{\mathrm{abs}}
\newcommand{\ann}{\mathrm{ann}}
\newcommand{\Exp}{\mathrm{Exp}}
\newcommand{\tot}{\mathrm{tot}}
\newcommand{\Inv}{\mathrm{\,Inv}}
\newcommand{\gnd}{\mathrm{gnd}}
\newcommand{\B}{\mathrm{B}}
\newcommand{\C}{\mathrm{C}}
\newcommand{\G}{\mathrm{G}}
\renewcommand{\L}{\mathrm{L}}
\newcommand{\can}{\mathrm{can}}
\newcommand{\Diff}{\mbox{Diff}}
\newcommand{\Id}{\mathrm{Id}}
\newcommand{\bb}{\mbox{\boldmath {$b$}}}
\newcommand{\bbf}{\mbox{\boldmath {$f$}}}
\newcommand{\bt}{\mbox{\boldmath {$t$}}}
\newcommand{\bn}{\mbox{\boldmath {$n$}}}
\newcommand{\br}{\mbox{\boldmath {$r$}}}
\newcommand{\bC}{\mbox{\boldmath {$C$}}}
\newcommand{\bD}{\mbox{\boldmath {$D$}}}
\newcommand{\bK}{\mbox{\boldmath {$K$}}}
\newcommand{\bp}{\mbox{\boldmath {$p$}}}
\newcommand{\bx}{\mbox{\boldmath {$x$}}}
\newcommand{\by}{\mbox{\boldmath {$y$}}}
\newcommand{\bz}{\mbox{\boldmath {$z$}}}
\newcommand{\bF}{\mbox{\boldmath {$F$}}}
\newcommand{\bJ}{\mbox{\boldmath {$J$}}}
\newcommand{\bT}{\mbox{\boldmath {$T$}}}
\newcommand{\bR}{\mbox{\boldmath {$R$}}}

\newcommand{\balpha}{\mbox{\boldmath {$\alpha$}}}
\newcommand{\bbeta}{\mbox{\boldmath {$\beta$}}}
\newcommand{\bgamma}{\mbox{\boldmath {$\gamma$}}}
\newcommand{\bomega}{\mbox{\boldmath {$\omega$}}}
\newcommand{\bsigma}{\mbox{\boldmath {$\sigma$}}}
\newcommand{\bDelta}{\mbox{\boldmath {$\Delta$}}}
\newcommand{\bPsi}{\mbox{\boldmath {$\Psi$}}}
\newcommand{\ve}{{\varepsilon}}
\newcommand{\e}{\mathrm{e}}
\newcommand{\dr}{\mathrm{d}}
\newcommand{\Dr}{\mathrm{D}}
\newcommand{\F}{\mathrm{F}}
\newcommand{\mbbD}{\mathbb{D}}
\newcommand{\mbbE}{\mathbb{E}}
\newcommand{\mbbH}{\mathbb{H}}
\newcommand{\mbbM}{\mathbb{M}}
\newcommand{\mbbP}{\mathbb{P}}
\newcommand{\mbbR}{\mathbb{R}}
\newcommand{\mbbT}{\mathbb{T}}
\newcommand{\mbbV}{\mathbb{V}}
\newcommand{\mbbZ}{\mathbb{Z}}
\newcommand{\tA}{\widetilde A}
\newcommand{\tB}{\widetilde B}
\newcommand{\tC}{\widetilde C}
\newcommand{\chX}{\check{X}}
\newcommand{\cA}{{\cal A}}
\newcommand{\cB}{{\cal B}}
\newcommand{\cC}{{\cal C}}
\newcommand{\cD}{{\cal D}}
\newcommand{\cE}{{\cal E}}
\newcommand{\cF}{{\cal F}}
\newcommand{\cG}{{\cal G}}
\newcommand{\cH}{{\cal H}}
\newcommand{\cI}{{\cal I}}
\newcommand{\cK}{{\cal K}}
\newcommand{\cL}{{\cal L}}
\newcommand{\cM}{{\cal M}}
\newcommand{\cN}{{\cal N}}
\newcommand{\cO}{{\cal O}}
\newcommand{\cP}{{\cal P}}
\newcommand{\cR}{{\cal R}}
\newcommand{\Reeb}{{\cal R}}
\newcommand{\cS}{{\cal S}}
\newcommand{\cT}{{\cal T}}
\newcommand{\cU}{{\cal U}}
\newcommand{\cV}{{\cal V}}
\newcommand{\cY}{{\cal Y}}
\newcommand{\cZ}{{\cal Z}}
\newcommand{\tcA}{\widetilde{\cal A}}
\newcommand{\cAone}{   \underset{(1)}{ {\cal A} }     }
\newcommand\wh[1]{\widehat{#1}}
\newcommand{\DD}{{\cal D}}
\newcommand{\scrH}{\mathscr{H}}

\newcommand{\bfe}{\boldsymbol e} 
\newcommand{\bfb}{{\boldsymbol b}}
\newcommand{\bfd}{{\boldsymbol d}}
\newcommand{\bfh}{{\boldsymbol h}}
\newcommand{\bfg}{{\boldsymbol g}}
\newcommand{\bfj}{{\boldsymbol j}}
\newcommand{\bfn}{{\boldsymbol n}}
\newcommand{\bfA}{{\boldsymbol A}}
\newcommand{\bfB}{{\boldsymbol B}}
\newcommand{\bfD}{{\boldsymbol D}}
\newcommand{\bfJ}{{\boldsymbol J}}

\newcommand{\sympro}{\overset{s}{\otimes}}
\newcommand{\hash}{\#}
\newcommand{\GamCLamM}[1]{{\Gamma\mathbb{C}\Lambda^{{#1}}\cal{M}}}
\newcommand{\GamLamA}[1]{{\Gamma\Lambda^{{#1}}\cal{A}}}
\newcommand{\GamLamB}[1]{{\Gamma\Lambda^{{#1}}\cal{B}}}
\newcommand{\GamLamC}[1]{{\Gamma\Lambda^{{#1}}\cal{C}}}
\newcommand{\GamLamS}[1]{{\Gamma\Lambda^{{#1}}\cal{S}}}
\newcommand{\GamLamM}[1]{{\Gamma\Lambda^{{#1}}\cal{M}}}
\newcommand{\GamLamN}[1]{{\Gamma\Lambda^{{#1}}\cal{N}}}
\newcommand{\GamLam}[2]{{\Gamma\Lambda^{{#1}}{#2}}}
\newcommand{\GTM}{{\Gamma T\cal{M}}}
\newcommand{\GTB}{{\Gamma T\cal{B}}}
\newcommand{\GTC}{{\Gamma T\cal{C}}}
\newcommand{\GT}[1]{{\Gamma T {#1}}}
\newcommand{\normM}[2]{\left(  {#1}\, , \, {#2} \right)}
\newcommand{\normU}[2]{\left\{ {#1}\, , \, {#2} \right\}}
\newcommand{\diag}[1]{\mbox{diag}\{\, {#1}\,\}}
\newcommand{\GtC}[2]{\Gamma T^{#1}_{#2}{\cal C}}
\newcommand{\GtM}[2]{\Gamma T^{#1}_{#2}{\cal M}}
\newcommand{\GtN}[2]{\Gamma T^{#1}_{#2}{\cal N}}
\newcommand{\Gt}[3]{\Gamma T^{#1}_{#2}{#3}}
\newcommand{\equp}[1]{\overset{\mathrm{#1}}{=}}
\newcommand{\lequp}[1]{\overset{\mathrm{#1}}{\leq}}
\newcommand{\gequp}[1]{\overset{\mathrm{#1}}{\geq}}
\newcommand{\lup}[1]{\overset{\mathrm{#1}}{<}}
\newcommand{\gup}[1]{\overset{\mathrm{#1}}{>}}
\newcommand{\rightup}[1]{\overset{\mathrm{#1}}{\longrightarrow}}
\newcommand{\leftup}[1]{\overset{\mathrm{#1}}{\Longleftarrow}}
\newcommand{\Leftrightup}[1]{\overset{\mathrm{#1}}{\Longleftrightarrow}}
\newcommand{\leftrightup}[1]{\overset{\mathrm{#1}}{\longleftrightarrow}}
\newcommand{\nLongleftup}{\Longleftarrow\hspace{-4.5mm}\diagup}
\newcommand{\fun}[1]{{#1}^{\hash}}
\newcommand{\wt}[1]{\widetilde{#1}}
\newcommand{\ol}[1]{\overline{#1}}
\newcommand{\sff}[1]{ {\sf{#1}}}
\newcommand{\sfL}{ {\sf L}}
\newcommand{\ddiv}{\mbox{div}}
\newcommand{\rank}{\mbox{rank}\,}
\newcommand{\bkt}[2]{\left\{\,{#1}\, ,\,{#2}\,\right\}}
\newcommand{\inp}[2]{\left\langle\,{#1}\, ,\,{#2}\,\right\rangle}
\newcommand{\inpr}[2]{\left(\,{#1}\, ,\,{#2}\,\right)}
\newcommand{\ave}[1]{\left\langle\,{#1}\, \right\rangle}
\newcommand{\avgg}[2]{\left\langle\,{#1}\, \right\rangle_{#2}}
\newcommand{\bktt}[2]{\left\langle\left\langle\,{#1}\, ,\,{#2}\,\right\rangle\right\rangle}
\newcommand{\Leg}{{\mathfrak{L}}}
\newcommand{\ic}{\sqrt{-1}\,}
\newcommand{\ii}{\imath}
\newcommand{\bracket}[3]{\left[\,{#1},{#2}\,\right]_{#3}}
\newcommand{\LM}[2]{{\Lambda^{{#1}}_{{#2}}\,\cal{M}}}
\newcommand{\TM}[2]{{T^{{#1}}_{{#2}}\,\cal{M}}}
\newcommand{\nequiv}{\equiv\hspace{-.85em}/\,\,}
%%%%%%%%%%%

%\renewcommand{\labelenumi}{\Roman{enumi}}
\renewcommand{\labelenumi}{{\bf\arabic{enumi}. }}
% 
% To be black
%\newcommand{\void}[1]{}
%\renewcommand{\color}{\void}

%\renewcommand{}
%%%%%%%%%%
 
%\def\man{{M}}
%\def\GamLamM#1{{\Gamma \Lambda^{#1}\cal{M}}}
%\begin{footnotesize} 

%%%%%%%%%%%%%%%%%%%%%%%%%%%%%%%%%%%%%%%%%%%%%%%%%%%%%%
\title{
Contact geometric description of distributed-parameter 
port-Hamiltonian systems 
with respect to Stokes-Dirac structures and its information geometry
%Implicit Hamiltonian systems as dynamical systems on Legendre submanifolds of 
%contact manifolds 
}
%%%%%%%%%%%%%%%%%%%%%%%%%%%%%%%%%%%%%%%%%%%%%%%%%%%%%%
%\author{First Author${}^\dag$ and Second Author${}^\ddag$}
\author{Shin-itiro Goto\\
 Department of Applied Mathematics and Physics, \\
Graduate School of Informatics, Kyoto University,\\
Yoshida-Honmachi, Sakyo-ku, Kyoto 606-8501, Japan
}

\date{ \today}%\today}%\quad 
%The date of resubmission to the revised version will be added}
\maketitle
%%%%%%%%%%%%%%%%%
\begin{abstract}%
%%%%%%%%%%%%%%%%%

% It is shown that  implicit generalized 
% Hamiltonian systems with respect to Dirac structures 
% are written in terms of contact Hamiltonian systems on contact manifolds. 
This paper studies distributed-parameter systems on Riemannian manifolds 
with respect to Stokes-Dirac structures in a language of contact geometry with 
fiber bundles.
For the class where energy functionals are quadratic,
it is shown that 
distributed-parameter port-Hamiltonian systems with respect to 
Stokes-Dirac structures on one, two, and three dimensional 
Riemannian manifolds are 
written  in terms of contact Hamiltonian vector fields 
on bundles. Their fiber spaces are contact manifolds and  base spaces are 
Riemannian manifolds. In addition, for a class of 
distributed-parameter 
port-Hamiltonian systems, information geometry 
induced from contact 
manifolds and convex energy functionals is introduced and briefly discussed.  
%%%%%%%%%%%%%%%%
\end{abstract}
%%%%%%%%%%%%%
\section{Introduction}
%%%%%%%%%%%%%%%%%%%

Contact geometry and 
symplectic one are often referred to as twins,   
and then they have been studied from viewpoints of pure 
mathematics\cite{Arnold1976} and applied mathematics. 
Applications of contact geometry are found in  
science and foundation of engineering. They include 
equilibrium thermodynamics\cite{Hermann1973,MrugalaX,Schaft2007}, 
nonequilibrium thermodynamics\cite{Bravetti-Sep-2014,Goto2015},   
statistical mechanics\cite{Mrugala1990,Jurkowski2000,Bravetti-Aug-2014}, 
fluid mechanics\cite{Ghrist2007}, 
electromagnetism\cite{Dahl2004}, 
control theory\cite{Ohsawa2015}, 
non-conservative system\cite{Bravetti-JPhysA2014,Bravetti-An2017}  
and so on. 
Also, some models for LC circuits in contact with a thermal environment  
can be written in terms of contact geometry\cite{Eberard2006}, and  
it was shown that a LC circuit model without any external source and  
Maxwell's equations in media without source 
can be formulated in terms 
of contact geometry\cite{Goto2016,Goto2017}. 
In Ref.\,\cite{Goto2017}, 
Maxwell's equations without source can be written 
in terms of a bundle whose fiber space is a contact manifold.
In Refs.\,\cite{Goto2015,Goto2017}
some links between information 
geometry and contact geometry were clarified. Here information geometry is a 
geometrization of parametric statistics\cite{AN}. With information 
geometry, dynamics on so-called statistical manifolds was discussed 
\cite{Fujiwara1995,Nakamura1994}  
and applications on optimization problems were studied\cite{Ohara2013}.

Dirac structures on manifolds 
are some sub-bundles known to be 
extensions of symplectic and Poisson manifolds\cite{Courant1990}, and they  
have been intensively studied from various viewpoints.  
These structures allow to describe constraint mechanical systems and 
interconnected systems as implicit Hamiltonian 
systems\cite{Dalsmo1998,Schaft1998,Schaft2014}. 
Engineering applications of these include  
electric circuit theory\cite{Schaft1998}. For example,   
LC circuit models can be viewed as a degenerate Lagrangian system and 
those can be analyzed by the use of a Dirac structure\cite{Yoshimura2006}. 
It was then shown that Dirac structures on manifolds can be extended 
so that distributed-parameter systems with boundary energy flow can be 
described\cite{Schaft-Maschke2002}. This extended structure is a sub-bundle  
and is referred to 
as the Stokes-Dirac structure.  
As examples of such, it was demonstrated that 
Maxwell's equations, telegraph equations, vibrating strings, and 
ideal fluids 
are formulated in terms of Stokes-Dirac 
structures\cite{Schaft-Maschke2002}. 
In addition, there are further extensions from the original Stokes-Dirac 
structures\cite{Nishida2004,Maschke2005SIAM} and it is 
expected that various sophisticated approaches to distributed-parameter 
systems will be provided.  
 
In this paper a partial overlap between contact geometry and the theory of 
Stokes-Dirac structures is shown.  
Since Maxwell's equations have been formulated 
on a bundle whose fiber space is a contact manifold\cite{Goto2017} and 
on a Stokes-Dirac structure\cite{Schaft-Maschke2002} as mentioned above, 
it can be expected that there is a link between these two geometric 
descriptions. 
We thus focus on this link by generalizing the work of Ref.\,\cite{Goto2017}. 
To this end, 
Maxwell's equations are treated as a distributed-parameter system with respect to a Stokes-Dirac structure in this paper.  
Then it will be shown how some class of distributed-parameter systems 
with respect to Stokes-Dirac structures 
are written as contact Hamiltonian vector fields. 
% This can be seen as 
% an extention of the work of Ref.\cite{Goto2017}. 
In addition, by generalizing Ref.\,\cite{Goto2017},  
information geometry for %port-Hamiltonian systems and 
a class of distributed-parameter port-Hamiltonian systems 
will be introduced and then discussed.

% In general, if geometric theories of mathematical disciplines are ascribed to
% the same geometry, then it can be expected that there are links among these 
% disciplines. These links may give a unified picture of such disciplines.  
% Such an example is found in contact geometry, where 
% information geometry is linked to 
% contact geometric thermodynamics\cite{Goto2015}\cite{Mrugala1990}.

% An approach to develop such contact geometric theories 
% of mathematical sciences 
% is therefore to give various relations on these three keys. 
% One subject expected to be developed with the three keys in 
% contact geometry is   
% information geometry, where 
% information geometry is a geometrization of mathematical statistics\cite{AN}.   
% As briefly mentioned earlier in this section, 
% it has been pointed out that 
% information geometry is connected to contact geometry. Although 
% ideas in information geometry and those in contact geometry 
% are well-developed,   
% they have not been communicated.  
% We then feel that a way to connect between them  should be explored. 

% In this paper, some basic notations are fixed in Section\,\ref{sec-review}.  
% By showing that electric circuit models  
% can be seen as dynamical systems  
% on dually flat spaces, we argue how the present contact geometric 
% and information geometric methodologies 
% apply to engineering problems. 

%%%%%%%%%%%%%%%%%%%%%%%%%%%%%%%%%
\section{Preliminaries}
\label{sec-review}
%%%%%%%%%%%%%%%%%%%%%%%%%%%%%%%%%
In this section a brief summary of several geometries is given. 
This summary is used throughout this paper   
in order to describe various statements in the following sections.
% that will be shown in the following sections. 
%%%%%%%%%%%%%%%%%%%%%%%%%%%%%%%%%%
\subsection{Mathematical symbols, definitions, and known theorems}
\label{sec-Mathematical-preliminaries}
%%%%%%%%%%%%%%%%%%%%%%%%%%%%%%%%%%%%%%%%%%%%%%%%%%%%
Throughout this paper, geometric objects are assumed smooth.  
A point on an $n$-dimensional manifold $\xi\in\cM$ 
is often identified with a set of values of 
local coordinates $x(\xi)=\{\,x^{\,1}(\xi),\ldots,x^{\,n}(\xi)\,\}$.  

The following definition of bundle and the definitions 
of related objects are used in this paper. 
More mathematically rigorous definitions can be found in 
Ref.\cite{Nakahara} and so on.   
%%%%%%%%%%%%
\begin{Def}
%%%%%%%%%%%%
(Bundle or fiber bundle): 
Let $\cB$ be a $d_{\cB}$-dimensional manifold 
with local coordinates $\zeta=\{\zeta_1,\ldots,\zeta_{d_{\cB}}\}$,
$\cF$ a $d_{\cF}$-dimensional manifold, 
$\cM$ a $(d_{\cB}+d_{\cF})$-dimensional manifold, $\pi:\cM\to\cB$ a projection, 
$G$ a group acting on $\cF$, and $\{U_{\,i}\}$ an open covering of $\cB$ 
with $\phi_{\,i}:U_{\,i}\times \cF\to\pi^{-1}(U_{\,i})$ such that 
$\pi\phi_{\,i}(\zeta,u)=\zeta$.  
Then the set $(\cM,\pi,\cB)$ or $(\cM,\pi,\cB,\cF,G)$ 
 is referred to as a bundle or a fiber bundle,  
$\cB$ a base space, $\cF$ a fiber space, $G$ a structure group, 
$\phi_{\,i}$ a local trivialization, 
and $\cM$ a total space. 
Furthermore, let    
$t_{\,ij}(\zeta):=\phi_{\,i,\zeta}^{-1}\circ\phi_{\,j,\zeta}$
be an element of $G$ ($ t_{\,ij}:U_{\,i}\cap U_{\,j}\to G$ ),  
where $\phi_{\,i,\zeta}(u)=\phi_{\,i}(\zeta,u)$ 
for $U_{\,i}\cap U_{\,j}\neq \emptyset$. Then $\{t_{\,ij}\}$ are referred to as 
transition functions. 
%%%%%%%%%
\end{Def}  
%%%%%%%%%%%
\begin{Def}
%%%%%%%%%%%%
(Trivial bundle and non-trivial bundle):  
If a transition function for a bundle can be chosen to 
be identical, then
the bundle is referred to as a trivial bundle. Otherwise, the bundle is 
referred to as a non-trivial bundle. 
%%%%%%%%%
\end{Def}
%%%%%%%%%%%

Non-identical transition functions are used 
for describing non-trivial bundles.   
For example, 
the M\"obius band can be constructed with this formulation\cite{Nakahara}.   
In this paper trivial bundles are only considered.

A special class of sub-space of a bundle is considered in this paper, and 
the definition is given as follows.
%%%%%%%%%%%
\begin{Def}
%%%%%%%%%%%%
(Sub-bundle):
Let $(\cM,\pi,\cB)$ and $(\cM',\pi',\cB)$ be bundles. 
If the two conditions, 
%%%%%%%%%%%%%%%%%%
\begin{enumerate}
\item
$\cM'$ is a submanifold of $\cM$, 
\item
$\pi'=\pi|_{\cM'}$
\end{enumerate}
%%%%%%%%%%%%%%%
are satisfied, then $(\cM',\pi',\cB)$ is referred to as a sub-bundle of 
$(\cM,\pi,\cB)$.
%%%%%%%%5
\end{Def}
%%%%%%%%%%

%%%%%%%%%%%
\begin{Def}
%%%%%%%%%%%%%
(Section): 
Let $(\cM,\pi,\cB)$ be a bundle, and $f:\cB\to\cM$ a map  
such that $\pi\circ f=\Id_{\,\cB}$. Then 
 $f$ is referred to as a section. 
Here $\Id_{\,\cB}:\cB\to\cB$ denotes the identity map on $\cB$.  
The space of sections is denoted as $\Gamma\cM$. 
%If $\cM=\Omega^{\,0}\cB$, then $f\in\Gamma\Lambda^{\,0}\cB$. 
%Similarly 
%The space of $q$-forms on $\cB$ is denoted by  
%$\Gamma\Lambda^{\,q}\cB$
% with $q\in\{\,0,\ldots,\dim\cB\,\}$, 
%and the space of vector fields on $\cB$ as 
%$\Gamma T\cB$.
%%%%%%%%%% 
\end{Def}
%%%%%%%%%%

A set of vector fields on a manifold $\cM$ is denoted $\GTM$,
the tangent space at $\xi\in\cM$ as $T_{\xi}\cM$, 
the cotangent space at $\xi\in\cM$ as $T_{\xi}^{\,*}\cM$, 
a set of $q$-form fields $\GamLamM{q}$ with $q\in\{\,0,\ldots,\dim\cM\,\}$,  
and a set of tensor fields $\GtM{q'}{q}$ with 
$q,q'\in\{\,0,1,\ldots\,\}$. 
%A $q$-dimensional distribution on a manifold $\cM$ is a choice of 
%$q$-dimensional linear space $D_{\,\xi}\subset T_{\,\xi}\cM$.  
To express tensor fields the direct product is denoted $\otimes$.  
Einstein notation, when an index variables appear twice in a single 
term it implies summation of all the values of the index, 
is used. The Kronecker delta is denoted 
$\delta_{\,a}^{\,b},\delta_{\,ab},\ldots,$ 
and its value is unity when $a=b$, and zero when $a\neq b$.   
The exterior derivative acting on $\GamLamM{q}$ is denoted 
$\dr:\GamLamM{q}\to\GamLamM{q+1}$, and 
the interior product operator with $X\in\GTM$ as $\ii_X:\GamLamM{q}\to\GamLamM{q-1}$.  
Given a map $\Phi$ between two manifolds, the pull-back is denoted $\Phi^*$, 
and the push-forward $\Phi_*$.  
Then one can define the Lie derivative acting on tensor fields 
with respect to 
$X\in\GTM$, which is denoted $\cL_{X}:\GtM{q'}{q}\to\GtM{q'}{q}$. 
It follows that $\cL_{X}\beta=(\ii_X\dr+\dr\ii_X)\beta,$ 
for any $\beta\in\GamLamM{q}$, which is referred to as the Cartan formula. 
One can also define a derivative along a given vector field $X$, called the 
covariant derivative, denoted $\nabla_X:\GtM{0}{q}\to\GtM{0}{q}$. The action is 
explicitly given by specifying the connection coefficients 
$\Gamma_{ab}^{\ \ c}, (a,b,c\in\{\,1,\ldots,n\,\})$ with $n=\dim\cM$ 
such that 
$\nabla_{X_{\,a}}X_{\,b}=\Gamma_{ab}^{\ \ c}X_{\,c}$ where 
$\{\,X_{\,1},\ldots,X_{\,n}\,\}\in\GTM$ 
is a basis. For a given $S\in\GtM{0}{q}$, an object $\nabla S\in\GtM{0}{q+1}$
is defined such that  
$(\nabla S)(X,Y_{\,1},\ldots,Y_{\,q})=(\nabla_{\,X}S)(Y_{\,1},\ldots,Y_{\,q})$, where  
$X,Y_{\,1},\ldots,Y_{\,q}\in \GTM$.  
For example it can be shown that $\nabla f=\dr f$ for $f\in\GtM{0}{0}$, 
and that $(\nabla\dr f)(Y,\partial/\partial x^{\,a})=Y^{\,b}(\partial^2\,f/\partial x^{\,a}\partial x^{\,b}-\Gamma_{ba}^{\ \ c}\,\partial f/\partial x^{\,c})$, where
$Y=Y^{\,b}\partial/\partial x^{\,b}\in\GTM$.

\subsection{Geometry of fiber bundles}
%%%%%%%%%%%%%%%%%%%%%%%%%%%%%%%%%%%%%%%%%
Given a bundle $(\cM,\pi,\cB)$, 
the space of $q$-forms on $\cB$ and 
the space of $q'$-forms on $\cM$ can be introduced as follows.
%%%%%%%%%%%%
\begin{Def}
%%%%%%%%%%%%
(Horizontal form): 
Let $(\cM,\pi,\cB)$ be a bundle with $\dim\cB=d_{\cB}$ and 
$\dim\cM=d_{\cB}+d_{\cF}$,  
$\zeta$ coordinates for $\cB$ with 
$\zeta=\{\zeta^{\,1},\ldots,\zeta^{\,d_{\cB}}\}$, 
$(\zeta,u)$ coordinates for $\cM$ with $u=\{\,u^{\,\,1},\ldots,u^{\,d_{\cF}}\,\}$, 
and 
$\{\alpha_{i_1\cdots i_{q}}\}\in\Gamma\Lambda^{0}\cM$ some functions. 
A $q$-form on the bundle $(\cM,\pi,\cB)$ of the form   
$$
\alpha_{\mbbH}
=\alpha_{i_1\cdots i_q}(\zeta,u)\,
\dr \zeta^{\,i_1}\wedge\cdots\wedge \dr\zeta^{i_q},
%\qquad 
%(\,0\leq q\leq d_{\,\cB}\,)
$$
is referred to as a horizontal $q$-form.
The space of horizontal $q$-forms is 
denoted as $\Gamma\Lambda_{\,\mbbH}^{\,q}\cM$.
%%%%%%%%%
\end{Def}  
%%%%%%%%%%%
%%%%%%%%%%%%
\begin{Def}
%%%%%%%%%%%%
(Vertical form): 
Let $(\cM,\pi,\cB)$ be a bundle , 
$\dim\cB=d_{\cB}$ and 
$\dim\cM=d_{\cB}+d_{\cF}$,  
$\zeta$ a set of coordinates for $\cB$, 
$(\zeta,u)$ a set of coordinates for $\cM$ 
with $\zeta=\{\zeta^{\,1},\ldots,\zeta^{\,d_{\,\cB}}\}$ and 
$u=\{u^{\,1},\ldots,u^{\,d_{\,{\cF}}}\}$,  and 
$\{\alpha_{\,i_1\cdots i_q}\}\in\GamLam{0}{\cM}$ some functions.
A $q$-form on the bundle $(\cM,\pi,\cB)$ 
of the form   
$$
\alpha_{\,\mbbV}
=\alpha_{\,i_1\cdots i_q}(\zeta,u)\,
\dr u^{\,i_1}\wedge\cdots\wedge \dr u^{\,i_q},
%\qquad (\,0\leq q\leq d_{\,\cF}\,)
$$
is referred to as a vertical $q$-form. The space of vertical $q$-forms is 
denoted as $\Gamma\Lambda_{\,\mbbV}^{\,q}\cM$. In addition, vertical $0$-forms are  
referred to as vertical functions.   
%%%%%%%%%
\end{Def}  
 %%%%%%%%%

The wedge product of a horizontal $q$-form and a vertical $q'$-form 
can be defined. Then one defines the following. 
%%%%%%%%%%%
\begin{Def}
%%%%%%%%%%%
(Mixed form):
If a $(q+q')$-form $\alpha_{\,\mbbM}\in\Gamma\Lambda^{q+q'}\cM$ can be written as 
$$
\alpha_{\,\mbbM}
=\beta_{\,\mbbH}\wedge\gamma_{\,\mbbV},
$$
with some $\beta_{\,\mbbH}\in\Gamma\Lambda_{\,\mbbH}^{\,q}\cM$ and 
$\gamma_{\,\mbbV}\in\Gamma\Lambda_{\,\mbbV}^{\,q'}\cM$, then 
$\alpha_{\,\mbbM}$ 
is referred to as a mixed $(q,q')$-form. 
The space of mixed $(q,q')$-forms on $\cM$ is denoted as 
$\Gamma\Lambda_{\,\mbbH,\mbbV}^{\,q,q'}\cM$.
%%%%%%%%%%
\end{Def}
%%%%%%%%%%
%%%%%%%%%%%%%%
% \begin{Remark}
% %%%%%%%%%%%%%
% The space $\Gamma\Lambda_{\,\mbbH,\mbbV}^{\,q,q'}\cM$ is decomposed as 
% $$
% \Gamma\Lambda_{\,\mbbH,\mbbV}^{\,q,q'}\cM
% =\Gamma\Lambda_{\,\mbbH}^{\,q}\cM\otimes
% \Gamma\Lambda_{\,\mbbV}^{\,q'}\cM.
% $$
% %%%%%%%%%%%%%
% \end{Remark}
%%%%%%%%%%%%
%%%%%%%%%%%
\begin{Def}
%%%%%%%%%%%
(Vertical derivative):
Let $\alpha_{\,\mbbM}\in\Gamma\Lambda_{\,\mbbH,\mbbV}^{\,q,q'}\cM$ be 
a mixed $(q,q')$-form whose 
local expression can be written as  
$$
\alpha_{\,\mbbM}
=\alpha_{\,i_1\cdots i_q,j_{1}\cdots j_{q'}}(\zeta,u)\,
\dr \zeta^{\,j_1}\wedge\cdots\wedge \dr \zeta^{j_{q}}\wedge
\dr u^{\,i_1}\wedge\cdots\wedge \dr u^{\,i_{q^{\,\prime}}}.
$$
The operator  
$\dr_{\,\mbbV}:\Gamma\Lambda_{\,\mbbH,\mbbV}^{\,q,q'}\cM\to \Gamma\Lambda_{\,\mbbH,\mbbV}^{\,q,q+1}\cM$
whose action is such that  
$$
\dr_{\,\mbbV}\alpha_{\,\mbbM}
=\frac{\partial\alpha_{\,i_1\cdots i_q,j_{1}\cdots j_{q'}}(\,\zeta,u)}{\partial u^{\,i_0}}
\dr u^{\,i_0}\wedge \dr \zeta^{\,i_{1}}\wedge\cdots\wedge \dr \zeta^{\,i_{q}}
\wedge
\dr u^{\,i_1}\wedge\cdots\wedge \dr u^{\,i_{q^{\,\prime}}}, 
$$
%with $\dr_{\,\mbbV}\alpha_{\,\mbbH}=0$  
%for any $\alpha_{\,\mbbH}\in\Gamma\Lambda_{\,\mbbH}^{q}\cM$ 
is referred to as the vertical derivative or the vertical exterior derivative. 
%%%%%%%%%%
\end{Def}
%%%%%%%%%%
% \begin{Remark}
% %%%%%%%%%%%%%%
% For a $(q,q')$-mixed form $\alpha_{\,\mbbM}=\beta_{\,\mbbH}\wedge\gamma_{\,\mbbV}$, 
% one has 
% $$
% \dr_{\,\mbbV}\,\alpha_{\,\mbbM}
% =\beta_{\,\mbbH}\,\wedge(\,\dr_{\,\mbbV}\,\gamma_{\,\mbbV}\,).
% $$
% %%%%%%%%%%%%
% \end{Remark}
%%%%%%%%%%%
\begin{Def}
%%%%%%%%%%%
(Functional):
Let $(\cM,\pi,\cB)$ be a bundle, 
$\alpha_{\,\mbbH}\in\Gamma\Lambda_{\,\mbbH}^{\,q}\cM$ a horizontal 
$q$-form, 
$h\in\Gamma\Lambda_{\,\mbbV}^{\,0}\cM$ a vertical $0$-form, and 
$\cB_{\,0}\subseteq\cB$ a $q$-dimensional space. Then  
the integral over $\cB_{\,0}$ 
$$
\wt{h}_{\,\cB_{0}}
=\int_{\cB_{\,0}}h\,\alpha_{\,\mbbH},
$$
is referred to as a functional. The space of 
functionals  is denoted as $\Gamma F\cM$.
%%%%%%%%%%
\end{Def}
%%%%%%%%%%
The functional derivative 
has been used in the infinite dimensional Hamiltonian 
systems\cite{Abraham-Marsden-Ratiu}, also in the theory of Stokes-Dirac
structures\cite{Schaft-Maschke2002}. 
In this paper this derivative is also used 
to describe distributed-parameter port-Hamiltonian systems.
%%%%%%%%%%%%
\begin{Def}
\label{definition-functional-derivative}
%%%%%%%%%%%%%
(Functional derivative): 
Let $(\cM,\pi,\cB)$ be a bundle, $\alpha_{\,\mbbM}$ a mixed $(q,0)$-form on 
a $q^{\,\prime}$-dimensional submanifold of $\cB$, $\cB_{\,0}\subseteq\cB$ a 
$q^{\,\prime}$-dimensional subspace, 
$\wt{h}_{\,\cB_{\,0}}$ a functional depending on $\alpha_{\,\mbbM}$, 
and $\eta\in\mathbb{R}$ a constant.  
Then the mixed 
$(q^{\,\prime}-q,0)$-form 
$\delta \wt{h}_{\,\cB_{\,0}}/\delta\alpha_{\,\mbbM}
\in\Gamma\Lambda_{\,\mbbH,\mbbV}^{\,q^{\,\prime}-q,0}\cM$ 
that is uniquely obtained by 
$$
\wt{h}_{\,\cB_{\,0}}
\left[\,\alpha_{\,\mbbM}+\eta\,\alpha_{\,\mbbM}^{\prime}\,\right]
=\wt{h}_{\,\cB_{\,0}}\left[\,\alpha_{\,\mbbM}\,\right]
+\eta\int_{\cB_{\,0}}\frac{\delta \wt{h}_{\,\cB_{\,0}}}{\delta\alpha_{\,\mbbM}}
\wedge\alpha_{\,\mbbM}^{\prime}
+\cO(\,\eta^2\,),\qquad \forall\, \alpha_{\,\mbbM}^{\prime}
\in\Gamma\Lambda_{\,\mbbH,\mbbV}^{\,q,0}\,\cM
$$
is referred to as 
the functional derivative of $\wt{h}_{\,\cB_{\,0}}$ 
with respect to $\alpha_{\,\mbbM}$. 
%%%%%%%%%
\end{Def}
%%%%%%%%%%%

Similar to the case of forms, 
let $(\cM,\pi,\cB)$ be a bundle. 
Then the space of vector fields on $\cB$ and 
the space of vector fields on $\cM$ can be introduced as follows.

%%%%%%%%%%%%
\begin{Def}
%%%%%%%%%%%%
(Horizontal vector field): 
Let $(\cM,\pi,\cB)$ be a bundle   
with $\dim\cB=d_{\cB}$ and $\dim\cM=d_{\cB}+d_{\cF}$,  
$\zeta$ a set of coordinates for $\cB$ with 
$\zeta=\{\zeta^{\,1},\ldots,\zeta^{\,d_{\cB}}\}$, 
$(\zeta,u)$ a set of coordinates for $\cM$ 
with $u=\{u^{\,1},\ldots,u^{\,d_{\cF}}\}$, and 
$\{Y_{\,1},\ldots,Y_{\,d_{\cB}}\}\in\GamLam{0}{\cM}$ some functions. 
A vector field on the bundle $(\cM,\pi,\cB)$ of the form   
$$
Y_{\mbbH}
=Y_{i}(\zeta,u) \frac{\partial}{\partial\zeta^{\,i}},
$$
is referred to as a horizontal vector field.
The space of horizontal vector fields is 
denoted as $\Gamma T_{\,\mbbH}\cM$.
%%%%%%%%%
\end{Def}  
%%%%%%%%%%%
\begin{Remark}
%%%%%%%%%%%%%%
The dual of a horizontal vector field is a horizontal $1$-form.
%%%%%%%%%%%%%
\end{Remark}
%%%%%%%%%%%%%

%%%%%%%%%%%%
\begin{Def}
%%%%%%%%%%%%
(Vertical vector field): 
Let $(\cM,\pi,\cB)$ be a bundle   
with $\dim\cB=d_{\,\cB}$ and $\dim\cM=d_{\,\cB}+d_{\,\cF}$,  
$\zeta$ a set of coordinates for $\cB$ with 
$\zeta=\{\,\zeta^{\,1},\ldots,\zeta^{\,d_{\cB}}\,\}$, 
$(\zeta,u)$ a set of coordinates for $\cM$ with 
$u=\{\,u^{\,\,1},\ldots,u^{\,d_{\,\cF}}\,\}$, and 
$\{Y_{\,1},\ldots,Y_{\,d_{\cF}}\}\in\Gamma\Lambda^{\,0}\cM$ some functions.  
A vector field on the bundle $(\cM,\pi,\cB)$ of the form   
$$
Y_{\,\mbbV}
=Y_{\,i}(\zeta,u) \frac{\partial}{\partial u^{\,i}},
$$
is referred to as a vertical vector field.
The space of vertical vector fields is 
denoted as $\Gamma T_{\,\mbbV}\cM$.
%%%%%%%%%
\end{Def}  
%%%%%%%%%%%
%%%%%%%%%%%
\begin{Remark}
%%%%%%%%%%%%%%
The dual of a vertical vector field is a vertical $1$-form.
%%%%%%%%%%%%%
\end{Remark}
%%%%%%%%%%%%%
%%%%%%%%%%%
% \begin{Remark}
% %%%%%%%%%%%%%%
% The vector space $\Gamma T\cM$ is decomposed as 
% $\Gamma T\cM=\Gamma T_{\mbbH}\cM\oplus \Gamma T_{\mbbV}\cM$.
% %%%%%%%%%%%%%
% \end{Remark}
%%%%%%%%%%%%%

For a vertical $q$-form $\alpha_{\mbbV}$, 
the action of the interior product with respect to a vertical 
vector field $Y_{\mbbV}$   
is denoted $\ii_{Y_{\mbbV}}\alpha_{\mbbV}$ and is similar to the action of a vector 
field $Y$ to a $q$-form $\alpha$, $\ii_Y\alpha$.  

% This operator 
% $\ii_{Y_\mbbV}:\Gamma\Lambda_{\mbbV}^{q'}\cM\to \Gamma\Lambda_{\mbbV}^{q'-1}\cM$ 
% is extended to that 
% $\ii_{Y_\mbbV}:\Gamma\Lambda_{\mbbH,\mbbV}^{q,q'}\cM\to \Gamma\Lambda_{\mbbH,\mbbV}^{q,q'-1}\cM$
% as follows.
 %%%%%%%%%%%%
 \begin{Def}
 %%%%%%%%%%%%
(Interior product associated with a vertical vector field for vertical form): 
Let $(\cM,\pi,\cB)$ be a bundle with $\dim\cB=d_{\cB}$ 
and $\dim\cM=d_{\cB}+d_{\cF}$, $\zeta$ a set of coordinates for 
$\cB$ with $\zeta=\{\zeta^{\,1},\ldots,\zeta^{\,d_{\cB}}\}$, 
$\beta_{\,\mbbH}\in\Gamma\Lambda_{\,\mbbH}^{\,q}\cM$ a horizontal $q$-form,
$\gamma_{\,\mbbV}\in\Gamma\Lambda_{\,\mbbV}^{\,q'}\cM$ a vertical $q'$-form,
$Y_{\,\mbbV}\in\Gamma T_{\,\mbbV}\cM$ a vertical vector field, and 
$\alpha_{\,\mbbM}\in\Gamma\Lambda_{\,\mbbH,\mbbV}^{\,q,q'}\cM$ a mixed $(q,q')$-form 
 written as 
$$
\alpha_{\,\mbbM}
=\gamma_{\,\mbbV}\wedge\beta_{\,\mbbH}.
$$
Then the  action of $\ii_{\,Y_{\mbbV}}$ to $\alpha_{\,\mbbM}$, 
$\ii_{\,Y_{\,\mbbV}}:\Gamma\Lambda_{\mbbH,\mbbV}^{q,q'}\cM\to \Gamma\Lambda_{\mbbH,\mbbV}^{q,q'-1}\cM$
 is defined as 
$$
\ii_{\,Y_{\mbbV}}\alpha_{\,\mbbM}
=(\,\ii_{\,Y_{\mbbV}}\gamma_{\,\mbbV}\,)\wedge\beta_{\,\mbbH}.
$$
% %%%%%%%%%
 \end{Def}
%%%%%%%%

%%%%%%%%%%%%%%%%%%%%%%%%%
\subsection{Riemannian geometry}
%%%%%%%%%%%%%%%%%%%%%%%%%%
In this subsection some of basics of Riemannian manifold are summarized.
Roughly speaking, Riemannian geometry is a geometry where a class of 
metric tensor fields is provided for manifolds. In this paper 
Riemannian manifolds are used as base spaces of fiber bundles 
so that distributed-parameter port-Hamiltonian systems are geometrically described.  
%%%%%%%%%%
\begin{Def}
%%%%%%%%%%%
(Riemannian manifold and Riemannian metric tensor field): 
Let $\cM$ be an $n$-dimensional manifold and $g\in\GtM{0}{2}$ a tensor field on 
$\cM$. 
If $g$ satisfies 
(i) $g|_{\xi}(X,Y)=g|_{\xi}(Y,X)$, (ii) $g|_{\xi}(X,X)\geq 0$, 
(iii) $g|_{\xi}(X,X)=0$ iff $X=0$, 
for $X,Y\in T_{\,\xi}\cM$, then $g$ is referred to as 
a Riemannian metric tensor field. An $n$-dimensional manifold together with a 
Riemannian metric tensor field $g$ is denoted $(\cM,g)$ 
and this is referred to as an $n$-dimensional Riemannian manifold.
%%%%%%%%%
\end{Def}
%%%%%%%%%
% \begin{Def}
% %%%%%%%%%%%
% (Pseudo-Riemannian manifold and Pseudo-Riemannian metric tensor field): 
% Let $\cM$ be an $n$-dimensioanl manifold and $g\in\GtM{0}{2}$ a tensor field on 
% $\cM$. 
% If  $g$ satisfies 
% (i) $g|_{\xi}(X,Y)=g|_{\xi}(Y,X)$, 
% (ii) $g|_{\xi}(Z,Y)=0,(\forall Z\in T_{\xi}\cM) \Longrightarrow Y=0$, 
% for $X,Y\in T_{\,\xi}\cM$, then $g$ is referred to as 
% a pseudo-Riemannian metric tensor field. 
% An $n$-dimensional manifold together with a 
% pseudo-Riemannian metric tensor field $g$ is denoted $(\cM,g)$ 
% and this is referred to as an $n$-dimensional pseudo-Riemannian manifold.
% %%%%%%%%%%
% \end{Def}
%%%%%%%%%%%
On an $n$-dimensional  
Riemannian manifold one can define various mathematical objects. 
%%%%%%%%%%%
\begin{Def}
%%%%%%%%%%
(Orthonormal frame and orthonormal co-frame): 
Let $(\cM,g)$ be an $n$-dimensional Riemannian manifold. 
If $\{X_{\,1},\ldots,X_{\,n}\}\in\GTM$ satisfies 
$g(X_{\,a},X_{\,b})=\delta_{\,ab}$, then  
$\{X_{\,1},\ldots,X_{\,n}\}$ is referred to as a $g$-orthonormal basis.
In addition, if $\{\sigma^{\,1},\ldots,\sigma^{\,n}\}\in\GamLamM{1}$ satisfies 
$\sigma^{\,a}(X_b)=\delta_{\,b}^{\,a}$, then $\{\sigma^{\,1},\ldots,\sigma^{\,n}\}$ 
is referred to as a $g$-orthonormal co-frame. 
%%%%%%%%
\end{Def}
%%%%%%%
\begin{Remark}
%%%%%%%%%%%%%%%%%
One can express a Riemannian metric tensor field $g$ in terms of 
a $g$-orthonormal co-frame $\{\sigma^{\,a}\}$ 
as $g=\delta_{\,ab}\,\sigma^{\,a}\otimes\sigma^{\,b}$.
%%%%%%%%%%%%
\end{Remark}
%%%%%%%%%%%%%

%%%%%%%%%%%%%
\begin{Def}
%%%%%%%%%5
(Volume form): 
Let $\cM$ be an $n$-dimensional manifold, and $\alpha$ an $n$-form.
If $\alpha$ does not vanish at any point of $\cM$, then $\alpha$ is 
referred to as a volume-form. 
%%%%%%%%%%
\end{Def}
%%%%%%%%%%%%

On Riemannian manifolds, a natural volume-form is given as below.
%%%%%%%%%%%%
\begin{Def}
%%%%%%%%%%
(Canonical volume-form): 
Let $(\cM,g)$ be an $n$-dimensional  Riemannian manifold, 
and $\{\sigma^{\,a}\}$ its $g$-orthonormal co-frame.
Then an $n$-form that does not vanish anywhere 
$\star 1:=\sigma^{\,1}\wedge\cdots\wedge\sigma^{\,n}$ is referred to as 
a canonical volume-form. 
%%%%%%%%
\end{Def}
%%%%%%%%%

On Riemannian manifolds, the following map is induced. 
%%%%%%%%%%
\begin{Def}
%%%%%%%%%%
(Hodge map): 
Let $(\cM,g)$ be an $n$-dimensional Riemannian manifold, 
and $\star 1$ its canonical volume-form.
Define the map $\star:\GamLamM{q}\to\GamLamM{n-q}$ that satisfies
$$
\star\, (\,\alpha\wedge\gamma\,)
=\ii_{\,Y}\,\star \alpha,\qquad 
\star\, (\,f\,\alpha\,)
=f\star\,\alpha,\qquad
\star\,(\,\alpha+\beta\,)
=(\star\,\alpha)+(\star\,\beta),
$$ 
for all $\alpha,\beta\in\GamLamM{q},(q\in\{0,\ldots,n\}), \gamma\in\GamLamM{1}$,
$f\in\GamLamM{0}$, and $Y$ is such that $\gamma=g(Y,-)$.   
Then, this map $\star$ is referred to as the Hodge map.
%%%%%%%%%%%
\end{Def}
%%%%%%%%%%
The following formulae will be used. 
%%%%%%%%%%%%%%%
\begin{Lemma}
\label{3-d-Riemannian-Hodge}
%%%%%%%%%%%%%
Let $(\cZ,g)$ be a $3$-dimensional Riemannian manifold. Then it follows that 
$$
\star\star \alpha
=\alpha,
$$ 
for all $\alpha\in\GamLam{q}{\cZ}$ with $q\in\{0,\ldots,3\}$.
Thus, the inverse of $\star$ denoted by $\star^{\,-1}$ can be shown to be 
the same as $\star$: 
$$
\star^{\,-1}\alpha
=\star\,\alpha,
$$
for all $q$-form $\alpha$.
%%%%%%%%%%%
\end{Lemma}
%%%%%%%%%%
%%%%%%%%%%%%%%%
\begin{Lemma}
\label{2-d-Riemannian-Hodge}
%%%%%%%%%%%%%
Let $(\cZ,g)$ be a $2$-dimensional Riemannian manifold. Then it follows that 
$$
\star\star \alpha
=(-1)^{\,q}\alpha,
$$ 
for all $\alpha\in\GamLam{q}{\cZ}$ with $q\in\{0,\ldots,2\}$.
Thus, the inverse of $\star$ denoted by $\star^{-1}$ is written as  
$$
\star^{\,-1}\alpha
=(-1)^{\,q}\star\alpha
$$
for all $q$-form $\alpha$.
%%%%%%%%%%%
\end{Lemma}
%%%%%%%%%%
%%%%%%%%%%%%%%%
\begin{Lemma}
\label{1-d-Riemannian-Hodge}
%%%%%%%%%%%%%
Let $(\cZ,g)$ be a $1$-dimensional Riemannian manifold. Then it follows that 
$$
\star\star \alpha
=\alpha,
$$ 
for all $\alpha\in\GamLam{q}{\cZ}$ with $q\in\{0,1\}$.
Thus, the inverse of $\star$ denoted by $\star^{-1}$ is the same as $\star$:   
$$
\star^{\,-1}\,\alpha
=\star\,\alpha
$$
for all $q$-form $\alpha$.
%%%%%%%%%%%
\end{Lemma}
%%%%%%%%%%

%%%%%%%%%%%%%%%%%%%%%%%%%
\subsection{Contact geometry}
%%%%%%%%%%%%%%%%%%%%%%%%%%
In this subsection some of basics of contact geometry is summarized.
Roughly speaking, contact geometry is a geometry where a 
class of $1$-forms are provided for manifolds. In this paper 
contact manifolds are used for describing implicit Hamiltonian systems.

%%%%%%%%%%%%
\begin{Def}
\label{definition-contact-manifold}
%%%%%%%%%%
(Contact manifold):  
Let $\cC$ be a $(2n+1)$-dimensional manifold, and $\lambda$ a 
$1$-form on $\cC$ such that
$$
\lambda\wedge\underbrace{\dr \lambda\wedge\dr\lambda\cdots\wedge\dr\lambda}_{n}
\neq 0,
$$
at any point on $\cC$.
If $\cC$ carries $\lambda$, then $(\,\cC,\lambda\,)$ 
is referred to as a contact manifold and $\lambda$ a contact form. 
%%%%%%%%%%
\end{Def}
%%%%%%%%%%
 \begin{Remark}
% %%%%%%%%%%%%
The $(2n+1)$-form $\lambda\wedge\dr\lambda\wedge\cdots\wedge\dr\lambda$ 
can be used for a volume form. 
% %%%%%%%%
 \end{Remark}
%%%%%%%%
It should be noted that there are other definitions of contact manifold. 
However the definition above is used in this paper.  
%The following definition is not commonly used in the literature. 
%%%%%%%%%%
% \begin{Def}
% %%%%%%%%%%%%
% (Standard volume-form) : 
% The $(2n+1)$-form in Definition\,\ref{definition-contact-manifold} 
% is referred to as the standard volume-form :  
% \beq
% \Omega_{\,\lambda}
% =\lambda\wedge\dr\lambda\wedge\cdots\wedge\dr\lambda.
% \label{definition-standard-volume-form-contact-manifold}
% \eeq
% %%%%%%%%%
% \end{Def}
%%%%%%%%%
 
%%%%%%%%

There is a standard local coordinate system.
%%%%%%%%%%%%%
\begin{Thm}
\label{theorem-canonical-coordinates}
%%%%%%%%%%%%
(Existence of particular coordinates): 
Given a contact manifold $(\cC,\lambda)$, 
there exist local $(2n+1)$ coordinates 
$(x,y,z)$ with $x=\{\,x^{\,1},\ldots,x^{\,n}\,\}$ and 
$y=\{\,y_{\,1},\ldots,y_{\,n}\,\}$,  
in which $\lambda$ has the form 
\beq
\lambda
=\dr z-y_{\,a}\,\dr x^{\,a}.
\label{contact-form-standard-Darboux}
\eeq
%%%%%%%%%%
\end{Thm} 
%%%%%%%%%%%
%%%%%%%%%%%%
\begin{Def}
%%%%%%%%%%%%
(Canonical coordinates or Darboux coordinates): 
The $(2n+1)$ coordinates $(x,y,z)$ introduced in 
Theorem \ref{theorem-canonical-coordinates}
are referred to as the canonical coordinates, 
or the Darboux coordinates. 
%%%%%%%%%% 
\end{Def}
%%%%%%%%%%
In addition to the above coordinates,
ones in which $\lambda$ has the 
form $\lambda=\dr z+y_{\,a}\,\dr x^{\,a}$ are also used in the literature.
In this paper \fr{contact-form-standard-Darboux} is used. 

Given a contact manifold, 
there exists a unique vector field that is defined as follows.
%%%%%%%
\begin{Def}
%%%%%%
(Reeb vector field or characteristic vector field):  
Let $(\,\cC,\lambda\,)$ be a contact manifold, and $\Reeb$ a vector field 
on $\cC$. 
If $\Reeb$ satisfies 
\beq
\ii_{\,\Reeb}\, \dr\lambda
=0,\qquad\mbox{and}\qquad 
\ii_{\Reeb}\,\lambda
=1,
\label{definition-Reeb-vector}
\eeq
then $\Reeb\in\GTC$ is referred to as the Reeb vector field, or  
the characteristic vector field.  
%%%%%%%
\end{Def}
%%%%%%

%%%%%%%
% \begin{Remark}
% %%%%%%%
% From the definition of $R$ one has 
% \beq
% \cL_{R}\lambda
% =0,
% \label{remark-Reeb-vector-Lie-derivative}
% \eeq
% where $\cL_{R}$ is the Lie derivative with respect to $R$. 
% To show \fr{remark-Reeb-vector-Lie-derivative},  
% one uses the Cartan formula.
% %%%%%%%
% \end{Remark}
%%%%%%%

%As mentioned, $R$ is uniquely determined 
When $\lambda$ is given, 
a coordinate expression for $\Reeb$ is given as follows.
%%%%%%%
\begin{Proposition}
%%%%%%
(Coordinate expression of the Reeb vector field):  
Let $(\,\cC,\lambda\,)$ be a contact manifold, and $\Reeb$ 
the Reeb vector field, $(x,y,z)$ 
the canonical coordinates  such that 
$\lambda=\dr z-y_{\,a}\,\dr x^{\,a}$  
with $x=\{\,x^{\,1},\ldots,x^{\,n}\,\}$ and $y=\{\,y_{\,1},\ldots,y_{\,n}\,\}$. 
Then the coordinate expression of the Reeb vector field $\Reeb$ is given by 
\beq
\Reeb
=\frac{\partial }{\partial z}.
\label{coordinate-expression-Reeb}
\eeq
%%%%%%%
\end{Proposition}
%%%%%%

The following submanifold plays various roles in applications of  
contact geometry.
%%%%%%%%%
\begin{Def}
\label{definition-Legendre-submanifold}
%%%%%%%
(Legendre submanifold):  
Let $(\,\cC,\lambda\,)$ be a contact manifold, and 
$\cA$  a submanifold of $\cC$. 
If $\cA$ is 
a maximal dimensional integral submanifold of $\lambda$,  
then $\cA$ is referred to as a Legendre submanifold.
%Let $(\,\cC,\lambda\,)$ be a contact manifold, 
%$\cA$  a submanifold of $\cC$, and $\Phi:\cA\to\cC$ an embedding.   
%If $\cA$ is 
%a maximal dimensional integral submanifold such that $\Phi^{\,*}\lambda=0$, 
%then $\cA$ is referred to as a Legendre submanifold.
%%%%%%%%
\end{Def}
%%%%%%%%

The following theorem states the dimension of a Legendre submanifold for 
a given contact manifold.
%%%%%%%%%%%%
\begin{Thm}
\label{thm:max-integral-submanifold-is-n}
(Maximal dimensional integral submanifold):  
On a $(2n+1)$-dimensional contact manifold $(\,\cC,\lambda\,)$,  
a maximal dimensional integral submanifold of $\lambda$  
is equal to $n$.  
%Let $(\,\cC,\lambda\,)$ be a $(2n+1)$-dimensional contact manifold, 
%$\cA$  a submanifold, and $\Phi:\cA\to\cC$ an embedding.  
%The maximal dimensional integral submanifolds such that $\Phi^{\,*}\lambda=0$ 
%is equal to $n$.  
\end{Thm}
%%%%%%%%%%

Combining Theorem \ref{thm:max-integral-submanifold-is-n}
 and Definition \ref{definition-Legendre-submanifold}, 
one concludes the following. 
%%%%%%%%
\begin{Thm} % corollary
%%%%%%%%%
(Dimension of Legendre submanifolds):  
The dimension of any Legendre submanifold of a 
$(2n+1)$-dimensional contact manifold is $n$.
%%%%%%%%
\end{Thm}
%%%%%%%

The following theorem shows
the explicit local expressions of Legendre submanifolds in terms of 
canonical coordinates. 
%%%%%%%%%%%
\begin{Thm}
\label{theorem-Legendre-submanifold-theorem-Arnold}
%%%%%%%%%%% 
(Local expressions of Legendre submanifolds, \cite{Arnold1976}):   
Let $(\,\cC,\lambda\,)$ be a $(2n+1)$-dimensional contact manifold,  
and $(x,y,z)$ the canonical coordinates such that 
$\lambda=\dr z-y_{\,a}\,\dr x^{\,a}$  
with $x=\{\,x^{\,1},\ldots,x^{\,n}\,\}$ and $y=\{\,y_{\,1},\ldots,y_{\,n}\,\}$. 
For any partition $I\cup J$ of the set of indices $\{\,1,\ldots,n\,\}$ into 
two disjoint subsets $I$ and $J$, and for a function $\phi(x^{\,J},y_{\,I})$ of 
$n$ variables $y_{\,i},i\in I$, and $x^{\,j},j\in J$ the $(n+1)$ equations
\beq
x^{\,i}=-\,\frac{\partial\phi}{\partial y_{\,i}},\qquad
y_{\,j}=\frac{\partial\phi}{\partial x^{\,j}},\qquad 
z=\phi-y_{\,i}\frac{\partial\phi}{\partial y_{\,i}},
\label{Legendre-submanifold-theorem-Arnold}
\eeq
define a Legendre submanifold. Conversely, every 
Legendre submanifold of $(\,\cC,\lambda\,)$ 
in a neighborhood of any point is    
defined by these equations for at least one of the $2^{n}$ possible choices 
of the subset $I$.
%%%%%%%%%%%
\end{Thm}
%%%%%%%%%%% 

The function $\phi$ in \fr{Legendre-submanifold-theorem-Arnold} is often used 
in this paper, and then that has a special name as follows.  
%%%%%%%%
\begin{Def}
%%%%%%%%
(Legendre submanifold generated by a function):  
The function $\phi$ used in 
Theorem\,\ref{theorem-Legendre-submanifold-theorem-Arnold}  
is referred to as a generating function of the Legendre submanifold. 
If a  Legendre submanifold $\cA$ is expressed 
as \fr{Legendre-submanifold-theorem-Arnold}, then $\cA$ is referred to as 
a Legendre submanifold generated by $\phi$. 
%%%%%%%%%
\end{Def}
%%%%%%%%%

The following are examples of local expressions for 
 Legendre submanifolds.  
%%%%%%%
\begin{Example}
%%%%%%%
\label{example-Arnold-Legendre-submanifold-psi}
Let $(\,\cC,\lambda\,)$ be a $(2n+1)$-dimensional contact manifold,
$(x,y,z)$ the canonical coordinates such that 
$\lambda=\dr z-y_{\,a}\,\dr x^{\,a}$  
with $x=\{\,x^{\,1},\ldots,x^{\,n}\,\}$ and $y=\{\,y_{\,1},\ldots,y_{\,n}\,\}$, 
and $\psi$ a function of $x$ only.   
The Legendre submanifold $\cA_{\,\psi}$ generated by $\psi$ with 
$\Phi_{\,\cC\cA\psi}:\cA_{\,\psi}\to\cC$ being the embedding 
is such that 
\beq
\Phi_{\,\cC\cA\psi}\cA_{\,\psi}
=\left\{\ (x,y,z)\in\cC \ \bigg|\ 
y_{\,j}=\frac{\partial\psi}{\partial x^{\,j}},\ \mbox{and}\ 
z=\psi(x),\quad j\in \{\,1,\ldots,n\,\}
\ \right\}. 
\label{example-psi-Legendre-submanifold}
\eeq
%The relation between this $\psi$ and $\phi$ of   
%\fr{Legendre-submanifold-theorem-Arnold} is $\psi(x)=\phi(x)$ with 
%$J=\{\,1,\ldots,n\,\}$.
One can verify that $\Phi_{\,\cC\cA\psi}^{\ \ \ \  *}\lambda=0$. 
%%%%%%%
\end{Example}
%%%%%%%
%%%%%
\begin{Example}
%%%%%
\label{example-Arnold-Legendre-submanifold-varphi}
Let $(\,\cC,\lambda\,)$ be a $(2n+1)$-dimensional contact manifold,
$(x,y,z)$ the canonical coordinates such that 
$\lambda=\dr z-y_{\,a}\,\dr x^{\,a}$ 
with $x=\{\,x^{\,1},\ldots,x^{\,n}\,\}$ and $y=\{\,y_{\,1},\ldots,y_{\,n}\,\}$, 
and $\varphi$ a function of $y$ only. 
The Legendre submanifold $\cA_{\,\varphi}$ generated by $-\,\varphi$ 
with $\Phi_{\,\cC\cA\varphi}:\cA_{\,\varphi}\to\cC$ being 
the embedding is such that 
\beq
\Phi_{\,\cC\cA\varphi}\cA_{\,\varphi}
=\left\{\ (x,y,z)\in\cC \ \bigg|\ 
x^{\,i}=\frac{\partial\varphi}{\partial y_{\,i}},\ \mbox{and}\ 
z=y_{\,i}\frac{\partial\varphi}{\partial y_{\,i}}-\varphi(y),\quad i\in 
\{\,1,\ldots,n\,\}
\ \right\}. 
\label{example-varphi-Legendre-submanifold}
\eeq
%The relation between this $\varphi$ and $\phi$ of   
%\fr{Legendre-submanifold-theorem-Arnold} is $\varphi(y)=-\,\phi(y)$ with
%$I=\{\,1,\ldots,n\,\}$.
One can verify that $\Phi_{\,\cC\cA\varphi}^{\ \ \ \ *}\lambda=0$.
%%%%%
\end{Example}
%%%%%

One can choose a function $\psi$ in 
Example\,\ref{example-Arnold-Legendre-submanifold-psi} to generate 
$\cA_{\,\psi}$,  and 
choose a function $\varphi$ in Example \ref{example-Arnold-Legendre-submanifold-varphi} to 
generate $\cA_{\,\varphi}$  independently, 
and in this case there is no relation between 
$\cA_{\,\psi}$ and $\cA_{\,\varphi}$ in general. 
On the other hand, when $\psi$ is strictly convex, and $\varphi$ 
is carefully chosen, it can be shown that 
there is a relation between $\cA_{\,\psi}$ and $\cA_{\,\varphi}$. 
To discuss such a case, 
the following transform should be introduced. The convention is 
adapted to that in information geometry. 
Note that several conventions exist in the literature. 
%%%%%%%%
\begin{Def}
\label{definition-total-Legendre-transform}
%%%%%%%%
(Total Legendre transform):  
Let $\cM$ be an $n$-dimensional manifold,
$x=\{\,x^{\,1},\ldots,x^{\,n}\,\}$ coordinates, 
and $\psi$ a function of $x$.
Then the total Legendre transform of $\psi$ with respect to $x$ 
is defined to be  
\beq
\Leg[\psi](y)
:=\sup_{x}\left[\,x^{\,a}y_{\,a}-\psi(x)\,\right],
%\label{total-Legendre-transform}
\eeq 
where $y=\{\,y_{\,1},\ldots,y_{\,n}\,\}$.
%%%%%%%%
\end{Def}
%%%%%%%%

From this definition, one has several formulae that will be used in the 
following sections. To state these formulae, one defines 
strictly convex function and convexity as follows. 
%%%%%%%%%%%%
\begin{Def}
%%%%%%%%%%
(Strictly convex function and convexity): 
Let $\cA_{\,0}\subseteq\cA$ be a convex domain, and $f$ a function of 
$\{x^{\,a}\}$ on $\cA_{\,0}$. If the matrix 
$$
\frac{\partial^{\,2}\,f}{\partial x^{\,a}\partial x^{\,b} },
$$ 
is strictly positive definite, then the function $f$ is referred to as 
a strictly convex function. In addition, the property that $f$ is 
strictly convex is 
referred to as convexity.
%%%%%%%%%
\end{Def}
%%%%%%%%%

Then one has the following. 
%%%%%%%%%
\begin{Thm}
%%%%%%%%
\label{theorem-Legendre-tranform-formula}
(Formulae involving the total Legendre transform): 
Let $\cM$ be an $n$-dimensional manifold, 
$x=\{\,x^{\,1},\ldots,x^{\,n}\,\}$ coordinates,
$\psi\in\GamLamM{0}$ a strictly convex function of $x$ only, and 
$\varphi$ the function of $y$ obtained by the total Legendre 
transform of $\psi$ with respect to $x$ where $y=\{\,y_{\,1},\ldots,y_{\,n}\,\}$ :
$\varphi(y)=\Leg[\psi](y)$. Then, for each $a$ and fixed $y$, the equation 
$$
y_{\,a}
=\left.\frac{\partial\psi (x)}{\partial x^{\,a}}\right|_{x=x_*}
=\frac{\partial\psi (x_{\,*})}{\partial x_{\,*}^{\,a}},
$$
has the unique solution
$x_{\,*}^{\,a}=x_{\,*}^{\,a}(y), (a\in\{\,1,\ldots,n\,\})$. 
In addition it follows that 
$$
\varphi(y)
=x_{\,*}^{\,a}y_{\,a}-\psi(x_{\,*}),\qquad
\frac{\partial\varphi}{\partial y_{\,a}}
=x_{\,*}^{\,a},\qquad
\delta_{\,b}^{\,a}
=\frac{\partial^2\psi}{\partial x_{\,*}^{\,b}\partial x_{\,*}^{\,l}}
\frac{\partial^2\varphi}{\partial y_{\,a}\partial y_{\,l}},
$$
and 
$$
\det\left(\frac{\partial^2\,\psi}{\partial x^{\,a}\partial x^{\,b}}\right)>0,
\qquad
\det\left(\frac{\partial^2\,\varphi}{\partial y_{\,a}\partial y_{\,b}}\right)>0.
$$
%%%%%%%%%
\end{Thm}
%%%%%%%%

A way to describe dynamics on a contact manifold is to introduce a 
continuous diffeomorphism with a parameter. 
First, one defines a class of diffeomorphisms on contact manifolds.
%%%%%%%%%%5
\begin{Def}
\label{definition-contact-diffeomorphism}
%%%%%%%% 
(Contact diffeomorphism):  
Let $(\,\cC,\lambda\,)$ be a $(2n+1)$-dimensional 
contact manifold, 
and $\Phi:\cC\to\cC$  a diffeomorphism. If it follows that 
$$
\Phi^*\lambda
=f\,\lambda,
$$  
where $f\in\GamLamC{0}$ is a function that does not vanish at  
any point of $\cC$, then the map $\Phi$ is referred to as 
a contact diffeomorphism. 
%%%%%%%%%%5
\end{Def}
%%%%%%%% 

%%%%%%%%
\begin{Remark}
%%%%%%%%%
It follows
that $\Phi$ preserves the contact structure, 
$\ker\lambda:=\{\,X\in\GTC\, |\,\ii_X\lambda=0\,\}$. 
%but does not preserve the original contact form.
%%%%%%%%
\end{Remark}
%%%%%%%%%

In addition to this diffeomorphism, one 
can introduce  one-parameter groups as follows.
%%%%%%%%%%5
\begin{Def}
%%%%%%%% 
(One-parameter group of continuous contact transformations):  
Let $(\,\cC,\lambda\,)$ be a $(2n+1)$-dimensional contact manifold, 
and $\Phi_{t}:\cC\to\cC$ a diffeomorphism with $t\in\mathbb{R}$ that satisfies
$\Phi_0=\Id_{\,\cC}$ and $\Phi_{t+s}=\Phi_{t}\circ\Phi_s, (t,s\in\mathbb{R})$,  
where $\Id_{\,\cC}$ is such that $\Id_{\,\cC}\xi=\xi$ for all $\xi\in\cC$. 
If it follows that  
$$
\Phi_{t}^*\lambda
=f_t\,\lambda,
$$  
where $f_{\,t}\in\GamLamC{0}$ is a function that does not vanish at  
any point of $\cC$, then the $\Phi_{t}$ is referred to as 
a one-parameter group of continuous contact transformations. 
If $t,s\in \mbbT$ with some $\mbbT\subset \mathbb{R}$, 
then $\Phi_{\,t}$ is referred to as 
a one-parameter local transformation group of continuous 
contact transformations.
%%%%%%%%%%5
\end{Def}
%%%%%%%% 
A contact vector field is defined as follows. 
%%%%%%%%%%
\begin{Def}
\label{definition-contact-vector}
%%%%%%%%
(Contact vector field):  
Let $(\,\cC,\lambda\,)$ be a contact manifold, and $X$ a vector field on $\cC$.
If $X$ satisfies 
$$
\cL_X\lambda
=f\,\lambda,
$$
where $f$ is a function on $\cC$, then 
$X$ is referred to as a contact vector field. 
%%%%%%%%%%
\end{Def}
%%%%%%%%

A  one-parameter (local) transformation groups
is realized by integrating the following vector field.
%%%%%%%%
\begin{Def}
%%%%%%%
(Contact vector field associated to a contact Hamiltonian):  
Let $(\,\cC,\lambda\,)$ be a contact manifold, 
$\Reeb$ the Reeb vector field, 
$h$  a function on $\cC$, and 
$X_{\,h}$  a vector field.   
If $X_{\,h}\in\GTC$ satisfies 
\beq
\ii_{\,X_{\,h}}\lambda
=h,\qquad\mbox{and}\qquad 
\ii_{\,X_{\,h}}\dr \lambda
=-\,(\,\dr h-(\,\Reeb h\,)\,\lambda\,),
\label{definition-contact-vector-Hamiltonian}
\eeq
then $X_{\,h}$ 
is referred to as a contact vector field \underline{associated} to 
a function $h$ or a contact Hamiltonian vector field. 
In addition $h$ is referred to as a contact Hamiltonian. 
%%%%%%% 
\end{Def}
%%%%%%
% Note that one cannot interpret contact Hamiltonian vector fields  
% as classical Hamiltonian vector fields on symplectic manifold  in general.  

%%%%%%%%%%%%%%
\begin{Remark}
%%%%%%%%%%%%%
The definition \fr{definition-contact-vector-Hamiltonian} 
and the Cartan formula give 
\beq
\cL_{\,X_{\,h}}\lambda
=(\,\Reeb h\,)\,\lambda.
\label{cartan-contact-Hamiltonian-vector}
\eeq
%%%%%%%%%%%%
\end{Remark}
%%%%%%%%%%%%
%%%%%%%%%%%%%%
\begin{Remark}
%%%%%%%%%%%%%
With \fr{definition-contact-vector-Hamiltonian}, 
\fr{cartan-contact-Hamiltonian-vector}, 
and the formula 
$\cL_{\,X}\ii_{\,X}\alpha=\ii_{\,X}\cL_{\,X}\alpha$  
for an arbitrary vector field $X$  and  an arbitrary $q$-form 
$\alpha$ with $q\in\{0,1,\ldots\}$,  
one has that 
\beq
X_{\,h}h
=\cL_{\,X_{\,h}}h
=\cL_{\,X_{\,h}}(\,\ii_{\,X_{\,h}}\lambda\,)
=\ii_{\,X_{\,h}}(\,\cL_{\,X_{\,h}}\lambda\,)
=(\,\Reeb\,h\,)(\,\ii_{\,X_{\,h}}\lambda\,)
=(\,\Reeb\,h\,)\,h.
\label{Xh=hRh}
\eeq
Thus, unlike the case of autonomous Hamiltonian systems, $h$ is not conserved. 
%%%%%%%%%%%%
\end{Remark}
%%%%%%%%%%%%

From \fr{cartan-contact-Hamiltonian-vector} and 
Definition\,\ref{definition-contact-vector}, one has the following.  
%%%%%%%
\begin{Thm}
%%%%% 
(Relation between a contact Hamiltonian vector field and a 
contact vector field):    
Let $(\,\cC,\lambda\,)$ be a contact manifold, $h$ a contact Hamiltonian,  
and $X_h$ a contact Hamiltonian vector field. 
Then $X_h$ is a contact vector field. 
%%%%%%%
\end{Thm}
%%%%% 

Local expressions of a contact Hamiltonian vector field
\fr{definition-contact-vector-Hamiltonian} are  calculated 
as follows.
%%%%%%%
\begin{Proposition}
%%%%% 
(Local expression of contact Hamiltonian vector field):  
Let $(\,\cC,\lambda\,)$ be a $(2n+1)$-dimensional contact manifold, 
$h$ a contact Hamiltonian,  
$X_h$ a contact Hamiltonian vector field, and $(x,y,z)$ 
the canonical coordinates 
such that $\lambda=\dr z-y_{\,a}\,\dr x^{\,a}$ 
with $x=\{\,x^{\,1},\ldots,x^{\,n}\,\}$ and $y=\{\,y_{\,1},\ldots,y_{\,n}\,\}$. 
Then  
$$
X_h
=\dot{x}^{\,a}\frac{\partial}{\partial x^{\,a}}
+\dot{y}_{\,a}\frac{\partial}{\partial y_{\,a}}
+\dot{z}\frac{\partial}{\partial z},
$$
where $\dot{}$ denotes the differential 
with respect to a parameter $t\in\mathbb{R}$, or $t\in \mbbT$ 
with some $\mbbT\subset\mathbb{R}$, 
and 
\beq
\dot{x}^{\,a}
=-\,\frac{\partial h}{\partial y_{\,a}},\qquad
\dot{y}_{\,a}
=\frac{\partial h}{\partial x^{\,a}}
+y_{\,a}\frac{\partial h}{\partial z},\qquad
\dot{z}
=h-y_{\,a}\frac{\partial h}{\partial y_{\,a}},\qquad a\in\{\,1,\ldots,n\,\}.
\label{contact-Hamiltonian-vector-components}
\eeq
%%%%%%%
\end{Proposition}
%%%%% 
\begin{Remark}
%%%%%%%%%%%%%%
When $\lambda=\dr z+y_{\,a}\,\dr x^{\,a}$,  signs in   
\fr{contact-Hamiltonian-vector-components} are changed. 
%%%%%%%%%%%%
\end{Remark}
%%%%%%%%%%
Let $\phi_{\,h}$ be an integral curve of $X_{\,h}$ such that 
$\phi_{\,h}:\mbbT\to\cC, (t\mapsto (x,y,z))$ with some $\mbbT\subseteq \mbbR$. 
Then $\dot{}$ denotes the derivative with respect to $t\in\mbbT$. 
Physically this $t$  is interpreted as time. In this case contact vector fields 
can be viewed as dynamical systems. 
Throughout this paper, vector 
fields are always identified with dynamical systems, and integral curves are 
focused when vector fields are given. 

The following theorem is well-known, and has been 
used in the literature of geometric thermodynamics.  
%%%%%%%%%%%%%
\begin{Thm}
\label{theorem-Mrugala-contact-Hamiltonian}
%%%%%%%%%%%%
(Tangent vector field of a Legendre submanifold realized 
by a contact Hamiltonian vector field, \cite{Mrugala1991}):   
Let $(\,\cC,\lambda\,)$ be a contact manifold, $\cA$ a Legendre submanifold, and 
$h$ a contact Hamiltonian. Then 
the contact Hamiltonian vector field is tangent to $\cA$ if and only if 
$h$ vanishes on $\cA$. 
%%%%%%%%%%%
\end{Thm}
%%%%%%%%%%

Introducing some symbols, 
one has other equivalent expressions for the Legendre submanifolds  
\fr{example-psi-Legendre-submanifold} and 
\fr{example-varphi-Legendre-submanifold}.
The following functions were introduced in Ref.\,\cite{Goto2016}, and  
it was shown that the introduced functions are tools to describe 
contact Hamiltonian vector fields concisely. They are as follows.
%%%%%%%%%%%%%%
\begin{Def}
\label{definition-adapted-function}
%%%%%%%%%%%%
(Adapted functions,\,\cite{Goto2016}):  
Let $(\cC,\lambda)$ be a $(2n+1)$-dimensional contact manifold, and $(x,y,z)$ 
canonical coordinates such that $\lambda=\dr z-y_{\,a}\,\dr x^{\,a}$ with 
$x=\{x^{\,1},\ldots,x^{\,n}\}$ and $y=\{y_{\,1},\ldots,y_{\,n}\}$. In addition let 
$\psi$ be a 
function on $\cC$ depending on $x$ only,   
and $\varphi$ a function on $\cC$ depending on $y$ only. Then
the functions 
$\Delta_{\,0}^{\,\psi},\{\Delta_{\,1}^{\,\psi},\ldots,\Delta_{\,n}^{\,\psi}\}:\cC\to\mbbR$
and 
$\Delta_{\,\varphi}^{\,0},\{\Delta_{\,\varphi}^{\,1},\ldots,\Delta_{\,\varphi}^{\,n}\}:\cC\to\mbbR$ 
such that 
$$
\Delta_{\,0}^{\,\psi}(x,z)
:=\psi(x)-z,\quad 
\Delta_{\,a}^{\,\psi}(x,y)
:=\frac{\partial\psi}{\partial x^{\,a}}-y_{\,a},\qquad 
a\in\{\,1,\ldots,n\,\}.
$$
$$
\Delta_{\,\varphi}^{\,0}(x,y,z)
:=x^{\,j}y_{\,j}-\varphi(y)-z,\quad 
\Delta_{\,\varphi}^{\,a}(x,y)
:=x^{\,a}-\frac{\partial\varphi}{\partial y_{\,a}},\qquad 
a\in\{\,1,\ldots,n\,\}.
$$  
are referred to as adapted functions. 
%%%%%%%%%%
\end{Def}
%%%%%%%%%%% 
In adapted functions, the local expressions of Legendre submanifolds
generated by $\psi$ and those by $-\varphi$ can be written as 
follows\cite{Goto2016}. 
%%%%%%%%%%%%%%%%%%%%
\begin{Proposition}
\label{proposition-Legendre-submanifold-adapted-functions}
%%%%%%%%%%%%%%%%%%%%%%
(Local expressions of Legendre submanifold with adapted functions,\, 
\cite{Goto2016}): 
The Legendre submanifold $\cA_{\,\psi}$ generated by $\psi$ as in 
\fr{example-psi-Legendre-submanifold}
is expressed as 
$$
\Phi_{\,\cC\cA\psi}\cA_{\,\psi}
=\left\{\ (x,y,z)\in\cC \ |\ 
\Delta_{\,0}^{\,\psi}
=0\ \mbox{and}\ 
\Delta_{\,1}^{\,\psi}
=\cdots
=\Delta_{\,n}^{\,\psi}
=0
\ \right\}, 
$$
where 
$\Phi_{\,\cC\cA\psi}\cA_{\,\psi}:\cA_{\,\psi}\to\cC$ is the embedding. Similarly, 
the Legendre submanifold $\cA_{\,\varphi}$ generated by $-\varphi$ as in 
\fr{example-varphi-Legendre-submanifold}
is expressed as 
$$
\Phi_{\,\cC\cA\varphi}\cA_{\,\varphi}
=\left\{\ (x,y,z)\in\cC \ |\ 
\Delta_{\,\varphi}^{\,0}
=0\ \mbox{and}\ 
\Delta_{\,\varphi}^{\,1}
=\cdots
=\Delta_{\,\varphi}^{\,n}
=0
\ \right\}, 
$$
where $\Phi_{\,\cC\cA\varphi}\cA_{\,\varphi}:\cA_{\,\varphi}\to\cC$ is the embedding. 
%%%%%%%%%%%%%%%%%%5
\end{Proposition} 
%%%%%%%%%%%%%%%%%%
From this proposition, a Legendre submanifold $\Phi_{\cC\cA\psi}\cA_{\,\psi}$ 
is a submanifold where the 
constraints $\Delta_{\,0}^{\,\psi}=\cdots=\Delta_{\,n}^{\,\psi}$ hold.  
Vector fields on $\Phi_{\cC\cA\psi}\cA_{\,\psi}$ 
%is the one 
%where relations $\Delta_{\,0}^{\,\psi}=\cdots=\Delta_{\,n}^{\,\psi}$ hold.
%This kind of a vector field 
can be 
constructed with a restricted contact Hamiltonian vector field.
It was shown in Ref.\cite{Goto2016} 
that contact Hamiltonian vector fields are 
also written in terms of adapted functions.

%%%%%%%%%%%%%%%%%%5
\begin{Proposition} 
\label{Mrugala-variant-psi}
%%%%%%%%%%%%
(Restricted contact Hamiltonian vector field 
as the push-forward of a vector field on 
the Legendre submanifold generated by $\psi$,\, 
\cite{Goto2016}): 
Let $\{\,F_{\,\psi}^{\,1},\ldots,F_{\,\psi}^{\,n}\,\}$ 
be a set of functions of $x$ on $\cA_{\,\psi}$ 
such that they do not identically vanish,   
and $\chX_{\,\psi}^{\,0}\in T_{\,x}\,\cA_{\,\psi}, ( x\in \cA_{\,\psi})$ 
the vector field given as   
$$
\chX_{\,\psi}^{\,0}
=\dot{x}^{\,a}\frac{\partial}{\partial x^{\,a}},\quad\mbox{where}\quad 
\dot{x}^{\,a}
=F_{\,\psi}^{\,a}(x),\qquad 
(a\in\{\,1,\ldots,n\,\}).
$$
In addition, let $X_{\,\psi}^{\,0}:=(\,\Phi_{\,\cC\cA\psi}\,)_{*}\chX_{\,\psi}^{\,0} 
\in T_{\,\xi}\cA_{\,\psi}^{\,\cC}, (\,\xi\in\cA_{\,\psi}^{\,\cC}\,)$ 
be the push-forward of
$\chX_{\,\psi}^{\,0}$,   
where $\cA_{\,\psi}^{\,\cC}:=\Phi_{\,\cC\cA\psi}\cA_{\,\psi}$ with   
$\Phi_{\,\cC\cA\psi}:\cA_{\,\psi}\to \cC$ being the embedding :  
%( see the diagrams below ). 
\beqa
\Phi_{\,\cC\cA\psi} &:& \cA_{\,\psi}\to 
\cA_{\,\psi}^{\,\cC},\qquad\qquad x\mapsto (\,x,y(x),z(x)\,)
\non\\
(\,\Phi_{\,\cC\cA\psi}\,)_{\,*} &:& T_{\,x}\,\cA_{\,\psi}\to  
T_{\,\xi}\,\cA_{\,\psi}^{\,\cC},\quad \chX_{\,\psi}^{\,0}\mapsto
X_{\,\psi}^{\,0}\,.
\non
\eeqa
Then it follows that   
\beq
X_{\,\psi}^{\,0}
=\dot{x}^{\,a}\frac{\partial}{\partial x^{\,a}}
+\dot{y}_{\,a}\frac{\partial}{\partial y_{\,a}}
+\dot{z}\frac{\partial}{\partial z},\quad\mbox{where}\quad 
\dot{x}^{\,a}
=F_{\,\psi}^{\,a}(x),\quad
\dot{y}_{\,a}
=\frac{\dr}{\dr t}\left(\,\frac{\partial\psi}{\partial x^{\,a}}\,\right),
\quad 
\dot{z}
=\frac{\dr \psi}{\dr t}.
\label{tangent-vector-Legendre-submanifold-psi-component}
\eeq
In addition, 
one has that $X_{\,\psi}^{\,0}=X_{\,h_{\,\psi}}|_{\,h_{\,\psi}=0}$. Here 
$X_{\,h_{\,\psi}}$ is the contact Hamiltonian vector field associated with  
\beq
h_{\,\psi}(x,y,z)
=\Delta_{\,a}(x,y) F_{\,\psi}^{\,a}(x) 
+\Gamma_{\,\psi}(\,\Delta_{\,0}(x,z)\,),
\label{tangent-vector-Legendre-submanifold-psi-Hamiltonian}
\eeq
where $\Gamma_{\,\psi}$ is a function of $\Delta_{\,0}$ such that 
$$
\Gamma_{\,\psi}(\,\Delta_{\,0}\,)
=\left\{
\begin{array}{cl}
0&\mbox{for}\quad\Delta_{\,0}=0\\
\mbox{non-zero}&\mbox{for}\quad \Delta_{\,0}\neq 0
\end{array}
\right..
% \quad\mbox{and}\qquad
% \left.\frac{\dr}{\dr \Delta_{\,0}}\right|_{\Delta_{\,0}=0}
% \Gamma_{\,\psi}(\,\Delta_{\,0}\,)\neq 0.
$$ 
%%%%%%%%%%%
\end{Proposition}
%%%%%%%%%%
%%%%%%%%%%%
\begin{Remark}
%%%%%%%%%%%%%%
The functions $\{\,F_{\,\psi}^{\,1},\ldots,F_{\,\psi}^{\,n}\,\}$ 
need not depend on $\psi$. 
%%%%%%%%%%%%%%
\end{Remark}
%%%%%%%%%%%%%
%%%%%%%%%%%
\begin{Remark}
\label{remark-dually-flat-space-psi-not-preserved}
%%%%%%%%%%%%%
The value of the generating function  
$\psi$ is not conserved along this restricted 
contact Hamiltonian vector field, since 
$$
\cL_{X_{\,\psi}^{\,0}}\psi
=\frac{\partial\psi}{\partial x^{\,a}}\frac{\dr x^{\,a}}{\dr t}
=\frac{\partial\psi}{\partial x^{\,a}}\,F_{\,\psi}^{\,a},
$$
does not identically vanish in general.
In some cases, this vanishes. Consider the system with $n=2$, and 
$F_{\,\psi}^{\,1}=\partial\psi/\partial x^{\,2},F_{\,\psi}^{\,2}=-\,\partial\psi/\partial x^{\,1} $. Then $\cL_{X_{\,\psi}^{\,0}}\psi=0$.
%%%%%%%%%%%
\end{Remark}
%%%%%%%%%%%%%

There exists a counterpart of Proposition\,\ref{Mrugala-variant-psi} as follows.
%%%%%%%%%%%
\begin{Proposition}
\label{Mrugala-variant-varphi}
%%%%%%%%%%%%
(Restricted contact Hamiltonian vector field as the push-forward of 
vector fields on the Legendre submanifold generated by $-\,\varphi$,\, 
\cite{Goto2016}): 
Let $\{\,F_{\,1}^{\,\varphi},\ldots,F_{\,n}^{\,\varphi}\,\}$ 
be a set of functions of $y$ on $\cA_{\,\varphi}$  
such that they do not identically vanish,   
and $\chX_{\,\varphi}^{\,0}\in T_{\,y}\cA_{\,\varphi}, ( y\in\cA_{\,\varphi} )$ 
given as   
$$
\chX_{\,\varphi}^{\,0}
=\dot{y}_{\,a}\frac{\partial}{\partial y_{\,a}},\quad\mbox{where}\quad 
\dot{y}_{\,a}
=F_{\,a}^{\,\varphi}(y).
$$
In addition, let $X_{\,\varphi}^{\,0}:=(\,\Phi_{\,\cC\cA\varphi}\,)_{*}\chX_{\,\varphi}^{\,0} 
\in T_{\,\xi}\cA_{\,\varphi}^{\,\cC}, (\,\xi\in\cA_{\,\varphi}^{\,\cC}\,)$ 
be the push-forward of
$\chX_{\,\varphi}^{\,0}$,   
where $\cA_{\,\varphi}^{\,\cC}:=\Phi_{\,\cC\cA\varphi}\cA_{\,\varphi}$ with   
$\Phi_{\,\cC\cA\varphi}:\cA_{\,\varphi}\to \cC$ being the embedding : 
\beqa
\Phi_{\,\cC\cA\varphi} &:& \cA_{\,\varphi}\to 
\cA_{\,\varphi}^{\,\cC},\qquad\qquad y\mapsto (\,x(y),y,z(y)\,)
\non\\
(\,\Phi_{\,\cC\cA\varphi}\,)_{\,*} &:& T_{\,y}\,\cA_{\,\varphi}\to  
T_{\,\xi}\,\cA_{\,\varphi}^{\,\cC},\quad \chX_{\,\varphi}^{\,0}\mapsto
X_{\,\varphi}^{\,0}\,.
\non
\eeqa
Then it follows that   
\beq
X_{\,\varphi}^{\,0}
=\dot{x}^{\,a}\frac{\partial}{\partial x^{\,a}}
+\dot{y}_{\,a}\frac{\partial}{\partial y_{\,a}}
+\dot{z}\frac{\partial}{\partial z},\quad\mbox{where}\quad 
\dot{x}_{\,a}
=\frac{\dr}{\dr t}\left(\,\frac{\partial\varphi}{\partial y_{\,a}}\,\right),
\quad 
\dot{y}^{\,a}
=F_{\,a}^{\,\varphi}(y),\quad
\dot{z}
=y_{\,j}F_{\,k}^{\,\varphi}\frac{\partial^2\,\varphi}{\partial y_{\,k}\partial y_{\,j}}.
\label{tangent-vector-Legendre-submanifold-varphi-component}
\eeq
In addition, one has that $X_{\,\varphi}^{\,0}=X_{\,h_{\,\varphi}}|_{\,h_{\,\varphi}=0}$. 
Here $X_{\,h_{\,\varphi}}$ is the contact Hamiltonian vector field associated with 
\beq
h_{\,\varphi}(x,y)
=\Delta^{\,a}(x,y) F_{\,a}^{\,\varphi}(y)
+\Gamma^{\,\varphi}(\,\Delta^{\,0}\,),
\label{tangent-vector-Legendre-submanifold-varphi-Hamiltonian}
\eeq
where $\Gamma^{\,\varphi}$ is a function of $\Delta^{\,0}$ such that 
$$
\Gamma^{\,\varphi}(\,\Delta^{\,0}\,)
=\left\{
\begin{array}{cl}
0&\mbox{for}\quad\Delta^{\,0}=0\\
\mbox{non-zero}&\mbox{for}\quad \Delta^{\,0}\neq 0
\end{array}
\right..
%\quad\mbox{and}\qquad
% \left.\frac{\dr}{\dr \Delta^{\,0}}\right|_{\Delta^{\,0}=0}
% \Gamma^{\,\varphi}(\,\Delta^{\,0}\,)\neq 0.
$$ 
%%%%%%%%%%%
\end{Proposition}
%%%%%%%%%%%
%%%%%%%%%%%
\begin{Remark}
%%%%%%%%%%%%%%%
The functions $\{\,F_{\,1}^{\,\varphi},\ldots,F_{\,n}^{\,\varphi}\,\}$   
need not depend on $\varphi$. 
%%%%%%%%%%%%%
\end{Remark}
%%%%%%%%%%%%
\begin{Remark}
\label{remark-dually-flat-space-varphi-not-preserved}
%%%%%%%%%%%%%
The value of the generating function $\varphi$  
is not conserved along this restricted 
contact Hamiltonian vector field, since 
$$
\cL_{X_{\,\varphi}^{\,0}}\varphi
=\frac{\partial\varphi}{\partial y_{\,a}}\frac{\dr y_{\,a}}{\dr t}
=\frac{\partial\varphi}{\partial y_{\,a}}\,F_{\,a}^{\,\varphi},
$$
does not identically vanish in general.
In some cases, this vanishes. Consider the system with $n=2$, and 
$F_{\,1}^{\,\varphi}=\partial\varphi/\partial y_{\,2},F_{\,2}^{\,\varphi}=-\,\partial\varphi/\partial y_{\,1} $. Then $\cL_{X_{\,\varphi}^{\,0}}\varphi=0$.
%%%%%%%%%%%
\end{Remark}
%%%%%%%%%%%%%
%%%%%%%%%%%%%%%%%%%%%%%%%%%%%%%%%%%%%%%%%%%%%%%%%%%
\subsection{Contact manifold over base space}
%%%%%%%%%%%%%%%%%%%%%%%%%%%%%%%%%%%%%%%%%%%%%%%%%%
In this paper it is shown that 
distributed port-Hamiltonian systems are written 
in terms of a contact manifold over some base space, and 
the contact manifold over a base space is treated as a bundle.  
To this end, some basic definitions and theorems obtained 
from the standard contact geometry are summarized below. 
The discussions in this subsection are based on Ref.\,\cite{Goto2017}.
%%%%%%%%%%%%
\begin{Def}
%%%%%%%%%%
(Contact manifold over base space or contact bundle): 
Let $\cB$ be a $d_{\cB}$-dimensional  manifold, 
$(\cK,\pi,\cB)$ a bundle over the base space $\cB$,
the fiber space $\pi^{-1}(\zeta)$ at a point $\zeta$ of $\cB$ 
a $(2n+1)$-dimensional manifold $\cC_{\zeta}$, 
$\cK=\bigcup_{\zeta\in\cB}\cC_{\zeta}$, and 
the structure group $G$ a contact transformation group.
If $\cK$ carries a vertical form $\lambda_{\,\mbbV}$ 
such that
$$
\lambda_{\,\mbbV}\wedge\underbrace{\dr_{\,\mbbV} \lambda_{\,\mbbV}\wedge
\cdots\wedge\dr_{\,\mbbV}
\lambda_{\,\mbbV}}_{n}
\neq 0,\qquad\mbox{at each point of $\pi^{-1}(\zeta)$ at each point $\zeta$ of 
$\cB$} 
$$
then 
$\cC_{\,\zeta}$ is referred to as a ($(2n+1)$-dimensional) contact manifold 
on the fiber space $\pi^{-1}(\zeta),(\zeta\in\cB)$, 
the quadruplet $(\cK,\lambda_{\,\mbbV},\pi,\cB)$ 
is referred to as a ($(2n+1)$-dimensional) 
contact manifold over the base space $\cB$ or a contact bundle,  
and $\lambda_{\,\mbbV}$ a contact vertical form.
%%%%%%%%%%
\end{Def}
%%%%%%%%%

In this paper trivial bundles are only considered,   
then the transition functions are always identical since this simple case is  
enough for our contact geometric formulation of 
distributed-parameter port-Hamiltonian systems.   
The contact geometry of the vertical space 
is the same as the standard contact geometry.    
Thus, all of the definitions and theorems for the standard contact geometry 
can be brought to vertical spaces. 
They are shown below.

At each base point $\zeta$ of $\cB$, 
one has Darboux's theorem for $\pi^{-1}(\zeta)$. Therefore one has the 
following.
%%%%%%%%%
\begin{Thm}
%%%%%%%%%%
\label{theorem-Darboux-bundle}
(Existence of Darboux coordinates on fiber space):
For the $(2n+1)$-dimensional contact manifold over a base space 
$(\cK,\lambda_{\,\mbbV},\pi,\cB)$, 
there exist local coordinates $(x,y,z)$ for $\pi^{-1}(\zeta)$ 
with $x=\{\,x^{\,1},\ldots,x^{\,n}\,\}$ and 
$y=\{\,y_{\,1},\ldots,y_{\,n}\,\}$ 
in which $\lambda_{\,\mbbM}^{\,q}\in\Gamma\Lambda_{\mbbH,\mbbV}^{q,1}\cK$ 
has the form
$$
\lambda_{\,\mbbM}^{\,q}
=\rho^{\,q}\wedge\lambda_{\,\mbbV},\qquad\mbox{where}\qquad
\lambda_{\,\mbbV}
=\dr_{\,\mbbV}z-y_{\,a}\dr_{\,\mbbV}x^{\,a},
$$  
with some $\rho^{\,q}\in\Gamma\Lambda_{\,\mbbH}^{\,q}\cK$ being nowhere vanishing.
%%%%%%%%%%
\end{Thm}
%%%%%%%%
%%%%%%%%%%%%
\begin{Def}
%%%%%%%%%%% 
(Canonical coordinates or Darboux coordinates): 
The $(2n+1)$ coordinates introduced in Theorem\,\ref{theorem-Darboux-bundle}
are referred to as the 
canonical coordinates for a fiber space 
or the Darboux coordinates for a fiber space.  
%%%%%%%%%%%
\end{Def}
%%%%%%%%%

%%%%%%%%%
\begin{Def}
%%%%%%%%%%
(Canonical contact mixed form and canonical contact vertical form):
Let $(\cK,\lambda_{\,\mbbV},\pi,\cB)$  be a contact manifold over  
a base space $\cB$, and $\rho^{\,q}$ a nowhere vanishing horizontal $q$-form. 
A mixed $(q,1)$-form $\lambda_{\,\mbbM}^{\,q}\in\Gamma\Lambda_{\,\mbbH,\mbbV}^{\,q,1}\cK$ 
written as 
$$
\lambda_{\,\mbbM}^{\,q}
=\rho^{\,q}\wedge\lambda_{\,\mbbV}\,,
\qquad\mbox{where}\quad
\lambda_{\,\mbbV}
=\dr_{\,\mbbV} z-y_{\,a}\dr_{\,\mbbV} x^{\,a},
$$
is referred to as the canonical contact mixed $(q,1)$-form associated with 
$\rho^{\,q}$, and $\lambda_{\,\mbbV}\in\Gamma\Lambda_{\,\mbbV}^{\,1}\cK$ 
the canonical contact vertical form.
%%%%%%%%%%
\end{Def}
%%%%%%%%
%%%%%%%%%%%%%%%%
%\begin{Remark}
%%%%%%%%%%%%%%%%5
%The canonical contact mixed $(q,1)$-form associated with $\rho^{\,q}$ is 
%a contact mixed $(q,1)$-form. 
%%%%%%%%%%%%%%%%
%\end{Remark}
%%%%%%%%%%%%%%%%5
%%%%%%%%%%%%%%%%
% \begin{Remark}
% %%%%%%%%%%%%%%%%5
% With the canonical contact vertical form 
% $\lambda_{\,\mbbV}\in\Gamma\Lambda_{\,\mbbV}^{\,1}\cK$ and a nowhere vanishing 
% horizontal $d_{\,\cB}$-form,  
% the mixed $(d_{\,\cB},2n+1)$-form 
% $$
% \rho^{\,d_{\,\cB}}\wedge\lambda_{\,\mbbV}\wedge 
% \underbrace{\dr_{\,\mbbV}\lambda_{\,\mbbV}\wedge\cdots\wedge
% \dr_{\,\mbbV}\,\lambda_{\,\mbbV}}_{n} 
% \qquad\in\Gamma\Lambda_{\,\mbbH,\mbbV}^{\,d_{\,\cB},2n+1}\cK
% $$
% is a volume-form on $\cK$.
% %%%%%%%%%%%5
% \end{Remark}
%%%%%%%%%%%%

%%%%%%%%%%%%%
\begin{Def}
%%%%%%%%%%%
(Reeb vertical vector field): 
Let $(\cK,\lambda_{\,\mbbV},\pi,\cB)$
be a contact manifold over a base space $\cB$, 
%$\rho^{\,q}$ a nowhere vanishing horizontal $q$-form, 
and $\cR_{\,\mbbV}$ a vertical vector field on $\cK$. If 
$\cR_{\,\mbbV}$ satisfies
$$
\ii_{\,\cR_{\,\mbbV}}\lambda_{\,\mbbV}
=1,\qquad\mbox{and}\quad
\ii_{\,\cR_{\,\mbbV}}\,\dr_{\,\mbbV}\lambda_{\,\mbbV}
=0,
$$
then $\cR_{\,\mbbV}$ is referred to as the Reeb 
vertical vector field on $\cK$. % associated with $\rho^{\,q}$.
%%%%%%%%%%
\end{Def}
%%%%%%%%%%
\begin{Proposition}
%%%%%%%%%%%%%
(Coordinate expression of the Reeb vertical vector field): 
Let 
$(\cK,\lambda_{\,\mbbV},\pi,\cB)$
be a contact manifold over a base space $\cB$, 
%$\rho^{\,q}$ a horizontal $q$-form, 
$\cR_{\,\mbbV}$ the Reeb vertical vector field on $\cK$, and $(x,y,z)$ the 
canonical coordinates for the fiber space such that $\lambda_{\,\mbbV}=\dr_{\mbbV}z-y_{\,a}\dr_{\,\mbbV}x^{\,a}$.
%The canonical 
%If $\lambda_{\,\mbbV}$ is the canonical contact form,  
%associated with $\rho^{\,q}$
Then the coordinate expression of the Reeb vertical vector field $\Reeb_{\mbbV}$ 
is 
$$
\cR_{\,\mbbV}
=\frac{\partial}{\partial z}.
$$
%%%%%%%%%%%
\end{Proposition}
%%%%%%%%%%

%%%%%%%%%%%%
\begin{Def}
%%%%%%%%%%5 
(Contact Hamiltonian vertical vector field): 
Let 
$(\cK,\lambda_{\mbbV},\pi,\cB)$
be a contact manifold over a base space $\cB$ with $\dim\cB=d_{\cB}$,  
$\cB_{\,0}\subseteq\cB$ a $d_{\cB}$-dimensional space,
$\rho^{\,d_{\cB}}\in\Gamma\Lambda_{\mbbH}^{d_{\cB}}\cK$ a nowhere vanishing 
horizontal form, 
$\cR_{\mbbV}$ the Reeb vertical vector field on $\cK$,  
$\wt{h}\in\Gamma F\cK$ the functional given by   
$$
\wt{h}
=\int_{\cB_{\,0}}h\,\rho^{\,d_{\cB}},
$$  
with some $h\in\Gamma\Lambda_{\,\mbbV}^{\,0}\cK$, and   
$X_{\,\wt{h}}$ a vertical vector field on $\cK$. 
If $X_{\,\wt{h}}$ satisfies 
\beq
\ii_{\,X_{\,\wt{h}}}\lambda_{\mbbV}
=h
\quad\mbox{and}\quad 
\ii_{\,X_{\,\wt{h}}}\,\dr_{\,\mbbV}\lambda_{\,\mbbV}
=-\,(\,\dr_{\,\mbbV}\, h-(\,\cR_{\,\mbbV}\,h\,)\,
\lambda_{\,\mbbV}\,),
\label{conditions-for-contact-Hamiltonian-vector-bundle}
\eeq
then $X_{\,\wt{h}}$ is referred to as 
the contact Hamiltonian vertical vector field, 
$\wt{h}$ a contact Hamiltonian functional, and 
$h$ a contact Hamiltonian vertical function.
%%%%%%%%%%
\end{Def}
%%%%%%%%%
%%%%%%%%%%%%%%%
\begin{Remark}
%%%%%%%%%%%%%%
With the Cartan formula, one has that 
$\cL_{\,X_{\,\wt{h}}}\lambda_{\,\mbbV}=(\,\cR_{\,\mbbV}h\,)\lambda_{\,\mbbV}$. Thus,   
$\cL_{\,X_{\,\wt{h}}}\lambda_{\,\mbbM}^{\,d_{\cB}}=\rho^{\,d_{\,\cB}}\wedge\cL_{\,X_{\,\wt{h}}}\lambda_{\,\mbbV}
=(\,\cR_{\,\mbbV}h\,)\lambda_{\,\mbbM}^{\,d_{\cB}}$.  
%%%%%%%%%%%%
\end{Remark}
%%%%%%%%%%%

In the following the coordinate expression of 
a contact Hamiltonian vertical vector field is shown.
%%%%%%%%%%%%%%%%%%
\begin{Proposition}
%%%%%%%%%%%%%%%%%%%%%%
(Coordinate expression of a contact vertical Hamiltonian vector field):
Let 
$(\cK,\lambda_{\,\mbbV},\pi,\cB)$
be a contact manifold over a  base space $\cB$ with $\dim\cB=d_{\,\cB}$,  
$\cB_{\,0}\subseteq\cB$ a $d_{\,\cB}$-dimensional space, 
$\rho^{\,d_{\,\cB}}\in\Gamma\Lambda_{\mbbH}^{\,d_{\cB}}\cK$ 
a nowhere vanishing form, $(x,y,z)$ 
the canonical coordinates for the fiber space such that 
%$\lambda_{\,\mbbM}^{\,d_{\cB}}=\rho^{\,d_{\cB}}\wedge\lambda_{\,\mbbV}$, 
$\lambda_{\,\mbbV}=\dr_{\,\mbbV}z-y_{\,a}\dr_{\,\mbbV}x^{\,a}$ with 
$x=\{\,x^{\,1},\ldots,x^{\,n}\,\}$ and $y=\{\,y_{\,1},\ldots,y_{\,n}\,\}$, 
$\wt{h}$ the contact Hamiltonian functional given by 
$$
\wt{h}
=\int_{\cB_{\,0}}h\,\rho^{\,d_{\,\cB}},
$$    
with some $h\in\Gamma\Lambda_{\,\mbbV}^{\,0}\cK$ depending on $(x,y,z)$, and
$X_{\,\wt{h}}$ the contact Hamiltonian vertical vector field on $\cK$. 
  
Then, the canonical coordinate expression of 
\fr{conditions-for-contact-Hamiltonian-vector-bundle} is given as 
$$
X_{\,\wt{h}}
=\dot{x}^{\,a}\frac{\partial}{\partial x^{\,a}}
+\dot{y}_{\,a}\frac{\partial}{\partial y_{\,a}}
+\dot{z}\frac{\partial}{\partial z},
$$
where 
% $$
% \dot{x}^{\,a}
% =-\,\frac{\delta \cH_{\mbbF}^{d} }{\delta p_{\,a}},\qquad 
% \dot{p}_{\,a}
% =\frac{\delta \cH_{\mbbF}^{q} }{\delta x^{\,a}}
% +p_{\,a}\frac{\delta \cH_{\mbbF}^{q}}{\delta z},
% \qquad 
% \dot{z}
% =\cH_{\mbbF}^{d}-p_{\,a}\frac{\delta \cH_{\mbbF}^{d}}{\delta p_{\,a}}.
% $$
% or equivalently, 
\beq
\dot{x}^{\,a}
=-\,\frac{\partial h }{\partial y_{\,a}},\qquad 
\dot{y}_{\,a}
=\frac{\partial h }{\partial x^{\,a}}
+y_{\,a}\frac{\partial h}{\partial z},
\qquad 
\dot{z}
=h-y_{\,a}\frac{\partial h}{\partial y_{\,a}},
\label{coordinate-expression-contact-Hamiltonian-vertical-vector}
\eeq
or equivalently, 
$$
\dot{x}^{\,a}
=-\,\frac{\delta \wt{h} }{\delta y_{\,a}},\qquad 
\dot{y}_{\,a}
=\frac{\delta \wt{h} }{\delta x^{\,a}}
+y_{\,a}\frac{\delta\wt{h} }{\delta z},
\qquad 
\dot{z}
=h-y_{\,a}\frac{\delta\wt{h}}{\delta y_{\,a}}.
$$
%%%%%%%%%%%%%%%%%%
\end{Proposition}
%%%%%%%%%%%%%%%%%
\begin{Remark}
%%%%%%%%%%%%%%
The coordinate expression 
\fr{coordinate-expression-contact-Hamiltonian-vertical-vector} 
is formally the same as that of 
\fr{contact-Hamiltonian-vector-components}.
%\fr{coordinate-expression-contact-Hamiltonian-vector}.
%%%%%%%%%%%%
\end{Remark}
%%%%%%%%%%%%%

Analogous to  
Definition\,\ref{definition-Legendre-submanifold}, 
Legendre submanifold on a  bundle is defined as follows.  
%%%%%%%%%
\begin{Def}
%%%%%%%%%%
(Legendre submanifold of vertical space and that of fiber space) :  
Let $(\cK,\lambda_{\,\mbbV},\pi,\cB)$ 
be a contact manifold over a base space $\cB$. 
If $\cA_{\,\zeta}$ is a maximal dimensional integral submanifold of 
$\lambda_{\,\mbbV}$ on $\pi^{-1}(\zeta)$,  $(\zeta\in\cB)$, 
then $\cA_{\,\zeta}$ 
is referred to as a Legendre submanifold
in the fiber space $\pi^{-1}(\zeta)$, and 
$\cA^{\,\cK}=\bigcup_{\,\zeta\in\cB}\cA_{\,\zeta}$ 
a Legendre submanifold in the fiber space.
%%%%%%%%%%%
\end{Def}
%%%%%%%%%%

An analogous theorem from 
Theorem\,\ref{theorem-Legendre-submanifold-theorem-Arnold}
holds for the present bundles. 
Examples of Legendre submanifolds on fiber spaces are as follows.  
%%%%%%%%%%%%%%%
\begin{Example}
%%%%%%%%%%%%%%
Let $(\,\cK,\lambda_{\,\mbbV},\pi,\cB\,)$ 
be a $(2n+1)$-dimensional contact manifold over a base space $\cB$ 
with $\dim\cB=d_{\,\cB}$, 
$(x,y,z)$ the canonical coordinates for the fiber space such that 
$\lambda_{\,\mbbV}=\dr_{\,\mbbV} z-y_{\,a}\,\dr_{\mbbV} x^{\,a}$  
with $x=\{\,x^{\,1},\ldots,x^{\,n}\,\}$ and $y=\{\,y_{\,1},\ldots,y_{\,n}\,\}$, 
$\psi\in\Gamma\Lambda_{\,\mbbV}^{\,0}\cK$ a vertical function of $x$, 
$\rho^{\,d_{\,\cB}}$ a nowhere vanishing horizontal $d_{\,\cB}$-form, 
$\cB_{\,0}\subseteq\cB$ a $d_{\cB}$-dimensional space, 
and $\wt{\psi}_{\,\cB_{\,0}}\in\Gamma F\cK$ 
the functional 
$$
\wt{\psi}_{\,\cB_0}
=\int_{\,\cB_0}\psi\,\rho^{\,d_{\,\cB}}.
$$   
Then, the Legendre submanifold $\cA_{\,\zeta\psi}$ generated 
by $\psi$ on $\pi^{-1}(\zeta),(\zeta\in\cB)$ with 
$\Phi_{\,\cC_{\zeta}\cA_{\,\zeta\psi}}:\cA_{\,\zeta\psi}\to\cC_{\zeta}$ 
being the embedding is such that 
\beq
\Phi_{\,\cC_{\zeta}\cA_{\,\zeta\psi}}\cA_{\,\zeta\psi}
=\left\{\ (x,y,z)\in\cC_{\zeta} \ \bigg|\ 
y_{\,j}=\frac{\partial\psi}{\partial x^{\,j}},\ \mbox{and}\ 
z=\psi(x),\quad j\in \{\,1,\ldots,n\,\}
\ \right\}. 
\label{example-psi-Legendre-submanifold-bundle}
\eeq
This can also be written as 
$$
\Phi_{\,\cC_{\zeta}\cA_{\,\zeta\psi}}\cA_{\,\zeta\psi}
=\left\{\ (x,y,z)\in\cC_{\zeta} \ \bigg|\ 
y_{\,j}=\frac{\delta\wt{\psi}_{\,\cB_0}}{\delta x^{\,j}},\ \mbox{and}\ 
z=\psi(x),\quad j\in \{\,1,\ldots,n\,\}
\ \right\}. 
$$
In addition, 
$(\cA_{\,\psi}^{\,\cK},\pi|_{\,\cA_{\,\psi}^{\,\cK}},\cB)$ is a sub-bundle of 
$(\cK,\pi,\cB)$, where
$$
\cA_{\,\psi}^{\,\cK}
=\bigcup_{\zeta\in\cB}\Phi_{\,\cC_{\zeta}\cA_{\,\zeta\psi}}\cA_{\,\zeta\psi}.
$$

%%%%%%%%%%%%%
\end{Example}
%%%%%%%%%%%%%%

%%%%%%%%%%%%%%%
\begin{Example}
%%%%%%%%%%%%%%
Let $(\,\cK,\lambda_{\mbbV},\pi,\cB\,)$ 
be a $(2n+1)$-dimensional contact manifold over a base space $\cB$, 
$(x,y,z)$ the canonical coordinates for the fiber space such that 
$\lambda_{\,\mbbV}=\dr_{\,\mbbV} z-y_{\,a}\,\dr_{\,\mbbV} x^{\,a}$  
with $x=\{\,x^{\,1},\ldots,x^{\,n}\,\}$ and $y=\{\,y_{\,1},\ldots,y_{\,n}\,\}$, 
$\varphi\in\Gamma\Lambda_{\,\mbbV}^{0}\cK$ a vertical function of $y$,  
$\rho^{\,d_{\,\cB}}$ a nowhere vanishing horizontal $d_{\,\cB}$-form, 
$\cB_{\,0}\subseteq\cB$ a $d_{\,\cB}$-dimensional space, 
and $\wt{\varphi}_{\,\cB_{\,0}}\in\Gamma F\cK$ 
the functional  
$$
\wt{\varphi}_{\,\cB_{\,0}}
=\int_{\,\cB_0}\varphi\,\rho^{\,d_{\cB}}.
$$   
Then, the  Legendre submanifold $\cA_{\,\zeta\varphi}$ generated 
by $-\varphi$ on $\pi^{-1}(\zeta),(\zeta\in\cB)$ with 
$\Phi_{\,\cC_{\zeta}\cA_{\,\zeta\varphi}}:\cA_{\,\zeta\varphi}\to\cC_{\,\zeta}$ 
being the embedding is such that 
\beq
\Phi_{\,\cC_{\,\zeta}\cA_{\,\zeta}\varphi}\cA_{\,\zeta\varphi}
=\left\{\ (x,y,z)\in\cC_{\zeta} \ \bigg|\ 
x_{\,i}=\frac{\partial\varphi}{\partial y_{\,i}},\ \mbox{and}\ 
z=y_{\,i}\frac{\partial\varphi}{\partial y_{\,i}}-\varphi(y),
\quad i\in \{\,1,\ldots,n\,\}
\ \right\}. 
\label{example-varphi-Legendre-submanifold-bundle}
\eeq
This can also be written as 
$$
\Phi_{\,\cC_{\,\zeta}\cA_{\,\zeta\varphi}}\cA_{\,\zeta\varphi}
=\left\{\ (x,y,z)\in\cC_{\zeta} \ \bigg|\ 
x_{\,i}=\frac{\delta\wt{\varphi}_{\,\cB_{\,0}}}{\delta y_{\,i}},\ \mbox{and}\ 
z=y_{\,i}\frac{\delta\wt{\varphi}_{\,\cB_{\,0}}}{\delta y_{\,i}}-\varphi(y),
\quad i\in \{\,1,\ldots,n\,\}
\ \right\}. 
$$
In addition, 
$(\cA_{\,\varphi}^{\,\cK},\pi|_{\,\cA_{\,\varphi}^{\,\cK}},\cB)$ is a sub-bundle of 
$(\cK,\pi,\cB)$, where 
$$
\cA_{\,\varphi}^{\,\cK}
=\bigcup_{\,\zeta\in\cB}\Phi_{\,\cC_{\,\zeta}\cA_{\,\zeta\varphi}}\cA_{\,\zeta\varphi}.
$$

%%%%%%%%%%%%%
\end{Example}
%%%%%%%%%%%%%%

%Although the following could not be commonly used in the literature,  
The total Legendre transform of a 
functional is defined as follows  in this paper. 
%%%%%%%%%%%
\begin{Def}
\label{definition-Legendre-transform-functional}
%%%%%%%%%%%
(Total Legendre transform of functional): 
Let 
$(\cK,\lambda_{\,\mbbV},\pi,\cB)$
be a contact manifold over a base space $\cB$ with $\dim\cB=d_{\,\cB}$,  
$\cB_{\,0}\subseteq\cB$ a $d_{\,\cB}$-dimensional space,
$\rho^{\,d_{\,\cB}}\in\Gamma\Lambda_{\,\mbbH}^{\,d_{\,\cB}}\cK$ 
a nowhere vanishing horizontal 
form, $\psi\in\Gamma\Lambda_{\,\mbbV}^{\,0}\cK$ a vertical $0$-form, 
and $\wt{\psi}_{\,\cZ_{\,0}}$ a functional such that  
$$
\wt{\psi}_{\,\cB_{\,0}}
=\int_{\cB_{\,0}}\,\psi\,\rho^{\,d_{\,\cB}}.
$$
Then, the total Legendre transform of $\Psi_{\cZ_{0}}$ is defined as 
$$
\wt{\psi}^{\,*}_{\,\cB_{0}}
=\int_{\cB_{\,0}}\Leg[\psi]\,\rho^{\,d_{\cB}},
$$
where $\Leg[\psi]$ is the total Legendre transform of $\psi$, 
( see Definition\,\ref{definition-total-Legendre-transform} ).  
%%%%%%%%%%  
\end{Def}
%%%%%%%%%%%

As shown in 
Propositions\,\ref{Mrugala-variant-psi} and \ref{Mrugala-variant-varphi},
vector fields on Legendre submanifolds of contact manifolds are concisely 
written as contact Hamiltonian vector fields with adapted functions introduced 
in Definition\,\ref{definition-adapted-function} 
for the standard contact geometry. 
Also, for contact geometry on fiber spaces,  
similar functions can be defined as follows. 

%%%%%%%%%%%%%%
\begin{Def}
\label{definition-adapted-function-on-bundle}
%%%%%%%%%%%%
(Adapted functions on fiber space):  
Let $(\cK,\lambda_{\,\mbbV},\pi,\cB)$ be a $(2n+1)$-dimensional contact manifold
over a base space $\cB$, $\cC_{\,\zeta}$ a fiber space on $\pi^{-1}(\zeta)$, 
$(x,y,z)$ 
canonical coordinates for $\pi^{-1}(\zeta),(\zeta\in\cB)$ such that 
$\lambda_{\,\mbbV}=\dr_{\,\mbbV} z-y_{\,a}\,\dr_{\mbbV} x^{\,a}$ with 
$x=\{\,x^{\,1},\ldots,x^{\,n}\,\}$ and $y=\{\,y_{\,1},\ldots,y_{\,n}\,\}$, 
and $\cK=\bigcup_{\,\zeta\in\cB}\cC_{\,\zeta}$. 
In addition let 
$\psi$ be a vertical 
function on $\cC_{\,\zeta}$ depending on $x$,    
and $\varphi$ a vertical function on $\cC_{\,\zeta}$ depending on $y$. Then
the functions 
%$\{\Delta_{\,0}^{\,\zeta\psi},\Delta_{\,a}^{\,\zeta\psi}\}\in\Gamma\Lambda_{\mbbV}^{\,0}\cK,$ 
$\Delta_{\,0}^{\,\zeta\psi},\{\Delta_{\,1}^{\,\zeta\psi},\ldots,\Delta_{\,n}^{\,\zeta\psi}\}:\cC_{\,\zeta}\to\mbbR,$ 
and 
%$\{\Delta_{\,\zeta\varphi}^{\,0},\Delta_{\,\zeta\varphi}^{\,a}\}\in\Gamma\Lambda_{\mbbV}^{\,0}\cK$
$\Delta_{\,\zeta\varphi}^{\,0},\{\Delta_{\,\zeta\varphi}^{\,1},\ldots,\Delta_{\,\zeta\varphi}^{\,n}\}:\cC_{\,\zeta}\to\mbbR$  
such that  
$$
\Delta_{\,0}^{\,\zeta\psi}(x,z)
:=\psi(x)-z,\quad 
\Delta_{\,a}^{\,\zeta\psi}(x,y)
:=\frac{\partial\psi}{\partial x^{\,a}}-y_{\,a},\qquad 
a\in\{\,1,\ldots,n\,\}.
$$
$$
\Delta_{\,\zeta\varphi}^{\,0}(x,y,z)
:=x^{\,j}p_{\,j}-\varphi(p)-z,\quad 
\Delta_{\,\zeta\varphi}^{\,a}(x,y)
:=x^{\,a}-\frac{\partial\varphi}{\partial y_{\,a}},\qquad 
a\in\{\,1,\ldots,n\,\}.
$$  
are referred to as adapted functions on the fiber space. 
%%%%%%%%%%
\end{Def}
%%%%%%%%%%% 

Similar to 
Proposition\,\ref{proposition-Legendre-submanifold-adapted-functions},
one has the following.
%%%%%%%%%%%%%%%%%%%%
\begin{Proposition}
\label{proposition-Legendre-submanifold-adapted-functions-on-fiber}
%%%%%%%%%%%%%%%%%%%%%%
(Local expressions of Legendre submanifold with adapted functions):  
The Legendre submanifold $\cA_{\,\zeta \psi}$ 
generated by $\psi$ on $\pi^{-1}(\zeta)$ 
as 
\fr{example-psi-Legendre-submanifold-bundle}
is expressed as 
\beq
\cA_{\,\zeta\psi}^{\,\cC}
:=\Phi_{\,\cC_{\,\zeta}\cA_{\,\zeta\psi}}\cA_{\,\zeta\psi}
=\left\{\ (x,p,z)\in\cC_{\zeta} \ |\ 
\Delta_{\,0}^{\,\zeta\psi}
=0\ \mbox{and}\ 
\Delta_{\,1}^{\,\zeta\psi}
=\cdots
=\Delta_{\,n}^{\,\zeta\psi}
=0
\ \right\}, 
\label{Legendre-submanifold-fiber-psi}
\eeq
where 
$\Phi_{\,\cC_{\,\zeta}\cA_{\,\zeta\psi}}\cA_{\,\zeta\psi}:\cA_{\,\zeta\psi}\to\cC_{\,\zeta}$ 
is the embedding. Similarly, 
the Legendre submanifold $\cA_{\,\zeta \varphi}$ generated by $-\varphi$ as 
\fr{example-varphi-Legendre-submanifold-bundle}
is expressed as 
\beq
\cA_{\,\zeta\varphi}^{\,\cC}
:=\Phi_{\,\cC_{\,\zeta}\cA_{\,\zeta\varphi}}\cA_{\,\zeta\varphi}
=\left\{\ (x,y,z)\in\cC_{\,\zeta} \ |\ 
\Delta_{\,\zeta\varphi}^{\,0}
=0\ \mbox{and}\ 
\Delta_{\,\zeta\varphi}^{\,1}
=\cdots
=\Delta_{\,\zeta\varphi}^{\,n}
=0
\ \right\}, 
\label{Legendre-submanifold-fiber-varphi}
\eeq
where 
$\Phi_{\,\cC_{\zeta}\cA_{\,\zeta\varphi}}\cA_{\,\zeta\varphi}:\cA_{\,\zeta\varphi}\to\cC_{\,\zeta}$ 
is the embedding. 
%%%%%%%%%%%%%%%%%%5
\end{Proposition} 
%%%%%%%%%%%%%%%%%%

Contact Hamiltonian vertical vector fields are 
also written in terms of adapted functions on fiber spaces.
%%%%%%%%%%%%%%%%%%5
\begin{Proposition} 
\label{Mrugala-variant-psi-bundle}
%%%%%%%%%%%%
(Restricted contact Hamiltonian vertical vector field  
as the push-forward of a vector field on 
the Legendre submanifold generated by $\psi$):  
Let $\{\,F_{\,\zeta\psi}^{\,1},\ldots,F_{\,\zeta\psi}^{\,n}\,\}$ 
be a set of functions of $x$ on $\cA_{\,\zeta\psi}$ 
such that they do not identically vanish,   
and 
$\chX_{\,\zeta\psi}^{\,0}\in T_{\,x}\,\cA_{\,\zeta\psi}, ( x\in \cA_{\,\zeta\psi})$ 
the vector field given as   
$$
\chX_{\,\zeta\psi}^{\,0}
=\dot{x}^{\,a}\,\frac{\partial}{\partial x^{\,a}},
\quad\mbox{where}\quad 
\dot{x}^{\,a}
=F_{\,\zeta\psi}^{\,a}(x),\qquad 
(a\in\{\,1,\ldots,n\,\}).
$$
In addition, let 
$X_{\,\zeta\psi}^{\,0}
:=(\,\Phi_{\,\cC_{\zeta}\cA_{\,\zeta\psi}}\,)_{*}\chX_{\,\zeta\psi}^{\,0} 
\in T_{\,\xi}\cA_{\,\zeta\psi}^{\,\cC}, (\,\xi\in\cA_{\,\zeta\psi}^{\,\cC}\,)$ 
be the push-forward of
$\chX_{\,\zeta\psi}^{\,0}$,   
where $\cA_{\,\zeta\psi}^{\,\cC}:=\Phi_{\,\cC_{\,\zeta}\cA_{\,\zeta\psi}}\cA_{\,\zeta\psi}$ with   
$\Phi_{\,\cC_{\,\zeta}\cA_{\,\zeta\psi}}:\cA_{\,\zeta\psi}\to \cC_{\,\zeta}$ 
being the embedding :  
%( see the diagrams below ). 
\beqa
\Phi_{\,\cC_{\,\zeta}\cA_{\,\zeta\psi}} &:& \cA_{\,\zeta\psi}\to 
\cA_{\,\zeta\psi}^{\,\cC},\qquad\qquad x\mapsto (\,x,y(x),z(x)\,),\quad
\mbox{on}\quad \pi^{-1}(\zeta),
\non\\
(\,\Phi_{\,\cC_{\,\zeta}\cA_{\,\zeta\psi}}\,)_{\,*} 
&:& T_{\,x}\,\cA_{\,\zeta\psi}\to  
T_{\,\xi}\,\cA_{\,\zeta\psi}^{\,\cC},\quad \chX_{\,\zeta\psi}^{\,0}\mapsto
X_{\,\zeta\psi}^{\,0}\,.
\non
\eeqa
Then it follows that   
\beq
X_{\,\psi}^{\,0}
=\dot{x}^{\,a}\frac{\partial}{\partial x^{\,a}}
+\dot{y}_{\, a}\frac{\partial}{\partial y_{\, a}}
+\dot{z}\frac{\partial}{\partial z},\quad\mbox{where}\quad 
\dot{x}^{\,a}
=F_{\,\zeta\psi}^{\,a}(x),\quad
\dot{y}_{\,a}
=\frac{\dr}{\dr t}\left(\,\frac{\partial\psi}{\partial x^{\,a}}\,
\right),
\quad 
\dot{z}
=\frac{\dr \psi}{\dr t}.
\label{tangent-vector-Legendre-submanifold-psi-component-bundle}
\eeq
In addition, 
one has that $X_{\,\zeta\psi}^{\,0}=X_{\,\wt{h}_{\,\psi}}|_{\,\wt{h}_{\,\psi}=0}$. Here 
$X_{\,\wt{h}_{\,\psi}}$ is the contact Hamiltonian vertical vector field associated 
with  
\beq
h_{\,\psi}(x,y,z)
=\Delta_{\,a}^{\,\zeta\psi}(x,y) F_{\,\zeta\psi}^{\,a}(x) 
+\Gamma_{\,\zeta\psi}(\,\Delta_{\,0}^{\,\zeta\psi}(x,z)\,),
\label{tangent-vector-Legendre-submanifold-psi-Hamiltonian-bundle}
\eeq
where $\Gamma_{\,\zeta\psi}$ is a function of $\Delta_{\,0}^{\,\zeta\psi}$ such that 
$$
\Gamma_{\,\zeta\psi}\left(\,\Delta_{\,0}^{\,\zeta\psi}\,\right)
=\left\{
\begin{array}{cl}
0&\mbox{for}\quad \Delta_{\,0}^{\,\zeta\psi}=0\\
\mbox{non-zero}&\mbox{for}\quad\Delta_{\,0}^{\,\zeta\psi}\neq 0
\end{array}
\right..
%,\qquad\mbox{and}\qquad
% \left.\frac{\dr}{\dr \Delta_{\,0}^{\,\zeta\psi}}
% \right|_{\Delta_{\,0}^{\,\zeta\psi}=0}\Gamma_{\,\zeta\psi}
% \left(\,\Delta_{\,0}^{\,\zeta\psi}\,\right)
% =0.
$$  
%%%%%%%%%%%
\end{Proposition}
%%%%%%%%%%
There exists 
a counterpart of Proposition\,\ref{Mrugala-variant-psi-bundle}, that 
is given as follows.
 
%%%%%%%%%%%%%%%%%%5
\begin{Proposition} 
\label{Mrugala-variant-varphi-bundle}
%%%%%%%%%%%%
(Restricted contact Hamiltonian vertical vector field  
as the push-forward of a vector field on 
the Legendre submanifold generated by $-\varphi$):  
Let $\{\,F_{\,\zeta,1}^{\,\varphi},\ldots,F_{\,\zeta,n}^{\,\varphi}\,\}$ 
be a set of functions of $y$ on $\cA_{\,\zeta\varphi}$ 
such that they do not identically vanish,   
and 
$\chX_{\,0}^{\,\zeta\varphi}\in T_{\,y}\,\cA_{\,\zeta\varphi}, ( y\in \cA_{\,\zeta\varphi})$ 
the vector field given as   
$$
\chX_{\,0}^{\,\zeta\varphi}
=\dot{y}_{\,a}\,\frac{\partial}{\partial y_{\,a}},
\quad\mbox{where}\quad 
\dot{y}_{\,a}
=F_{\,a}^{\,\zeta\varphi}(y),\qquad 
(a\in\{\,1,\ldots,n\,\}).
$$
In addition, let 
$X_{\,0}^{\,\zeta\varphi}
:=(\,\Phi_{\,\cC_{\zeta}\cA_{\,\zeta\varphi}}\,)_{*}\chX_{\,0}^{\,\zeta\varphi} 
\in T_{\,\xi}\cA_{\,\zeta\varphi}^{\,\cC}, (\,\xi\in\cA_{\,\zeta\varphi}^{\,\cC}\,)$ 
be the push-forward of
$\chX_{\,0}^{\,\zeta\varphi}$,   
where $\cA_{\,\zeta\varphi}^{\,\cC}:=\Phi_{\,\cC_{\,\zeta}\cA_{\,\zeta\varphi}}\cA_{\,\zeta\varphi}$
 with $\Phi_{\,\cC_{\,\zeta}\cA_{\,\zeta\varphi}}:\cA_{\,\zeta\varphi}\to \cC_{\,\zeta}$ 
being the embedding :  
%( see the diagrams below ). 
\beqa
\Phi_{\,\cC_{\,\zeta}\cA_{\,\zeta\varphi}} &:& \cA_{\,\zeta\varphi}\to 
\cA_{\,\zeta\varphi}^{\,\cC},\qquad\qquad y\mapsto (\,x(y),y,z(y)\,),\quad
\mbox{on}\quad \pi^{-1}(\zeta),
\non\\
(\,\Phi_{\,\cC_{\,\zeta}\cA_{\,\zeta\psi}}\,)_{\,*} 
&:& T_{\,y}\,\cA_{\,\zeta\varphi}\to  
T_{\,\xi}\,\cA_{\,\zeta\varphi}^{\,\cC},\quad \chX_{\,0}^{\,\zeta\varphi}
\mapsto
X_{\,0}^{\,\zeta\varphi}\,.
\non
\eeqa
Then, it follows that   
\beq
X_{\,0}^{\,\zeta\varphi}
=\dot{x}^{\,a}\frac{\partial}{\partial x^{\,a}}
+\dot{y}_{\, a}\frac{\partial}{\partial y_{\, a}}
+\dot{z}\frac{\partial}{\partial z},\ \mbox{where}\quad  
\dot{x}^{\,a}
=\frac{\dr}{\dr t}\left(\,\frac{\partial\varphi}{\partial y_{\,a}}\,
\right),\quad
\dot{y}_{\,a}
=F_{\,a}^{\,\zeta\varphi}(y)
,\quad 
\dot{z}
=y_{\,j}F_{\,k}^{\,\zeta\varphi}
\frac{\partial^2 \varphi}{\partial y_{\,k}\partial y_{\,j}}.
\label{tangent-vector-Legendre-submanifold-varphi-component-bundle}
\eeq
In addition, 
one has that $X_{\,0}^{\,\zeta\varphi}=X_{\,\wt{h}_{\,\varphi}}|_{\,\wt{h}_{\,\varphi}=0}$. Here 
$X_{\,\wt{h}_{\,\varphi}}$ is the contact Hamiltonian vertical vector field 
associated with  
\beq
h_{\,\varphi}(x,y,z)
=\Delta_{\,\zeta\varphi}^{\,a}(x,y) F_{\,a}^{\,\zeta\varphi}(y) 
+\Gamma^{\,\zeta\varphi}(\,\Delta_{\,\zeta\varphi}^{\,0}(x,y,z)\,),
\label{tangent-vector-Legendre-submanifold-varphi-Hamiltonian-bundle}
\eeq
where $\Gamma^{\,\zeta\varphi}$ is a function of $\Delta_{\,\zeta\varphi}^{\,0}$ 
such that 
$$
\Gamma^{\,\zeta\varphi}\left(\,\Delta_{\,\zeta\varphi}^{\,0}\,\right)
=\left\{
\begin{array}{cl}
0&\mbox{for}\quad \Delta_{\,\zeta\varphi}^{\,0}=0\\
\mbox{non-zero}&\mbox{for}\quad\Delta_{\,\zeta\varphi}^{\,0}\neq 0
\end{array}
\right..
%,\qquad\mbox{and}\qquad
% \left.\frac{\dr}{\dr \Delta_{\,\zeta\varphi}^{\,0}}
% \right|_{\Delta_{\,\zeta\varphi}^{\,0}=0}\Gamma^{\,\zeta\varphi}\left(\,\Delta_{\,\zeta\varphi}^{\,0}
% \,\right)
% =0.
$$  
%%%%%%%%%%%
\end{Proposition}
%%%%%%%%%%

%%%%%%%%%%%%%%%%%%%%%%%%%%%%%%%%%%%%%%%
\subsection{Information geometry}
\label{subsection-information-geometry}
%%%%%%%%%%%%%%%%%%%%%%%%%%%%%%%%%%%%%%
Information geometry is a geometrization of parametric statistics\cite{AN},
 and it was shown that there are some overlap between contact geometry and 
information geometry\cite{Goto2015} (see also Refs.\,\cite{Bravetti-Sep-2014,Mrugala1990}). Later on 
in this paper it will be shown 
that a class of Hamiltonian functionals 
for distributed-parameter port-Hamiltonian systems
with respect to Stokes-Dirac structures 
can induce information geometry.
% that there exist overlaps 
% between contact geometry and Dirac structures.   
To this end, definitions and known theorems are summarized below. 
The following definitions follow the standard information geometry.
%%%%%%%%% 
\begin{Def}
\label{definition-affine-coordinate-flat-connection}
%%%%%%%%%
(Affine-coordinate and flat connection, \cite{AN}): 
Let $\cM$ be an $n$-dimensional manifold, 
$x=\{\,x^{\,1},\ldots,x^{\,n}\,\}$
coordinates, $\nabla$ a connection, $\{\Gamma_{ab}^{\ \ c}\}$ 
connection coefficients such that 
$\nabla_{\partial_a}\partial_b=\Gamma_{ab}^{\ \ c}\partial_c,$ 
$(\partial_{\,a}:=\partial/\partial x^{\,a})$. If 
$\{\,\Gamma_{ab}^{\ \ c }\,\}\equiv 0$ hold for all $\xi\in\cM$, then
$x$ is referred to as a $\nabla$-affine coordinate system, 
or affine coordinates. 
If it is the case, then $\nabla$ 
is referred to as a flat connection. 
%%%%%%%%%
\end{Def}
%%%%%%%%%
%%%%%%%%%
\begin{Def}
\label{definition-dual-connection}
%%%%%%%%%
(Dual connection, \cite{AN}):  
Let $(\cM,g)$ be an $n$-dimensional Riemannian or pseudo-Riemannian manifold,
 and $\nabla$ a connection. 
If a connection $\nabla^{\, *}$ satisfies 
\beq
Z\,\left[\,g(\,X,Y\,)\,\right]
=g(\,\nabla_ZX,Y\,)+g(\,X,\nabla_Z^{\,*}Y\,),\qquad \forall\, X,Y,Z\in\GTM
\label{dual-connection-general}
\eeq
then $\nabla$ and $\nabla^{\,*}$ are referred to as dual connections, also 
$\nabla^{\,*}$ 
is referred to as a dual connection of $\nabla$ with respect to $g$. 
%%%%%%%%%
\end{Def}
%%%%%%%%%

%%%%%%%%%%%%%%%
\begin{Lemma}
\label{lemma-uniquely-detemined-dual-connection}
%%%%%%%%%%%%%% 
(Existence of a unique dual connection, \cite{AN}): 
Given a metric tensor field and a connection, 
there exists a unique dual connection .  
%%%%%%%%%%%
\end{Lemma}
%%%%%%%%%%%
%%%%%%%%%%%%%
% \begin{Proof}
% %%%%%%%%%%%%%
% Let $g$ and $\nabla$ be the given metric tensor field and connection, 
% respectively. 
% Then in what follows the explicit form of the dual connection $\nabla^{\,\prime}$ 
% is shown.  
% Let $x=\{\,x^{\,1},\ldots,x^{\,n}\,\}$ be local coordinates, $\{\,g_{\,ab}\,\}$ 
% the components of the metric tensor field such that 
% $g=g_{\,ab}\,\dr x^{\,a}\otimes\dr x^{\,b}$, 
% $\{\,\Gamma_{ab}^{\ \ c}\,\}$ connection coefficients for $\nabla$ 
% such that $\nabla_{\partial_{\,a}}\partial_{\,b}=\Gamma_{ab}^{\ \ c}\partial_{\,c}$ with 
% $\partial_{\,a}:=\partial/\partial x^{\,a}$, and $\{\,\Gamma_{ab}^{\,\prime \ c}\,\}$ 
% connection coefficients for $\nabla^{\,\prime}$ such that 
% $\nabla_{\partial_{\,a}}^{\,\prime}\partial_{\,b}=\Gamma_{ab}^{\,\prime \ c}\partial_{\,c}$.
% Then, defining $\Gamma_{abc}:=g_{\,ck}\Gamma_{ab}^{\ \ k}$ and 
% $\Gamma_{abc}^{\,\prime}:=g_{\,ck}\Gamma_{ab}^{\,\prime \ k}$, one can determine  
% $\{\,\Gamma_{abc}^{\,\prime}\,\}$ uniquely with \fr{dual-connection-general} 
% as 
% $\Gamma_{abc}^{\,\prime}=\partial_{\,c}\,g_{\,ab}-\Gamma_{abc}$.    
% \qed
% \end{Proof}
%%%%%%%%%%%
%%%%%%%%%
\begin{Def}
\label{definition-dual-coordinate}
%%%%%%%%%
(Dual coordinates, \cite{AN}):  
Let $(\cM,g)$ be an $n$-dimensional Riemannian or pseudo-Riemannian manifold, 
$x=\{\,x^{\,1},\ldots,x^{\,n}\,\}$ a set of local coordinates, and 
$y=\{\,y_{\,1},\ldots,y_{\,n}\,\}$ 
another set of local coordinates.  
If 
\beq
g\left(\,\frac{\partial}{\partial x^{\,a}},\frac{\partial}{\partial y_{\,b}}\,\right)
=\delta_{\,a}^{\,b},
\label{dual-coordinate-general} 
\eeq
then $y$ is referred to as the dual coordinate system. If it is the case, 
then $x$ and $y$ are referred to as being mutually dual with respect to $g$.  
%%%%%%%%%
\end{Def}
%%%%%%%%%

Combining Definitions\,\ref{definition-affine-coordinate-flat-connection}, 
\ref{definition-dual-connection} and 
\ref{definition-dual-coordinate}, one has the following.
%%%%%%%%%%%%%5
\begin{Lemma}
\label{lemma-dual-affine-coordinates}
%%%%%%%%%%%%%
(Dual coordinates and affine coordinates) : 
Let $(\cM,g)$ be an $n$-dimensional Riemannian or pseudo-Riemannian manifold,
$\nabla$ a connection, $x=\{\,x^{\,1},\ldots,x^{\,n}\,\}$ 
a set of $\nabla$-affine coordinates, and 
$y=\{\,y_{\,1},\ldots,y_{\,n}\,\}$ another set of coordinates. 
If $x$ and $y$ are mutually dual with respect to $g$, then  
$y$ is a $\nabla^{\,*}$-affine coordinate system. 
%%%%%%%%%%%%
\end{Lemma}
%%%%%%%%%%%%
% \begin{Proof}
% %%%%%%%%%%%%%
% There exists the connection $\nabla^{\,*}$ being dual of $\nabla$ due to 
% Lemma\,\ref{lemma-uniquely-detemined-dual-connection}.   
% It follows from \fr{dual-coordinate-general} that 
% $$
% Z\left[\,g\left(\,\frac{\partial}{\partial x^{\,a}},\frac{\partial }{\partial y_{\,b}}\,\right)\,\right]
% =0,
% $$
% for $\forall\,Z\in\GTM$.
% Then, with this and \fr{dual-connection-general}, one has 
% $$
% Z\left[\,
% g\left(\,\frac{\partial}{\partial x^{\,a}},\frac{\partial}{\partial y_{\,b}}\right)\,
% \right]
% =g\left(\,\nabla_{\,Z}\frac{\partial}{\partial x^{\,a}},
% \frac{\partial}{\partial y_{\,b}}\right)
% +g\left(\,\frac{\partial}{\partial x^{\,a}},
% \nabla_{\,Z}^{\,*}\frac{\partial}{\partial y_{\,b}}\right)
% =0.
% $$
% Since $x$ is a $\nabla$-affine coordinate system, 
% $\nabla_{\,Z}(\,\partial/\partial x^{\,a}\,)=0,(a\in\{\,1,\ldots,n\,\})$, 
% one can conclude that 
% $\nabla_{\,Z}^{\,*}(\,\partial/\partial y_{\,b}\,)=0, (b\in\{\,1,\ldots,n\,\})$, 
% from which $y$ is a $\nabla^{\,*}$-affine coordinate system.   
% \qed
% %%%%%%%%%%%%
% \end{Proof}
%%%%%%%%%%%

The following space plays various roles in information geometry.
%%%%%%%%%
\begin{Def}
\label{definition-dually-flat-space}
%%%%%%%%%
(Dually flat space, \cite{AN}): 
Let $(\cM,g)$ be an $n$-dimensional Riemannian or pseudo-Riemannian manifold, 
$\nabla$ and $\nabla^{\, *}$ dual connections.
If there exist $\nabla$-affine coordinates 
and $\nabla^{\,*}$-affine ones,
then $(\cM,g,\nabla,\nabla^*)$ is 
referred to as a dually flat space. 
%%%%%%%%%
\end{Def}
%%%%%%%%%

From these definitions, one can show the following relation between 
pairings and inner products.
%%%%%%%%%%%%%%%%%%%%
\begin{Proposition}
\label{proposition-inner-product-paring-dually-flat-space}
%%%%%%%%%%%%%%%%%%%
(Inner products and pairings on a dually flat space):  
Let $(\cM,g,\nabla,\nabla^{\,*})$ be an $n$-dimensional 
dually flat space, $x=\{\,x^{\,1},\ldots,x^{\,n}\,\}$ a set of 
$\nabla$-affine coordinates, $y=\{\,y_{\,1},\ldots,y_{\,n}\,\}$ a set of 
$\nabla^{\,*}$-affine coordinates.
%such that 
%$g(\partial/\partial x^{\,a},\partial/\partial y_{\,b})=\delta_{\,a}^{\,b}$
If the inner products 
$T_{\xi}\cM\times T_{\xi}\cM\to\mbbR, (\xi\in\cM)$ between the bases
$\{\,\partial/\partial x^{\,1},\ldots,\partial/\partial x^{\,n}\,\}$ and 
$\{\,\partial/\partial y_{\,1},\ldots,\partial/\partial y_{\,n}\,\}$ are given as 
\beqa
g\left(\,\frac{\partial}{\partial x^{\,a}},\frac{\partial}{\partial x^{\,b}}\,\right)
&=&g_{\,ab},\qquad
g\left(\,\frac{\partial}{\partial x^{\,a}},\frac{\partial}{\partial y_{\,b}}\,\right)
=\delta_{\,a}^{\,b},
\non\\
g\left(\,\frac{\partial}{\partial y_{\,a}},\frac{\partial}{\partial x^{\,b}}\,\right)
&=&\delta_{\,b}^{\,a},\qquad
g\left(\,\frac{\partial}{\partial y_{\,a}},\frac{\partial}{\partial y_{\,b}}\,\right)
=g^{\,ab},
\non
\eeqa
( i.e., $x$ and $y$ are mutually dual with respect to $g$ ), then 
one has 
the following pairings  
$T_{\xi}^{\,*}\cM\times T_{\xi}\cM\to\mbbR, (\xi\in\cM)$
\beqa
\dr x^{\,a}\left(\,\frac{\partial}{\partial x^{\,b}}\,\right)
&=&\delta^{\,ab},\qquad
\dr x^{\,a}\left(\,\frac{\partial}{\partial y_{\,b}}\,\right)
=g^{\,ab},
\non\\
\dr y_{\,a}\left(\,\frac{\partial}{\partial x^{\,b}}\,\right)
&=&g_{\,ab},\qquad
\dr y_{\,a}\left(\,\frac{\partial}{\partial y_{\,b}}\,\right)
=\delta_{\,ab}.
\non
\eeqa
%%%%%%%%%%%%%%%
\end{Proposition}
%%%%%%%%%%%%%%%%

As shown in Ref.\cite{Goto2015}, 
a contact manifold and a strictly convex function 
induce a dually flat space. 
%%%%%%%%%%%
\begin{Thm}
\label{theorem-goto-2015}
(Contact manifold and a function induce a dually flat space,\,\cite{Goto2015}) : 
Let $(\,\cC,\lambda\,)$ be a $(2n+1)$-dimensional contact manifold, 
$(x,y,z)$ the canonical coordinates such that $\lambda=\dr z-y_{\,a}\,\dr x^{\,a}$
with $x=\{\,x^{\,1},\ldots,x^{\,n}\,\}$ and $y=\{\,y_{\,1},\ldots,y_{\,n}\,\}$, 
and $\psi$ a strictly convex function of $x$ only. 
If the Legendre submanifold generated by $\psi$ is simply connected, then 
$(\,(\,\cC,\lambda\,),\,\psi\,)$ induces an $n$-dimensional dually flat space
$(\,\Phi_{\,\cC\cA\psi}\cA_{\,\psi},g,\nabla,\nabla^{\,*}\,)$ 
with $\Phi_{\,\cC\cA\psi}:\cA_{\,\psi}\to\cC$ being the embedding.
% %%%%a%%%%%
\end{Thm}

The following is often used in information geometry.
%%%%%%%%%%%%5
\begin{Def}
\label{definition-canonical-divergence-information-geometry}
%%%%%%%%%%%%
(Canonical divergence,\, \cite{AN}): 
Let $(\cM,g,\nabla,\nabla^{\,*})$ be an $n$-dimensional dually flat space,
$\{\,x^{\,1},\ldots,x^{\,n}\,\}$ $\nabla$-affine coordinates, 
$\{\,y_{\,1},\ldots,y_{\,n}\,\}$ $\nabla^{\,*}$-affine coordinates, 
$\xi$ and $\xi^{\,\prime}$ two points of $\cM$. 
Then, the function %on $\cM\times\cM$,  
$\mbbD:\cM\times\cM\to\mbbR $, 
\beq
\mbbD\,(\,\xi\,\|\,\xi^{\,\prime}\,)
:=\psi(\,\xi\,)+\varphi(\,\xi^{\,\prime}\,)
-\left.x^{\,a}\right|_{\,\xi}\left.y_{\,a}\right|_{\,\xi^{\,\prime}},
\label{canonical-divergence}
\eeq
is referred to as the canonical divergence. 
%%%%%%%%%5
\end{Def}
%%%%%%%%%%%%
\begin{Remark}
%%%%%%%%%%%%%%
There is another convention for the canonical divergence
(\,see Ref.\,\cite{Fujiwara1995}\,). 
%%%%%%%%%%%%%
\end{Remark}
%%%%%%%%%%%%
% \begin{Remark}
% %%%%%%%%%%%%%%
% The canonical divergence 
% \fr{canonical-divergence} 
% should be  distinguished from the volume-form divergence 
% that has been in Definition\,\ref{definition-volume-form-compressibility}.  
% %%%%%%%%%%%%%
% \end{Remark}
%%%%%%%%%%%%

In information geometry, the following theorem is well-known.
%%%%%%%%%%%%
\begin{Thm}
\label{theorem-Pythagorean}
%%%%%%%%%% 
(Generalized Pythagorean theorem,\,\cite{AN}):  
Let $(\,\cM,g,\nabla,\nabla^{\,*}\,)$ be a dually flat space,
%$\xi$ and $\xi^{\,\prime}$ be two points of $\cM$, 
$\mbbD:\cM\times\cM\to\mbbR $ the canonical divergence, 
$\xi^{\,\prime},\xi^{\,\prime\prime},\xi^{\,\prime\prime\prime}$ be three points of 
$\cM$ such that 
1. $\xi^{\,\prime}$ and $\xi^{\,\prime\prime}$ are 
connected with the $\nabla^{\,*}$-geodesic curve and 
2. $\xi^{\,\prime\prime}$ and $\xi^{\,\prime\prime\prime}$ are connected with 
the $\nabla$-geodesic curve.
Then, it follows that 
$$
\mbbD\,(\,\xi^{\,\prime\prime\prime}\,\|\,\xi^{\,\prime}\,)
=\mbbD\,(\,\xi^{\,\prime\prime\prime}\,\|\,\xi^{\,\prime\prime}\,)
+\mbbD\,(\,\xi^{\,\prime\prime}\,\|\,\xi^{\,\prime}\,).
$$
%%%%%%%%%%
\end{Thm}
\subsection{Stokes-Dirac structure}
%%%%%%%%%%%%%%%%%%%%%%%%%%%%%%%%%%%%%%%5
To discuss the Hamiltonian formulation of distributed-parameter 
systems on bounded spatial domain, 
Dirac structure was extended. That extended structure 
is referred to as Stokes-Dirac structure\cite{Schaft-Maschke2002}.  
After  some spaces 
are introduced, the definition of Stokes-Dirac structure and a fundamental 
theorem are given. Also some definitions are given so that 
distributed-parameter systems can be discussed in the following sections. 
%%%%%%%%%%%%%
\begin{Def}
%%%%%%%%%%%
(Spaces of flow variables and effort variables): 
Let $\cZ$ be an $n$-dimensional connected manifold,  
and $p,q$ natural numbers satisfying $0\leq p,q\leq n$ and $p+q=n+1$. 
Then  
\beq
\cF_{\,p,q}
:=\Gamma\Lambda^{\,p}\cZ\times \Gamma\Lambda^{\,q}\cZ\times
\Gamma\Lambda^{\,n-p}\partial\cZ,\qquad 
\cE_{\,p,q}
:=\Gamma\Lambda^{\,n-p}\cZ\times \Gamma\Lambda^{\,n-q}\cZ\times
\Gamma\Lambda^{\,n-q}\partial\cZ,
\label{definition-flow-effort-forms}
\eeq
are referred to as the space of flow variables on $\cZ$, and 
the space of effort variables on $\cZ$, respectively. In addition, 
elements of $\cF_{\,p,q}$ and those of $\cE_{\,p,q}$ are referred to as 
flow variables and effort variables, respectively. 
%%%%%%%%%5
\end{Def}
%%%%%%%%%
\begin{Def}
%%%%%%%%%%%
(Bilinear form for Stokes-Dirac structure,\cite{Schaft-Maschke2002}):
Let $\cZ$ be an $n$-dimensional manifold, $\cF_{\,p,q}$ the space of 
flow variables, $\cE_{\,p,s}$ the space of effort variables. Then,  
the map $\bktt{-}{-}:\cF_{\,p,q}\times\cE_{\,p,q}\to\mbbR$ given by 
\beqa
&&\bktt{(\,\bbf_{\,p},\bbf_{\,q},\bbf_{\,\partial},\bfe_{\,p},\bfe_{\,q},\bfe_{\,\partial})}{(\,\bbf_{\,p}^{\,\prime},\bbf_{\,q}^{\,\prime},\bbf_{\,\partial}^{\,\prime},\bfe_{\,p}^{\,\prime},\bfe_{\,q}^{\,\prime},\bfe_{\,\partial}^{\,\prime})}
\non\\
&&\quad 
=\int_{\cZ}
\bfe_{\,p}\wedge\bbf_{\,p}^{\,\prime}+\bfe_{\,q}\wedge\bbf_{\,q}^{\,\prime}
+\bfe_{\,p}^{\,\prime}\wedge\bbf_{\,p}+\bfe_{\,q}^{\,\prime}\wedge\bbf_{\,q}
+\int_{\partial\cZ}
\bfe_{\partial}\wedge\bbf_{\partial}^{\,\prime}
+\bfe_{\partial}^{\,\prime}\wedge\bbf_{\partial},
\label{bilinear-form-Stokes-Dirac}
\eeqa
for 
$$
\bbf_{\,p},\bbf_{\,p}^{\,\prime}\,\in\,\Gamma\Lambda^{\,p}\cZ,\quad 
\bbf_{\,q},\bbf_{\,q}^{\,\prime}\,\in\,\Gamma\Lambda^{\,q}\cZ,\qquad 
\bfe_{\,p},\bfe_{\,p}^{\,\prime}\,\in\,\Gamma\Lambda^{\,n-p}\cZ,\quad
\bfe_{\,q},\bfe_{\,q}^{\,\prime}\,\in\,\Gamma\Lambda^{\,n-q}\cZ,
$$
and 
$$
\bbf_{\,\partial},\bbf_{\,\partial}^{\,\prime}\,\in\,\Gamma\Lambda^{\,n-p}\partial\cZ,
\qquad
\bfe_{\,\partial},\bfe_{\,\partial}^{\,\prime}\,\in\,\Gamma\Lambda^{\,n-q}\partial\cZ,
$$
%%%%%%%%%%
\end{Def}
%%%%%%%%%%
Then the following theorem holds.
%%%%%%%%%%%%
\begin{Thm}
\label{theorem-Stokes-Dirac-structure}
%%%%%%%%%%%
(Stokes-Dirac structure,\cite{Schaft-Maschke2002}):
Let $\cZ$ be an $n$-dimensional connected manifold, 
$\cF_{\,p,q}$ and $\cE_{\,p,q}$ the space of flow variables on $\cZ$ 
and that of 
effort variables on $\cZ$ given in 
\fr{definition-flow-effort-forms} with $p,q$ satisfying $0\leq p,q\leq n$ and 
$p+q=n+1$. 
Define the relations 
\beq
\left(
\begin{array}{c}
\bbf_{\,p}\\
\bbf_{\,q}
\end{array}
\right)
=\left(
\begin{array}{cc}
0&(-1)^{\,r}\dr\\
\dr&0
\end{array}
\right)
\left(
\begin{array}{c}
\bfe_{\,p}\\
\bfe_{\,q}
\end{array}
\right),\qquad
\left(
\begin{array}{c}
\bbf_{\,\partial}\\
\bfe_{\,\partial}
\end{array}
\right)
=
\left(
\begin{array}{c}
\bfe_{\,p}|_{\partial\cZ}\\
-(-1)^{n-q}\,\bfe_{\,q}|_{\partial\cZ}
\end{array}
\right),\quad 
r=p\,q+1.
\label{conditions-Stokes-Dirac-general}
\eeq
Then, the following linear subspace $\bD_{\,\cZ}$ of 
$\cF_{\,p,q}\times\cE_{\,p,q}$ 
\beq
\bD_{\,\cZ}
:=\left\{\,\left(\,
\bbf_{\,p},\bbf_{\,q},\bbf_{\,\partial},\bfe_{\,p},\bfe_{\,q},\bfe_{\,\partial}
\,\right)\,\bigg|\,\mbox{
The set of equations \fr{conditions-Stokes-Dirac-general} 
hold. }
\right\},
\label{Stokes-Dirac-general}
\eeq
satisfies $\bD_{\,\cZ}=\bD_{\,\cZ}^{\,\perp}$, with $\perp$ denoting the orthogonal 
complement with respect to the bilinear form given by 
\fr{bilinear-form-Stokes-Dirac}. 
%%%%%%%%%%%
\end{Thm}
%%%%%%%%%%%
 %%%%%%%%%%%%
\begin{Def}
%%%%%%%%%%%
The linear subspace $\bD_{\,\cZ}$ in 
Theorem\,\ref{theorem-Stokes-Dirac-structure} 
is referred to as  
a Stokes-Dirac structure.
%%%%%%%%%%%
\end{Def}
%%%%%%%%%%%
This paper studies the following class of Stokes-Dirac structures. 
%%%%%%%%%%%
\begin{Def}
\label{definition-distributed-parameter-port-Hamiltonian-system}
%%%%%%%%%%%
(Distributed-parameter port-Hamiltonian system,\cite{Schaft-Maschke2002}): 
Let $\cZ$ be an $n$-dimensional manifold, 
$\wt{\psi}\in\Gamma F\cZ$ a functional,   
$\Gamma\Lambda^{p}\cZ\times \Gamma\Lambda^{\,q}\cZ$ 
a state space with $p+q=n+1$ and $0\leq p,q\leq n$,  
and $\bD_{\,\cZ}$ the Stokes-Dirac structure  
given by \fr{Stokes-Dirac-general}. Then,   
for $(\,\balpha_{\,p},\balpha_{\,q}\,)\in\Gamma\Lambda^{\,p}\cZ\times\Gamma\Lambda^{\,q}\cZ$ with $r=p\,q+1$, identify 
\beq
\bbf_{\,p}
=-\frac{\partial}{\partial t}\balpha_{\,p},\quad
\bbf_{\,q}
=-\frac{\partial}{\partial t}\balpha_{\,q},\quad
\bfe_{\,p}
=\frac{\delta\wt{\psi}}{\delta\balpha_{\,p}},\quad
\bfe_{\,q}
=\frac{\delta\wt{\psi}}{\delta\balpha_{\,q}},
\label{flow-effort-variables-Stokes-Dirac-structure}
\eeq
such that 
$$
\left(
\begin{array}{c}
-\,\frac{\partial}{\partial t}\balpha_{\,p}\\
-\,\frac{\partial}{\partial t}\balpha_{\,q}
\end{array}
\right)
=\left(
\begin{array}{cc}
0&(-1)^{\,r}\dr\\
\dr&0
\end{array}
\right)
\left(
\begin{array}{c}
\frac{\delta}{\delta\balpha_{\,p}}\wt{\psi}\\
\frac{\delta}{\delta\balpha_{\,q}}\wt{\psi}
\end{array}
\right),
\qquad
\left(
\begin{array}{c}
\bbf_{\,\partial}\\
\bfe_{\,\partial}
\end{array}
\right)
=\left(
\begin{array}{cc}
1&0\\
0&-(-1)^{\,n-q}
\end{array}
\right)
\left(
\begin{array}{c}
\left.\frac{\delta}{\delta\balpha_{\,p}}\wt{\psi}\right|_{\partial\cZ}\\
\left.\frac{\delta}{\delta\balpha_{\,q}}\wt{\psi}\right|_{\partial\cZ}
\end{array}
\right).
$$
This system is 
referred to as the distributed-parameter port-Hamiltonian systems with 
$\cZ$.
%%%%%%%%%
\end{Def}
%%%%%%%%%%
\begin{Remark}
%%%%%%%%%%%%%%
By the power-conserving property, it follows for any 
$(\bbf_{\,p},\bbf_{\,q},\bbf_{\,\partial},\bfe_{\,p},\bfe_{\,q},\bfe_{\,\partial})$ 
in the Stokes-Dirac structure that 
\beq
\int_{\cZ}\left(\,\bfe_{\,p}\wedge \bbf_{\,p}+\bfe_{\,q}\wedge \bbf_{\,q}
\,\right)
+\int_{\partial\cZ}\bfe_{\,\partial}\wedge \bbf_{\,\partial}
=0.
\label{from-power-conserving-structure}
\eeq
Thus, one has from 
%\fr{energy-functional-time-derivative-general}
%and  
\fr{flow-effort-variables-Stokes-Dirac-structure} and 
\fr{from-power-conserving-structure} 
that 
$$
\frac{\dr}{\dr t}\wt{\psi}
=\int_{\cZ}
\frac{\delta\,\wt{\psi}}{\delta \balpha_{\,p}}
\wedge\frac{\partial\balpha_{\,p}}{\partial t}
+\frac{\delta\,\wt{\psi}}{\delta \balpha_{\,q}}
\wedge\frac{\partial\balpha_{\,q}}{\partial t}
=-\int_{\cZ}\left(\,
\bfe_{\,p}\wedge\bbf_{\,p}+\bfe_{\,q}\wedge\bbf_{\,q}
\,\right)
=\int_{\partial\cZ}\bfe_{\,\partial}\wedge\bbf_{\,\partial}.
$$
%%%%%%%%%%%%
\end{Remark}
%%%%%%%%%%%%
Physically the manifold $\cZ$ is used for expressing spatial domain. 
Thus, the cases $n=1,2$, and $3$ are focused in this paper.

Examples of distributed-parameter port-Hamiltonian systems 
have been given in Ref.\,\cite{Schaft-Maschke2002}. They are  
Maxwell's equations, telegraph equation, vibrating string, ideal fluid. 
Furthermore, various models can be described such as shallow water 
equations\cite{Schaft2014}.

%%%%%%%%%%%
\begin{Def}
\label{definition-energy-functional-energy-density-function}
%%%%%%%%%%%
(Energy functional and energy density function): 
A functional $\wt{\psi}\in\Gamma F\cZ$ used in Definition\,  
\ref{definition-distributed-parameter-port-Hamiltonian-system}
that can depend on $\balpha_{\,p}\in\Gamma\Lambda^{\,p}\cZ$ and 
$\balpha_{\,q}\in\Gamma\Lambda^{\,q}\cZ$ 
is referred to as an energy functional. 
On a Riemannian manifold $(\cZ,g)$ with $g$ being a Riemannian metric tensor 
field, let $\star 1$ be a canonical volume-form. If $\wt{\psi}$ can be written
as 
$$
\wt{\psi}
=\int_{\cZ}\psi\,\star 1,
$$
with 
$\psi\in\Gamma\Lambda^{\,0}\cZ$ being a function, 
then $\psi$ is referred to as an 
energy density function.
%%%%%%%%%%
\end{Def}
%%%%%%%%%%
% \begin{Remark}
% %%%%%%%%%%%%%%%%
% It follows that 
% \beq
% \frac{\dr \wt{\psi}}{\dr t}
% =\int_{\cZ}\frac{\delta\,\wt{\psi}}{\delta \balpha_{\,p}}
% \wedge\frac{\partial\balpha_{\,p}}{\partial t}
% +\frac{\delta\,\wt{\psi}}{\delta \balpha_{\,q}}
% \wedge\frac{\partial\balpha_{\,q}}{\partial t}.
% \label{energy-functional-time-derivative-general}
% \eeq
% %%%%%%%%%%%%
% \end{Remark}
%%%%%%%%%%%%%

%%%%%%%%%%%%
\begin{Def}
%%%%%%%%%%%
(Energy mixed form):
Let $\wt{\psi}$ be an energy functional for a Stokes-Dirac structure. Then 
$\balpha_{\,p}$ and $\balpha_{\,q}$ are referred to as energy mixed forms. 
%%%%%%%%%5
\end{Def}
%%%%%%%%%
%%%%%%%%%%%
\begin{Def}
\label{definition-co-energy-mixed-form}
%%%%%%%%%%%
(Co-energy mixed forms): 
Let $\wt{\psi}$ be an energy functional for a Stokes-Dirac structure. Then 
$\bfe_{\,p}\in\Gamma\Lambda^{n-p}\cZ$ and 
$\bfe_{\,q}\in\Gamma\Lambda^{n-q}\cZ$ derived from $\wt{\psi}$,
as appeared in \fr{flow-effort-variables-Stokes-Dirac-structure} as  
$$
\bfe_{\,p}
=\frac{\delta\wt{\psi}}{\delta\balpha_{\,p}},\quad\mbox{and}\quad
\bfe_{\,q}
=\frac{\delta\wt{\psi}}{\delta\balpha_{\,q}},
$$
are referred to as co-energy mixed forms.
%%%%%%%%%
\end{Def}
%%%%%%%%%5
The relations between 
$\bfe_{\,p},\bfe_{\,q}$ and $\balpha_{\,p},\balpha_{\,q}$
in Definition\,\ref{definition-co-energy-mixed-form} 
% $\bfe_{\,p}=\bfe_{\,p}(\balpha_{\,p},\balpha_{\,q})$ and 
% $\bfe_{\,q}=\bfe_{\,q}(\balpha_{\,p},\balpha_{\,q})$ 
 can be seen as constitutive relations. For example,  
for the case of $(3+1)$ decomposed Maxwell's equations 
on a Riemannian manifold, 
these constitutive relations are 
$\bfe=\varepsilon^{-1}\star \bfD,\bfh=\mu^{-1}\star\bfB$ where  $\bfD$ is 
a $2$-form displacement field, $\bfB$ a $2$-form magnetic induction field, 
$\bfe$ a $1$-form electric field, $\bfh$ a $1$-form magnetic field, 
$\varepsilon$ a permittivity, and $\mu$ a permeability   
(\,see Ref.\,\cite{Goto2017}\,).  

Given a Riemannian manifold, 
the following energy functionals are mainly focused in this paper.
%%%%%%%%%%%%%%
\begin{Def}
%%%%%%%%%%%%%
(Quadratic energy functional): 
Let $(\cZ,g)$ be a Riemannian manifold with $g$ being 
a Riemannian metric field, and 
$\star 1$ a canonical volume form. 
If $\wt{\psi}$ is quadratic in the sense that 
\beq
\wt{\psi}\left[\,\balpha_{\,p},\balpha_{\,q}\,\right]
=\frac{1}{2}\int_{\cZ}\left(\,
\balpha_{\,p}\wedge\star\,\balpha_{\,p}
+\balpha_{\,q}\wedge\star\,\balpha_{\,q}\,\right),
\label{quadratic-energy-functional-p-q}
\eeq 
then $\wt{\psi}$ is referred to as a quadratic energy functional.
%%%%%%%%%%%5
\end{Def}
%%%%%%%%%
%%%%%%%%%%%%%%%
\begin{Remark}
%%%%%%%%%%%%%%%
If $\wt{\psi}$ is quadratic, then it follows that 
$$
\bfe_{\,p}=\star\,\balpha_{\,p},\quad\mbox{and}\quad 
\bfe_{\,q}=\star\,\balpha_{\,q}.
$$
%%%%%%%%%%%%%%5
\end{Remark}
%%%%%%%%%%%%%%5

For the energy functional $\wt{\psi}$ given in   
\fr{quadratic-energy-functional-p-q}, one can introduce 
the co-energy functional. 
%%%%%%%%%%%%%
\begin{Def}
%%%%%%%%%%%5
(Co-energy functional): 
Given an energy functional, its total Legendre transform of functional 
( see Definition\,\ref{definition-Legendre-transform-functional} ) 
is referred to as a co-energy functional. 
%%%%%%%%%%
\end{Def}
%%%%%%%%%%%
In this paper the following class of co-energy functionals is focused.
%%%%%%%%%%%%%
\begin{Def}
%%%%%%%%%%%
(Quadratic co-energy functional): 
Given a functional $\wt{\psi}$ in \fr{quadratic-energy-functional-p-q}, 
the functional 
\beq
\wt{\varphi}\left[\,\bfe_{\,p},\bfe_{\,q}\,\right]
=\frac{1}{2}\int_{\cZ}
\left(\,\bfe_{\,p}\wedge\star\,\bfe_{\,p}
+\bfe_{\,q}\wedge\star\,\bfe_{\,q}
\,\right),\quad 
\bfe_{\,p}
=\frac{\delta\wt{\psi}}{\delta\balpha_{\,p}},\quad
\bfe_{\,q}
=\frac{\delta\wt{\psi}}{\delta\balpha_{\,q}},
\label{quadratic-co-energy-functional-p-q}
\eeq
is referred to as the co-energy functional. 
%%%%%%%%%55
\end{Def}
%%%%%%%%%%
The functional $\wt{\psi}$ in \fr{quadratic-energy-functional-p-q}
depends on $\balpha_{\,p}$ and $\balpha_{\,q}$.  
On the other hand the functional $\wt{\varphi}$ in 
\fr{quadratic-co-energy-functional-p-q} depends on the effort variables,
$\bfe_{\,p}$ and $\bfe_{\,q}$. 
In 
Section\,\ref{section-information-geometry-distributed-parameter-port-Hamiltonian},  
it will be shown that this $\wt{\varphi}$ is the total Legendre transform of 
the $\wt{\psi}$ in the sense of 
Definition\,\ref{definition-Legendre-transform-functional} 
%a link between these two functionals will be discussed 
(\,see 
Proposition\,\ref{proposition-total-Legendre-transform-quadratic-functional}\,). 
% To this end, one defines convex function.

%%%%%%%%%%%%%%%%%%%%%%%%%%%%%%%%%%%%%%%%
\section{Systems with respect to Stokes-Dirac structures 
described  as 
contact Hamiltonian vector fields}
\label{section-distributed-Hamiltonian-systems-Dirac}

In this section it is shown that 
a class of distributed-parameter port-Hamiltonian systems 
with respect to Stokes-Dirac structures are written in terms of 
contact geometry, where such systems are assumed to be described on 
Riemannian manifolds. 
To this end, contact bundles are used where 
base spaces are Riemannian manifolds.

To describe distributed-parameter port-Hamiltonian systems 
in a contact geometric language, 
one defines the following.
%%%%%%%%%%%%
\begin{Def}
%%%%%%%%%%
(Adapted mixed forms): 
Let $(\cZ,g)$ be an $n$-dimensional Riemannian manifold, 
$(\cK,\lambda_{\,\mbbV},\pi,\cZ)$ a contact manifold over $\cZ$, 
$p$ and $q$ natural numbers satisfying $0\leq p,q\leq n$ and $p+q=n+1$, 
$\cC_{\,\zeta}$ a 
$(2 ({}_{n}C_{\,p}+{}_{\,n}C_{\,q})+1)$-dimensional contact manifold over 
a point $\zeta\in\cZ$ such that $\cK=\bigcup_{\,\zeta\in\cB}\cC_{\,\zeta}$ with 
${}_{n}C_{\,p}=n!/(p!(n-p)!)$,
$\wt{\psi}\in\Gamma F\cK$  an energy functional consisting of 
$\balpha_{\,p}\in\Gamma\Lambda_{\,\mbbH,\mbbV}^{\,p,0}\cK$ 
and $\balpha_{\,q}\in\Gamma\Lambda_{\,\mbbH,\mbbV}^{\,q,0}\cK$, 
$\psi$ an energy density function on $\cZ$,  
$\bfe_{\,p}\in\Gamma\Lambda_{\,\mbbH,\mbbV}^{\,n-p,0}\cK$,   
$\bfe_{\,q}\in\Gamma\Lambda_{\,\mbbH,\mbbV}^{\,n-p,0}\cK$, and 
$\cE$ the coordinate such that the Reeb vertical vector field is 
$\partial/\partial \cE$.   
Then 
\beq
\Delta_{\,0}^{\,\zeta \psi}
:=\psi-\cE,\quad
\bDelta_{\,p}^{\,\zeta \psi}
:=\frac{\delta \wt{\psi}}{\delta\,\balpha_{\,p}}-\bfe_{\,p},\quad
\bDelta_{\,q}^{\,\zeta \psi}
:=\frac{\delta\,\wt{\psi}}{\delta\balpha_{\,q}}-\bfe_{\,q},\qquad
(1\leq p,q\leq n)
\label{adapted-mixed-form}
\eeq
are referred to as adapted mixed forms.
%%%%%%%%%%
\end{Def}
%%%%%%%%%%
\begin{Remark}
%%%%%%%%%%%%%%
The dimension of the space for $\balpha_{\,p}$ is ${}_{n}C_{\,p}$, 
which is the same as that for $\bfe_{\,p}$ due to 
${}_{n}C_{\,n-p}={}_{n}C_{\,p}$. Similarly  
the dimension of the space for $\balpha_{\,q}$ 
is ${}_{n}C_{\,q}$, which is the same as 
that for $\bfe_{\,q}$ due to ${}_{n}C_{\,n-q}={}_{n}C_{\,q}$.  
The explicit correspondences between 
canonical coordinates $(x,y,z)$ for $\cC_{\,\zeta}$ and 
$\balpha_{\,p},\balpha_{\,q},\bfe_{\,p},\bfe_{\,q}$ are discussed after 
$n,p,q$ are fixed. 
%%%%%%%%%%%%%
\end{Remark}
%%%%%%%%%%%%%

With the adapted mixed forms, the distributed-parameter port-Hamiltonian 
systems can be formulated in the following space.
%%%%%%%%%%%
\begin{Def}
\label{definition-phase-space-distributed-parameter-port-Hamiltonian-system}
%%%%%%%%%%
(Phase space for distributed-parameter port-Hamiltonian system): 
Let $\cA_{\,\zeta \psi}^{\,\cC}$ be the Legendre submanifold of 
the vertical space generated by $\psi$ as 
\beq
\cA_{\,\zeta \psi}^{\,\cC}
=\left\{\,(x,y,z)\in\pi^{-1}(\zeta)\,|\,
\Delta_{\,0}^{\,\zeta \psi}
=\bDelta_{\,p}^{\,\zeta \psi}
=\bDelta_{\,q}^{\,\zeta \psi}
=0
\,\right\}.
\label{phase-space-distributed-parameter-port-Hamiltonian-system}
\eeq
Then the sub-bundle $(\,\cA_{\,\psi}^{\,\cK},\pi|_{\,\cA_{\,\psi}^{\,\cK}},\cZ)$ 
with $\cA_{\,\psi}^{\,\cK}=\bigcup_{\,\zeta\in\cB}\cA_{\,\zeta \psi}^{\,\cC}$
is referred to as the phase space 
for the distributed-parameter port-Hamiltonian systems.
%%%%%%%%
\end{Def}
%%%%%%%%%%%

%In what follows the main theorems in this paper are stated.
%%%%%%%%%%%%%%%%%%%%%%%%
\subsection{Tree-dimensional Riemannian manifold}
%%%%%%%%%%%%%%%%%%%%%%% 
Before stating the main theorems in this paper, 
we note the following. 
On the phase space for the distributed-parameter port-Hamiltonian systems,
one has the distributed-parameter port-Hamiltonian systems.
For the case of $n=3$, the possible combinations of $0\leq p,q\leq n$ that 
satisfy $p+q=n+1$ are  
%%%%%%%%%%%%%%
\begin{itemize}
\item
$\{\,p=1,q=3\,\}$,\quad
$\{\,p=3,q=1\,\}$,
\item
$\{\,p=q=2\,\}$.
\end{itemize}
%%%%%%%%%%%%%% 
Since the case $\{\,p=3,q=1\,\}$ can reduce to 
$\{\,p=1,q=3\,\}$, attention is concentrated on the cases $\{\,p=1,q=3\,\}$ 
and $\{\,p=q=2\,\}$.   
Taking into account these combinations, one has the 
following one of main theorems.  
%%%%%%%%%%%
\begin{Thm}
\label{theorem-n-3-dirac-contact}
%%%%%%%%%%5 
(Distributed-parameter port-Hamiltonian system 
as contact Hamiltonian vertical vector field): 
Let $(\cZ,g)$ be a connected $3$-dimensional Riemannian manifold, 
and 
$\psi$ an energy density function specified later.  
%%%%%%%%%%%%%%%%
\begin{itemize}
\item
For the case $\{p=1,q=3\}$, let 
$(\cK,\lambda_{\,\mbbV},\pi,\cZ)$ be a $9$-dimensional contact 
manifold over the base space $\cZ$ with 
$\lambda_{\,\mbbV}=\dr_{\mbbV} z-p_{\,a}\,\dr_{\,\mbbV} x^{\,a}$,   
$\{\bsigma^{\,a}\}\in\Gamma\Lambda_{\,\mbbH}^{1}\cK$ an orthonormal co-frame,  
$$
\balpha_{\,p}
=(\,\balpha_{\,p}\,)_{\,a}\bsigma^{\,a}\ 
\in\Gamma\Lambda_{\,\mbbH,\mbbV}^{1,0}\cK,\qquad   
\balpha_{\,q}
=(\,\balpha_{\,q}\,)_{\,0}\,\star1\ \in\Gamma\Lambda_{\,\mbbH,\mbbV}^{3,0}\cK,
$$
$$
\bfe_{\,p}
=\frac{1}{2}(\,\bfe_{\,p}\,)_{\,ab}\,\bsigma^{\,a}\wedge\bsigma^{\,b}\ 
\in\Gamma\Lambda_{\,\mbbH,\mbbV}^{2,0}\cK,\qquad
\bfe_{\,q}
=(\,\bfe_{\,q}\,)_{\,0}
\in\Gamma\Lambda_{\,\mbbH,\mbbV}^{0,0}\cK,\qquad
\mbox{with}\quad
(\,\bfe_{\,p}\,)_{\,ba}
=-\,(\,\bfe_{\,p}\,)_{\,ab}
$$ 
mixed forms,   
$(x,y,z)$ canonical coordinates with $x=\{\,x_{\,(p)},x_{\,(q)}\,\}$, 
$y=\{\,y^{\,(p)},y^{\,(q)}\,\}$,  
\beqa
x&=&\{\,(\,\balpha_{\,p}\,)^{\,1},
(\,\balpha_{\,p}\,)^{\,2},
(\,\balpha_{\,p}\,)^{\,3},
\star\,\balpha_{\,q}
\,\},\quad
x_{\,(p)}^{\,a}
=\delta^{\,ab}(\,\balpha_{\,p}\,)_{\,b},\quad
x_{\,(q)}
=\star\,\balpha_{\,q}
=(\,\balpha_{\,q}\,)_{\,0},
\non\\
y&=&\{\,
(\,\star\,\bfe_{\,p}\,)_{\,1},
(\,\star\,\bfe_{\,p}\,)_{\,2},
(\,\star\,\bfe_{\,p}\,)_{\,3},
\bfe_{\,q}
\,\},\quad 
y_{\,a}^{\,(p)}
=(\,\star\,\bfe_{\,p}\,)_{\,a},\quad 
y^{\,(q)}
=(\,\bfe\,)_{\,0},\quad 
\star\,\bfe_{\,p}
=(\,\star\,\bfe_{\,p}\,)_{\,a}\bsigma^{\,a},
\non\\
z&=&\cE,
\non
\eeqa
where 
$\cE$ is the value of the energy density such that $\cE=\psi$ on  
$\cA_{\,\zeta \psi}^{\,\cC}$ with $\psi$ depending only on $x$. 
%%%%%
\item
%%%%%%
For the case $\{\,p=q=2\,\}$, let 
$(\cK,\lambda_{\,\mbbV},\pi,\cZ)$ be a $13$-dimensional contact 
manifold over the base space $\cZ$ with 
$
\lambda_{\,\mbbV}
=\dr_{\mbbV} z-p_{\,a}\,\dr_{\,\mbbV} x^{\,a}$,
$\{\bsigma^{\,a}\}\in\Gamma\Lambda_{\,\mbbH}^{1}\cK$ an orthonormal co-frame,  
$$
\balpha_{\,p}
=\frac{1}{2}(\,\balpha_{\,p}\,)_{\,ab}\,\bsigma^{\,a}\wedge\bsigma^{\,b}\,
\in\Gamma\Lambda_{\,\mbbH,\mbbV}^{2,0}\cK,\quad   
\balpha_{\,q}
=\frac{1}{2}(\,\balpha_{\,q}\,)_{\,ab}\,\bsigma^{\,a}\wedge\bsigma^{\,b}\,
\in\Gamma\Lambda_{\,\mbbH,\mbbV}^{2,0}\cK,
$$
with 
$(\,\balpha_{\,p}\,)_{\,ba}=-\,(\,\balpha_{\,p}\,)_{\,ab}$,
$(\,\balpha_{\,q}\,)_{\,ba}=-\,(\,\balpha_{\,q}\,)_{\,ab}$, 
$$ 
\bfe_{\,p}
=(\bfe_{\,p})_{\,a}\,\bsigma^{\,a}\ 
\in\Gamma\Lambda_{\,\mbbH,\mbbV}^{1,0}\cK,\qquad
\bfe_{\,q}
=(\bfe_{\,q})_{\,a}\,\bsigma^{\,a}\ 
\in\Gamma\Lambda_{\,\mbbH,\mbbV}^{1,0}\cK 
$$
mixed forms,   
$(x,y,z)$ be canonical coordinates with $x=\{\,x_{\,(p)},x_{\,(q)}\,\}$, 
$y=\{\,y^{\,(p)},y^{\,(q)}\,\}$,  
\beqa
x&=&\{\,(\,\star\,\balpha_{\,p}\,)^{\,1},
(\,\star\,\balpha_{\,p}\,)^{\,2},
(\,\star\,\balpha_{\,p}\,)^{\,3},
(\,\star\,\balpha_{\,q}\,)^{\,1},
(\,\star\,\balpha_{\,q}\,)^{\,2},
(\,\star\,\balpha_{\,q}\,)^{\,3}\},
\non\\
&&
\quad
x_{\,(p)}^{\,a}
=\delta^{\,ab}(\,\star\,\balpha_{\,p}\,)_{\,b},\quad 
x_{\,(q)}^{\,a}
=\delta^{\,ab}(\,\star\,\balpha_{\,q}\,)_{\,b},
\non\\ 
y&=&
\{\,(\,\bfe_{\,p}\,)_{\,1},(\,\bfe_{\,p}\,)_{\,2},(\,\bfe_{\,p}\,)_{\,3},
(\,\bfe_{\,q}\,)_{\,1},(\,\bfe_{\,q}\,)_{\,2},(\,\bfe_{\,q}\,)_{\,3}\}, 
\qquad
y_{\,a}^{\,(p)}
=(\,\bfe_{\,p}\,)_{\,a},\qquad 
y_{\,a}^{\,(q)}
=(\,\bfe_{\,q}\,)_{\,a},
\non\\
z&=&\cE,
\non
\eeqa
where $\cE$ is the value of the energy density such that $\cE=\psi$ on 
$\cA_{\,\zeta \psi}^{\,\cC}$ 
with $\psi$ depending only on $x$. 
%%%%%%%%%%%%%
\end{itemize}
%%%%%%%%%%%%%
For the both of cases, 
let $\wt{\psi}$ be a quadratic energy functional such that 
$$
\wt{\psi}
=\frac{1}{2}\int_{\cZ}\left(\,\balpha_{\,p}\wedge\star\,\balpha_{\,p}
+\balpha_{\,q}\wedge\star\,\balpha_{\,q}\,\right)
=\int_{\cZ}\psi\star 1,\quad\mbox{so that }\quad 
\bfe_{\,p}
=\frac{\delta\wt{\psi}}{\delta\balpha_{\,p}}\quad \mbox{and}\quad 
\bfe_{\,q}
=\frac{\delta\wt{\psi}}{\delta\balpha_{\,q}}.
$$
In addition, choose the contact Hamiltonian functional as  
$$
\wt{h}_{\,\psi}
=\int_{\cZ}h_{\,\psi}\star 1
=\int_{\cZ}\,\left[\,
\bDelta_{\,p}^{\,\zeta \psi}\wedge \left(\,-\bF_{\,\psi}^{\,\zeta p}\,\right)
+\bDelta_{\,q}^{\,\zeta \psi}\wedge \left(\,-\bF_{\,\psi}^{\,\zeta q}\,\right)
+\Gamma_{\,\zeta \psi}\left(\,\bDelta_{\,0}^{\,\zeta \psi}\,\right)\,\star 1
\,\right],
$$
where $\Delta_{\,0}^{\,\zeta \psi},\bDelta_{\,p}^{\,\zeta \psi},\bDelta_{\,q}^{\,\zeta \psi}$
are defined in \fr{adapted-mixed-form},  
$h_{\,\psi}\in\Gamma\Lambda_{\,\mbbH,\mbbV}^{\,0,0}\cK$, 
$\bF_{\,\psi}^{\,\zeta p}\in\Gamma\Lambda_{\,\mbbH,\mbbV}^{\,p,0}\cK$,
$\bF_{\,\psi}^{\,\zeta q}\in\Gamma\Lambda_{\,\mbbH,\mbbV}^{\,q,0}\cK$ 
are 
$$
h_{\,\psi}
=\star^{\,-1}\left[\,
\bDelta_{\,p}^{\,\zeta \psi}\wedge\left(\,-\, \bF_{\,\psi}^{\,\zeta p}\,\right)
\,\right]
+\star^{\,-1}\left[\,
\bDelta_{\,q}^{\,\zeta \psi}\wedge\left(\,-\, \bF_{\,\psi}^{\,\zeta q}\,\right)
\,\right]+\Gamma_{\,\zeta \psi}\left(\,\bDelta_{\,0}^{\,\zeta \psi}\,\right),
$$
$$
\bF_{\,\psi}^{\,\zeta p}
:=(-1)^{\,r}\dr\left(\,\frac{\delta\wt{\psi}}{\delta \balpha_{\,q}}
\right)
=(-1)^{\,r}\,\dr\star\balpha_{\,q},\qquad 
\bF_{\,\psi}^{\,\zeta q}
:=\dr\left(\frac{\delta\wt{\psi}}{\delta\balpha_{\,p}}\right)
=\dr\star\balpha_{\,p},
$$
respectively, and $\Gamma_{\,\zeta \psi}$ is such that 
$$
\Gamma_{\,\zeta \psi}\left(\,\Delta_{\,0}^{\,\zeta \psi}\,\right)
=\left\{
\begin{array}{cc}
0&\mbox{for}\quad \Delta_{\,0}^{\,\zeta \psi}=0\\
\mbox{non-zero}&\mbox{for}\quad \Delta_{\,0}^{\,\zeta \psi}\neq 0.
\end{array}
\right.
$$
% and 
% $$
% \left.
% \frac{\dr}{\dr\Delta_{\,0}^{\,\zeta \psi}}\right|_{\Delta_{\,0}^{\,\zeta \psi}=0}
% \Gamma_{\,\zeta \psi}\left(\,\Delta_{\,0}^{\,\zeta \psi}\,\right)
% =0.
% $$
Also, define 
$$
\bbf_{\,\partial}
=\left.\frac{\delta\wt\psi}{\delta\balpha_{\,p}}\right|_{\partial\cZ},
\quad\mbox{and}\quad
\bfe_{\,\partial}
=(-1)^{\,p}\left.\frac{\delta\wt\psi}{\delta\balpha_{\,q}}\right|_{\partial\cZ}.
$$ 
Then, the restricted contact Hamiltonian vertical vector field 
$X_{\,\wt{h}_{\,\psi}}|_{\,\cA_{\,\psi}^{\,\cK}}$ onto 
the phase space for distributed-parameter port-Hamiltonian system 
(\,see Definition\,
\ref{definition-phase-space-distributed-parameter-port-Hamiltonian-system}\,)
gives distributed-parameter 
port-Hamiltonian systems defined in 
Definition\,\ref{definition-distributed-parameter-port-Hamiltonian-system}. 
In addition, 
\fr{from-power-conserving-structure} is satisfied.
%%%%%%%%%%5
\end{Thm} 
%%%%%%%%%
\begin{Proof}
%%%%%%%%%%%%%%
Throughout this proof, $\star\star\balpha=\balpha$ and 
$\star^{-1}\balpha=\star\,\balpha$ 
for any $p$-form $\balpha$ (\,see Lemma\,\ref{3-d-Riemannian-Hodge}\,), 
and $\cA_{\,\zeta \psi}^{\cC}=\{\,\wt{h}_{\,\psi}=0\,\}$ are used. 
The case of $\{\,p=1,q=3\,\}$ is proven first and then the case of $\{\,p=q=2\,\}$ is 
proven.  

\noindent
(Proof for the case of $p=1,q=3$): 
% Write 
% $$
% \balpha_{\,p}
% =(\,\balpha_{\,p}\,)_{\,a}\,\bsigma^{\,a},\quad
% \balpha_{\,q}
% =(\,\balpha_{\,q}\,)_{\,0}\,\star 1
% \quad\mbox{and}\quad 
% \bfe_{\,p}=\frac{1}{2}(\,\bfe_{\,p}\,)_{\,ab}\,\bsigma^{\,a}\wedge\bsigma^{\,b},\quad
% \bfe_{\,q}=(\,\bfe_{\,q}\,)_{\,0}, 
% $$
% with $(\,\bfe_{\,p}\,)_{\,ba}=-(\,\bfe_{\,p}\,)_{\,ab}$, and 
% Let $x=\{x_{\,(p)},x_{\,(q)}\}=\{x_{\,(p)}^{\,1},x_{\,(p)}^{\,2},x_{\,(p)}^{\,3},x_{\,(q)}\}$ 
% and $y=\{y^{\,(p)},y^{\,(q)}\}=\{y_{\,1}^{\,(p)},y_{\,2}^{\,(p)},y_{\,3}^{\,(p)},y^{\,(q)}\}$ 
% such that 
% \beqa
% x&=&\{\,(\,\balpha_{\,p}\,)^{\,1},
% (\,\balpha_{\,p}\,)^{\,2},
% (\,\balpha_{\,p}\,)^{\,3},
% (\,\balpha_{\,q}\,)^{\,0}
% \,\},\qquad
% x_{\,(p)}^{\,a}
% =\delta^{\,ab}(\,\balpha_{\,p}\,)_{\,b},\qquad
% x_{\,(q)}
% =(\,\balpha_{\,q}\,)^{\,0}
% =(\,\balpha_{\,q}\,)_{\,0},
% \non\\
% y&=&\{\,
% (\,\star\,\bfe_{\,p}\,)_{\,1},
% (\,\star\,\bfe_{\,p}\,)_{\,2},
% (\,\star\,\bfe_{\,p}\,)_{\,3},
% (\,\bfe_{\,q}\,)_{\,0},
% \,\},\qquad 
% y_{\,a}^{\,(p)}
% =(\,\star\,\bfe_{\,p}\,)_{\,a},\qquad 
% y^{\,(q)}
% =(\,\bfe\,)_{\,0}.
% \non
% \eeqa
Since 
$$
%\int_{\cZ}\bDelta_{\,p}^{\,\zeta \psi}\wedge\left(\,-\bF_{\,\psi}^{\,\zeta p}\,\right)
%=\int_{\cZ}\star
\star^{-1}\left[\,\bDelta_{\,p}^{\,\zeta \psi}\wedge\left(\,
-\bF_{\,\psi}^{\,\zeta p}\,\right)\,\right]
=%\int_{\cZ}
g^{-1}\left(\,\star\,\bDelta_{\,p}^{\,\zeta \psi},-\,\bF_{\,\psi}^{\,\zeta p}\,\right),%\,\star 1,
\quad\mbox{and}\quad 
\star^{-1}\left[
\bDelta_{\,q}^{\,\zeta \psi}\wedge\left(-\bF_{\,\psi}^{\,\zeta q}\right)
\right]
=\bDelta_{\,q}^{\,\zeta \psi}\wedge\left(-\star\bF_{\,\psi}^{\,\zeta q}\right),
$$
%and so on, 
the contact Hamiltonian density function $h_{\,\psi}$ is written as 
\beqa
h_{\,\psi}
&=&g^{-1}
\left(\,\star\,\bDelta_{\,p}^{\,\zeta \psi},-\,\bF_{\,\psi}^{\,\zeta p}\,\right)
+\bDelta_{\,q}^{\,\zeta \psi}\left(\,-\,\star\,\bF_{\,\psi}^{\,\zeta q}\,\right)
+\Gamma_{\,\zeta \psi}\left(\,\bDelta_{\,0}^{\,\zeta \psi}\,\right),
\non\\
&=&\delta^{\,ab}\left(\,\star\,\bDelta_{\,p}^{\,\zeta \psi}\,\right)_{\,a}
\left(\,-\,\bF_{\,\psi}^{\,\zeta p}\,\right)_{\,b}
+\bDelta_{\,q}^{\,\zeta \psi}\left(\,-\,\star\,\bF_{\,\psi}^{\,\zeta q}\,\right)
+\Gamma_{\,\zeta \psi}\left(\,\bDelta_{\,0}^{\,\zeta \psi}\,\right),
\non
\eeqa
where 
$$
\star\,\bDelta_{\,p}^{\,\zeta \psi}
=\left(\,\star\,\bDelta_{\,p}^{\,\zeta \psi}\,\right)_{\,a}\,\bsigma^{\,a},
\quad\mbox{and}\quad
\bF_{\,\psi}^{\,\zeta p}
=\left(\,\bF_{\,\psi}^{\,\zeta p}\,\right)_{\,a}\,\bsigma^{\,a}.
$$
In addition, one has from \fr{adapted-mixed-form} that 
$$
\left(\,\star\,\bDelta_{\,p}^{\,\zeta \psi}\right)_{\,a}
=(\,\balpha_{\,p}\,)_{\,a}-(\,\star\,\bfe_{\,p})_{\,a},\quad
\left(\,\bDelta_{\,q}^{\,\zeta \psi}\right)
=(\,\star\,\balpha_{\,q}\,)-(\,\bfe_{\,q}\,),\qquad 
\mbox{where}\quad \star\,\bfe_{\,p}
=(\,\star\,\bfe_{\,p}\,)_{\,a}\bsigma^{\,a}.
$$
The component expression of the restricted contact vertical vector field is 
obtained from \fr{coordinate-expression-contact-Hamiltonian-vertical-vector} 
as 
\beqa
\left.\dot{x}_{\,(p)}^{\,a}\right|_{\,\cA_{\,\zeta \psi}^{\cC}}
&=&-\,\left.\frac{\partial h_{\,\psi}}{\partial y_{\,a}^{\,(p)}}
\right|_{\,\cA_{\,\zeta \psi}^{\cC}}
=-\,\left.\delta^{\,ab}\left(\,\bF_{\,\psi}^{\,\zeta p}\,\right)_{\,b}
\right|_{\,\cA_{\,\zeta \psi}^{\cC}},
\non\\
\left.\dot{x}_{\,(q)}\right|_{\,\cA_{\,\zeta \psi}^{\cC}}
&=&-\,\left.\frac{\partial h_{\,\psi}}{\partial y^{\,(q)}}
\right|_{\,\cA_{\,\zeta \psi}^{\cC}}
=-\,\left.\star\,\bF_{\,\psi}^{\,\zeta q}
\right|_{\,\cA_{\,\zeta \psi}^{\cC}},
\non\\
\left.\dot{y}_{\,a}^{\,(p)}\right|_{\,\cA_{\,\zeta \psi}^{\cC}}
&=&\left.\left(\frac{\partial h_{\,\psi}}{\partial x_{\,(p)}^{\,a}}
+y_{\,a}^{\,(p)}\frac{\partial h_{\,\psi}}{\partial z}\,\right)
\right|_{\,\cA_{\,\zeta \psi}^{\cC}}
=-\,\left.\left(\,\bF_{\,\psi}^{\,\zeta p}\,\right)_{\,a}
\right|_{\,\cA_{\,\zeta \psi}^{\cC}},
\non\\
\left.\dot{y}^{\,(q)}
\right|_{\,\cA_{\,\zeta \psi}^{\cC}}
&=&\left.\left(\frac{\partial h_{\,\psi}}{\partial x_{\,(q)}}
+y^{\,(q)}\frac{\partial h_{\,\psi}}{\partial z}\right)
\right|_{\,\cA_{\,\zeta \psi}^{\cC}}
=-\,\left.\star\,\bF_{\,\psi}^{\,\zeta q}
\right|_{\,\cA_{\,\zeta \psi}^{\cC}},
\non\\
\left.\dot{z}\right|_{\,\cA_{\,\zeta \psi}^{\cC}}
&=&\left.
\left(h_{\,\psi}-y_{\,a}^{\,(p)}\frac{\partial h_{\,\psi}}{\partial y_{\,a}^{\,(p)}}
-y_{\,a}^{\,(q)}\frac{\partial h_{\,\psi}}{\partial y_{\,a}^{\,(q)}}
\right)\right|_{\,\cA_{\,\zeta \psi}^{\cC}}
=-\,\left.\left[g^{\,-1}\left(\,\star\,\bfe_{\,p},\bF_{\,\psi}^{\,\zeta p}\right)
+\bfe_{\,q}\,\left(\,\star\,\bF_{\,\psi}^{\,\zeta q}\,\right)\,\right]
\right|_{\,\cA_{\,\zeta \psi}^{\cC}}.
\non
\eeqa
These are equivalent to write 
$$
-\frac{\partial\balpha_{\,p}}{\partial t}
=\bF_{\,\psi}^{\,\zeta p},\qquad 
-\frac{\partial\balpha_{\,q}}{\partial t}
=\bF_{\,\psi}^{\,\zeta q},
$$
$$
\frac{\partial\bfe_{\,p}}{\partial t}
=-\,\star\,\bF_{\,\psi}^{\,\zeta p}
=\frac{\partial}{\partial t}\left(\,\star\balpha_{\,p}\,\right),\quad
\frac{\partial\bfe_{\,q}}{\partial t}
=-\,\star\,\bF_{\,\psi}^{\,\zeta q}
=\frac{\partial}{\partial t}\left(\,\star\balpha_{\,q}\,\right),
\quad\mbox{on}\ \cA_{\,\zeta \psi}^{\cC},
$$
and 
\beqa
\dot{z}
&=&\dot{\psi}
=-\,\star\,\left(\,\bfe_{\,p}\wedge \bF_{\,\psi}^{\,\zeta p}\,\right)
-\bfe_{\,q}\left(\,\star\, \bF_{\,\psi}^{\,\zeta q}\,\right)
=-\,\star\,\left[\,%-\,
\frac{\delta\wt{\psi}}{\delta\balpha_{\,p}}\wedge
\dr\left(\frac{\delta\wt{\psi}}{\delta\balpha_{\,q}}
\right)+\frac{\delta\wt{\psi}}{\delta\balpha_{\,q}}\wedge
\dr\left(\frac{\delta\wt{\psi}}{\delta\balpha_{\,p}}\right)
\,\right]
\non\\
&=&-\,\star\,\dr\left(\,
\frac{\delta\wt{\psi}}{\delta\balpha_{\,p}}\wedge
\frac{\delta\wt{\psi}}{\delta\balpha_{\,q}}
\,\right)
\quad
\mbox{on}\ \cA_{\,\zeta \psi}^{\cC}.
\non
\eeqa
The last equation above yields the following 
$$
\frac{\dr\wt{\psi}}{\dr t}
=\int_{\cZ}\dot{\psi}\star 1
=-\,\int_{\cZ}\dr\,\left(\,
\frac{\delta\wt{\psi}}{\delta\balpha_{\,p}}\wedge
\frac{\delta\wt{\psi}}{\delta\balpha_{\,q}}
\,\right)
=-\,\int_{\partial\cZ}\left(\,
\frac{\delta\wt{\psi}}{\delta\balpha_{\,p}}\wedge
\frac{\delta\wt{\psi}}{\delta\balpha_{\,q}}
\,\right)
=\int_{\partial\cZ}\bbf_{\,\partial}\wedge\bfe_{\,\partial}
=\int_{\partial\cZ}\bfe_{\,\partial}\wedge\bbf_{\,\partial},
$$
where 
$$
\bbf_{\,\partial}
=\left.\frac{\partial\wt{\psi}}{\delta\balpha_{\,p}}\right|_{\,\partial\cZ},
\qquad\mbox{and}\qquad 
\bfe_{\,\partial}
=-\,\left.\frac{\partial\wt{\psi}}{\delta\balpha_{\,q}}\right|_{\,\partial\cZ}
$$
have been used.
Thus, one obtains 
$$
\frac{\dr\wt{\psi}}{\dr t}
=\int_{\cZ}\left(\,
\frac{\delta\wt{\psi}}{\delta\balpha_{\,p}}
\wedge\frac{\partial\balpha_{\,p}}{\partial t}
+\frac{\delta\wt{\psi}}{\delta\balpha_{\,q}}
\wedge\frac{\partial\balpha_{\,q}}{\partial t}
\,\right)
=-\,\int_{\cZ}\left(\,
\bfe_{\,p}\wedge\bbf_{\,p}+\bfe_{\,q}\wedge\bbf_{\,q}
\,\right)
=\int_{\partial\cZ}\bfe_{\,\partial}\wedge\bbf_{\,\partial},
$$
from which one has \fr{from-power-conserving-structure}.

\noindent
(Proof for the case of $\{\,p=q=2\,\}$): 
% Write 
% $$
% \balpha_{\,p}
% =\frac{1}{2}
% (\,\balpha_{\,p}\,)_{\,ab}\,\bsigma^{\,a}\wedge\bsigma^{\,b},\quad
% \balpha_{\,q}
% =\frac{1}{2}
% =(\,\balpha_{\,q}\,)_{\,ab}\,\bsigma^{\,a}\wedge\bsigma^{\,b},
% \quad\mbox{and}\quad 
% \bfe_{\,p}=(\,\bfe_{\,p}\,)_{\,a}\,\bsigma^{\,a},\quad
% \bfe_{\,q}=(\,\bfe_{\,q}\,)_{\,a}\,\bsigma^{\,a}, 
% $$
% with $(\,\balpha_{\,q}\,)_{\,ba}=-(\,\balpha_{\,q}\,)_{\,ab}$,  
% and 
% $$
% \star\,\balpha_{\,p}
% =(\,\star\,\balpha_{\,p})_{\,a}\,\bsigma^{\,a},\quad
% \quad\mbox{where}\quad
% %(\balpha_{\,p})_{\,a}=
% (\star\,\balpha_{\,p})_{\,a}
% =\frac{1}{2}\epsilon_{\,a}^{\ bc}(\,\balpha_{\,p}\,)_{\,bc}.
% $$
% Let $x=\{\,x_{\,(p)}^{\,a},x_{\,(q)}^{\,a}\,\}$ and 
% $y=\{\,y_{\,a}^{\,(p)},y_{\,a}^{\,(q)}\,\}$ such that 
% \beqa
% x&=&\{\,(\,\star\,\balpha_{\,p}\,)^{\,1},
% (\,\star\,\balpha_{\,p}\,)^{\,2},
% (\,\star\,\balpha_{\,p}\,)^{\,3},
% (\,\star\,\balpha_{\,q}\,)^{\,1},
% (\,\star\,\balpha_{\,q}\,)^{\,2},
% (\,\star\,\balpha_{\,q}\,)^{\,3}\},
% \non\\
% &&x_{\,(p)}^{\,a}
% =\delta^{\,ab}(\,\star\,\balpha_{\,p}\,)_{\,b},\qquad 
% x_{\,(q)}^{\,a}
% =\delta^{\,ab}(\,\star\,\balpha_{\,q}\,)_{\,b},
% \non\\ 
% y&=&
% \{\,(\,\bfe_{\,p}\,)_{\,1},(\,\bfe_{\,p}\,)_{\,2},(\,\bfe_{\,p}\,)_{\,3},
% (\,\bfe_{\,q}\,)_{\,1},(\,\bfe_{\,q}\,)_{\,2},(\,\bfe_{\,q}\,)_{\,3}\}. 
% \non\\
% &&
% y_{\,a}^{\,(p)}
% =(\,\bfe_{\,p}\,)_{\,a},\qquad 
% y_{\,a}^{\,(q)}
% =(\,\bfe_{\,q}\,)_{\,a},
% \non
% \eeqa
Since 
$$
\star^{-1}\left[\, \left(\,-\,\bF_{\,\psi}^{\,\zeta p}\,\right)\wedge
\bDelta_{\,p}^{\,\zeta \psi}
\right]
=g^{-1}\left(\,\bDelta_{\,p}^{\,\zeta \psi},-\star\,\bF_{\,\psi}^{\,\zeta p}\,\right),
$$
the contact Hamiltonian density function $h_{\,\psi}$ is written as 
\beqa
h_{\,\psi}
&=&g^{-1}
\left(\,\bDelta_{\,p}^{\,\zeta \psi},-\star\,\bF_{\,\psi}^{\,\zeta p}\,\right)
+g^{-1}
\left(\,\bDelta_{\,q}^{\,\zeta \psi},-\star\,\bF_{\,\psi}^{\,\zeta q}\,\right)
+\Gamma_{\,\zeta \psi}\left(\,\bDelta_{\,0}^{\,\zeta \psi}\,\right)
\non\\
&=&\delta^{\,ab}\left(\bDelta_{\,p}^{\,\zeta \psi}\right)_{\,a}
\left(\,-\star\,\bF_{\,\psi}^{\,\zeta p}\right)_{\,b}
+\delta^{\,ab}
\left(\,\bDelta_{\,q}^{\,\zeta \psi}\,\right)_{\,a}
\left(\,-\star\,\bF_{\,\psi}^{\,\zeta q}\,\right)_{\,b}
+\Gamma_{\,\zeta \psi}\left(\,\bDelta_{\,0}^{\,\zeta \psi}\,\right),
\non
\eeqa
where 
$$
\bDelta_{\,p}^{\,\zeta \psi}
=\left(\bDelta_{\,p}^{\,\zeta \psi}\right)_{\,a}\,\bsigma^{\,a},\quad
\bDelta_{\,q}^{\,\zeta \psi}
=\left(\bDelta_{\,q}^{\,\zeta \psi}\right)_{\,a}\,\bsigma^{\,a},\quad 
\star\,\bF_{\,p}^{\,\zeta \psi}
=\left(\star\,\bF_{\,p}^{\,\zeta \psi}\right)_{\,a}\bsigma^{\,a},\quad
\star\,\bF_{\,q}^{\,\zeta \psi}
=\left(\star\,\bF_{\,q}^{\,\zeta \psi}\right)_{\,a}\bsigma^{\,a}.
$$
In addition, one has from \fr{adapted-mixed-form} that
$$
\left(\,\bDelta_{\,p}^{\,\zeta \psi}\,\right)_{\,a}
=\left(\,\star\,\balpha_{\,p}\,\right)_{\,a}-(\,\bfe_{\,p}\,)_{\,a},\quad
\left(\,\bDelta_{\,q}^{\,\zeta \psi}\,\right)_{\,a}
=\left(\,\star\,\balpha_{\,q}\,\right)_{\,a}-(\,\bfe_{\,q}\,)_{\,a},\quad
$$
where 
$$
(\,\star\,\balpha_{\,p}\,)_{\,a}
=\frac{1}{2}\epsilon_{\,a}^{\ bc}(\,\balpha_{\,p}\,)_{\,bc},\qquad 
(\,\star\,\balpha_{\,q}\,)_{\,a}
=\frac{1}{2}\epsilon_{\,a}^{\ bc}(\,\balpha_{\,p}\,)_{\,bc}. 
$$
The component expression of the restricted contact vertical vector field is 
obtained from \fr{coordinate-expression-contact-Hamiltonian-vertical-vector} 
as 
\beqa
\left.\dot{x}_{\,(p)}^{\,a}\right|_{\,\cA_{\,\zeta \psi}^{\cC}}
&=&-\,\left.\frac{\partial h_{\,\psi}}{\partial y_{\,a}^{\,(p)}}
\right|_{\,\cA_{\,\zeta \psi}^{\cC}}
=-\,\left.\delta^{\,ab}\left(\star\,\bF_{\,\psi}^{\,\zeta p}\right)_{\,b}
\right|_{\,\cA_{\,\zeta \psi}^{\cC}},
\non\\
\left.\dot{x}_{\,(q)}^{\,a}\right|_{\,\cA_{\,\zeta \psi}^{\cC}}
&=&-\,\left.\frac{\partial h_{\,\psi}}{\partial y_{\,a}^{\,(q)}}
\right|_{\,\cA_{\,\zeta \psi}^{\cC}}
=-\,\left.\delta^{\,ab}\left(\star\,\bF_{\,\psi}^{\,\zeta q}\right)_{\,b}
\right|_{\,\cA_{\,\zeta \psi}^{\cC}},
\non\\
\left.\dot{y}_{\,a}^{\,(p)}\right|_{\,\cA_{\,\zeta \psi}^{\cC}}
&=&\left.\left(\frac{\partial h_{\,\psi}}{\partial x_{\,(p)}^{\,a}}
+y_{\,a}^{\,(p)}\frac{\partial h_{\,\psi}}{\partial z}\right)
\right|_{\,\cA_{\,\zeta \psi}^{\cC}}
=-\,\left.\left(\star\,\bF_{\,\psi}^{\,\zeta p}\right)_{\,a}
\right|_{\,\cA_{\,\zeta \psi}^{\cC}},
\non\\
\left.\dot{y}_{\,a}^{\,(q)}\right|_{\,\cA_{\,\zeta \psi}^{\cC}}
&=&\left.\left(\frac{\partial h_{\,\psi}}{\partial x_{\,(q)}^{\,a}}
+y_{\,a}^{\,(q)}\frac{\partial h_{\,\psi}}{\partial z}\right)
\right|_{\,\cA_{\,\zeta \psi}^{\cC}}
=-\,\left.\left(\star\,\bF_{\,\psi}^{\,\zeta q}\right)_{\,a}
\right|_{\,\cA_{\,\zeta \psi}^{\cC}},
\non\\
\left.\dot{z}\right|_{\,\cA_{\,\zeta \psi}^{\cC}}
&=&\left.
\left(h_{\,\psi}-y_{\,a}^{\,(p)}\frac{\partial h_{\,\psi}}{\partial y_{\,a}^{\,(p)}}
-y_{\,a}^{\,(q)}\frac{\partial h_{\,\psi}}{\partial y_{\,a}^{\,(q)}}
\right)\right|_{\,\cA_{\,\zeta \psi}^{\cC}}
=-\,\left.\left[\,g^{\,-1}\left(\bfe_{\,p},\star\,\bF_{\,\psi}^{\,\zeta p}\right)
+g^{\,-1}\left(\bfe_{\,q},\star\,\bF_{\,\psi}^{\,\zeta q}\right)\,\right]
\right|_{\,\cA_{\,\zeta \psi}^{\cC}}.
\non
\eeqa
These are equivalent to write 
$$
-\,\frac{\partial\,\balpha_{\,p}}{\partial t}
=\bF_{\,\psi}^{\,\zeta p},\qquad 
-\,\frac{\partial\,\balpha_{\,q}}{\partial t}
=\bF_{\,\psi}^{\,\zeta q},
$$
$$
\frac{\partial\, \bfe_{\,p}}{\partial t}
=-\,\star\,\bF_{\,\psi}^{\,\zeta p}
=\frac{\partial\, }{\partial t}\,\left( \star\,\balpha_{\,p}\,\right),\qquad 
\frac{\partial\, \bfe_{\,q}}{\partial t}
=-\,\star\,\bF_{\,\psi}^{\,\zeta q}
=\frac{\partial\,}{\partial t}\left(\,\star\, \balpha_{\,q}\,\right)\quad
\mbox{on}\ \cA_{\,\zeta \psi}^{\cC},
$$
and 
\beqa
\dot{z}
&=&\dot{\psi}
=-\,\star\,\left(\,\bfe_{\,p}\wedge \bF_{\,\psi}^{\,\zeta p}
+\bfe_{\,q}\wedge \bF_{\,\psi}^{\,\zeta q}\,\right)
=-\,\star\,\left[\,-\,\frac{\delta\wt{\psi}}{\delta\balpha_{\,p}}\wedge
\dr\left(\frac{\delta\wt{\psi}}{\delta\balpha_{\,q}}
\right)+\frac{\delta\wt{\psi}}{\delta\balpha_{\,q}}\wedge
\dr\left(\frac{\delta\wt{\psi}}{\delta\balpha_{\,p}}\right)
\,\right]
\non\\
&=&-\,\star\,\dr\left(\,
\frac{\delta\wt{\psi}}{\delta\balpha_{\,p}}\wedge
\frac{\delta\wt{\psi}}{\delta\balpha_{\,q}}
\,\right)
\quad
\mbox{on}\ \cA_{\,\zeta \psi}^{\cC}.
\non
\eeqa
The last equation above yields the following 
$$
\frac{\dr\wt{\psi}}{\dr t}
=\int_{\cZ}\dot{\psi}\star 1
=-\,\int_{\cZ}\dr\,\left(\,
\frac{\delta\wt{\psi}}{\delta\balpha_{\,p}}\wedge
\frac{\delta\wt{\psi}}{\delta\balpha_{\,q}}
\,\right)
=-\,\int_{\partial\cZ}\left(\,
\frac{\delta\wt{\psi}}{\delta\balpha_{\,p}}\wedge
\frac{\delta\wt{\psi}}{\delta\balpha_{\,q}}
\,\right)
=-\,\int_{\partial\cZ}\bbf_{\,\partial}\wedge\bfe_{\,\partial}
=\int_{\partial\cZ}\bfe_{\,\partial}\wedge\bbf_{\,\partial},
$$
where 
$$
\bbf_{\,\partial}
=\left.\frac{\partial\wt{\psi}}{\delta\balpha_{\,p}}\right|_{\,\partial\cZ},
\qquad\mbox{and}\qquad 
\bfe_{\,\partial}
=\left.\frac{\partial\wt{\psi}}{\delta\balpha_{\,q}}\right|_{\,\partial\cZ}
$$
have been used.
Thus, one obtains 
$$
\frac{\dr\wt{\psi}}{\dr t}
=-\int_{\cZ}\left(\,\bfe_{\,p}\wedge\bbf_{\,p}+\bfe_{\,q}\wedge\bbf_{\,q}\,
\right)
=\int_{\partial\cZ}\bfe_{\,\partial}\wedge\bbf_{\,\partial},
$$ 
from which one has \fr{from-power-conserving-structure}.

So far the discussion above is for $\cA_{\,\zeta \psi}^{\cC}$ and that is 
valid for a covering $U_{\,i}$ containing $\zeta$. Taking into account this, 
one completes the proof. 
\qed
%%%%%%%%%%%
\end{Proof}
%%%%%%%%%%%
%%%%%%%%%%%%
\begin{Remark}
%%%%%%%%%%%%%%
The dimension of the contact manifold $\cC_{\,\zeta}$ over the $n$-dimensional 
base space $\cZ$ with $\zeta\in\cZ$ is given by 
$\dim\,\cC_{\zeta}=2({}_{n}C_{\,p}+{}_{n}C_{\,q})+1$, with ${}_{n}C_{\,p}=n!/(p!(n-p)!)$.
%%%%%%%%%%%%
\end{Remark}
%%%%%%%%%%%%%%

%%%%%%%%%%%%
\begin{Remark}
%%%%%%%%%%%%%%
As an example of the case of $\{\,p=q=2\,\}$,  
Maxwell's equations in media without source 
can be formulated in this approach (\,see Ref.\,\cite{Goto2017}\,). 
%%%%%%%%%%%%
\end{Remark}
%%%%%%%%%%%%%

The theorem above is about 
the case where the dimension of base spaces is $n=3$.
In addition,  the quadratic energy functionals given by 
\fr{quadratic-energy-functional-p-q} have only been considered. 
In what follows, the case of $n=2$ is stated. 
%%%%%%%%%%%%%%%%%%%%%%%%
\subsection{Two-dimensional Riemannian manifold}
%%%%%%%%%%%%%%%%%%%%%%% 
For the case $n=2$, the possible combinations of $0\leq p,q\leq n$ that 
satisfy $p+q=n+1$ are  
%%%%%%%%%%%%%%%%
\begin{itemize}
\item
$\{\,p=1,q=2\,\}$,\quad
$\{\,p=2,q=1\,\}$. 
\end{itemize} 
%%%%%%%%%%%%
Since the case $\{\,p=2,q=1\,\}$ can reduce to 
$\{\,p=1,p=2\,\}$, attention is concentrated on the case $\{\,q=1,p=2\,\}$.  
Taking into account this combination, one has the 
following theorem.  
%%%%%%%%%%%
\begin{Thm}
\label{theorem-n-2-dirac-contact}
%%%%%%%%%%5
(Distributed-parameter port-Hamiltonian system 
as contact Hamiltonian vertical vector field): 
Let $(\cZ,g)$ be a $2$-dimensional Riemannian manifold, and $\psi$ an 
energy density function specified later. 
%%%%%%%%%%%%%%%%
%\begin{itemize}
%%%%%
%\item
%%%%%
Fix $\{p=1,q=2\}$, let $(\cK,\lambda_{\,\mbbV},\pi,\cZ)$ 
be a $7$-dimensional contact 
manifold over the base space $\cZ$ with 
$\lambda_{\,\mbbV}=\dr_{\mbbV} z-p_{\,a}\,\dr_{\,\mbbV} x^{\,a}$,   
$\{\bsigma^{\,a}\}\in\Gamma\Lambda_{\,\mbbH}^{1}\cK$ an orthonormal co-frame,  
$$
\balpha_{\,p}
=(\,\balpha_{\,p}\,)_{\,a}\,\bsigma^{\,a}\ 
\in\Gamma\Lambda_{\,\mbbH,\mbbV}^{1,0}\cK,\qquad
\balpha_{\,q}
=(\,\balpha_{\,q}\,)_{\,0}\,\star 1\
\in\Gamma\Lambda_{\,\mbbH,\mbbV}^{2,0}\cK,
$$
$$
\bfe_{\,p}
=(\,\bfe_{\,p}\,)_{\,a}\,\bsigma^{\,a}\
\in\Gamma\Lambda_{\,\mbbH,\mbbV}^{1,0}\cK,\qquad
\bfe_{\,q}
=(\,\bfe_{\,q}\,)_{\,0}\
\in\Gamma\Lambda_{\,\mbbH,\mbbV}^{0,0}\cK
$$ 
mixed forms,   
$(x,y,z)$ canonical coordinates with $x=\{\,x_{\,(p)},x_{\,(q)}\,\}$, 
$y=\{\,y^{\,(p)},y^{\,(q)}\,\}$,  
% Let $x=\{\,x_{\,(p)}^{\,a},x_{\,(q)}\,\}$ and 
% $y=\{\,y_{\,a}^{\,(p)},y^{\,(q)}\,\}$ such that 
\beqa
x&=&\{\,(\,\star\,\balpha_{\,p}\,)^{\,1},
(\,\star\,\balpha_{\,p}\,)^{\,2},
\star\,\balpha_{\,q}\,
\},\qquad
x_{\,(p)}^{\,a}
=\delta^{\,ab}(\,\star\,\balpha_{\,p}\,)_{\,b},\quad 
x_{\,(q)}
=\star\,\balpha_{\,q}
=(\,\balpha_{\,q}\,)_{\,0},
\non\\ 
y&=&
\{\,(\,\bfe_{\,p}\,)_{\,1},(\,\bfe_{\,p}\,)_{\,2},
\bfe_{\,q}\,\}. \qquad
y_{\,a}^{\,(p)}
=(\,\bfe_{\,p}\,)_{\,a},\qquad 
y_{\,a}^{\,(q)}
=\bfe_{\,q}
=(\,\bfe_{\,q}\,)_{\,0}.
\non\\
z&=&\cE,
\non
\eeqa
where 
$\cE$ is the value of the energy density such that $\cE=\psi$ on  
$\cA_{\,\zeta \psi}^{\,\cC}$ with $\psi$ depending only on $x$. 
% %%%%%%%
% \item
% %%%%%
% For the case $\{p=2,q=1\}$, let $(\cK,\lambda_{\,\mbbV},\pi,\cZ)$ 
% be a $7$-dimensional contact 
% manifold over the base space $\cZ$ with 
% $\lambda_{\,\mbbV}=\dr_{\mbbV} z-p_{\,a}\,\dr_{\,\mbbV} x^{\,a}$,   
% $$
% \balpha_{\,p}
% =(\,\balpha_{\,p}\,)_{\,0}\,\star 1
% \ \in\Gamma\Lambda_{\,\mbbH,\mbbV}^{2,0}\cK,\qquad   
% \balpha_{\,q}
% =(\,\balpha_{\,q}\,)_{\,a}\,\bsigma^{\,a}\ 
% \in\Gamma\Lambda_{\,\mbbH,\mbbV}^{1,0}\cK
% $$
% $$
% \bfe_{\,p}
% =(\,\bfe_{\,p}\,)_{\,0}\ \in\Gamma\Lambda_{\,\mbbH,\mbbV}^{0,0}\cK,\qquad
% \bfe_{\,q}
% =(\,\bfe_{\,q}\,)_{\,a}\,\bsigma^{\,a}\ 
% \in\Gamma\Lambda_{\,\mbbH,\mbbV}^{1,0}\cK
% $$ 
% mixed forms,   
% $(x,y,z)$ canonical coordinates with $x=\{\,x_{\,(p)},x_{\,(q)}\,\}$, 
% $y=\{\,y^{\,(p)},y^{\,(q)}\,\}$,  
% % Let $x=\{\,x_{\,(p)}^{\,a},x_{\,(q)}\,\}$ and 
% % $y=\{\,y_{\,a}^{\,(p)},y^{\,(q)}\,\}$ such that 
% \beqa
% x&=&\{\,
% \star\,\balpha_{\,p}\,,
% (\,\star\,\balpha_{\,q}\,)^{\,1},
% (\,\star\,\balpha_{\,q}\,)^{\,2}
% \},\qquad
% x_{\,(p)}
% =(\,\balpha_{\,p}\,)_{\,0},\quad 
% x_{\,(q)}^{\,a}
% =\delta^{\,ab}(\,\star\,\balpha_{\,q}\,)_{\,b},
% \non\\ 
% y&=&
% \{\,
% \bfe_{\,p}\,,
% (\,\bfe_{\,q}\,)_{\,1},
% (\,\bfe_{\,q}\,)_{\,2}\}. \qquad
% y^{\,(p)}
% =(\,\bfe_{\,p}\,)_{\,0},\qquad 
% y_{\,a}^{\,(q)}
% =(\,\bfe_{\,q}\,)_{\,a}.
% \non\\
% z&=&\cE,
% \non
% \eeqa
% where 
% $\cE$ is the value of the enrgy density such that $\cE=\psi$ on  
% $\cA_{\,\zeta \psi}^{\,\cC}$ with $\psi$ depending only on $x$. 
% %%%%%%%%%%%%
% \end{itemize}
% %%%%%%%%%%%%%
%For the both of the cases, 
Furthermore, let $\wt{\psi}$ be a quadratic energy functional 
such that 
$$
\wt{\psi}
=\frac{1}{2}\int_{\cZ}\left(\,\balpha_{\,p}\wedge\star\,\balpha_{\,p}
+\balpha_{\,q}\wedge\star\,\balpha_{\,q}\,\right)
=\int_{\cZ}\psi\star 1, \quad\mbox{so that}\quad 
\bfe_{\,p}
=\frac{\delta\wt{\psi}}{\delta\balpha_{\,p}}\quad\mbox{and}\quad
\bfe_{\,q}
=\frac{\delta\wt{\psi}}{\delta\balpha_{\,q}}.
$$
In addition, choose the contact Hamiltonian functional as  
$$
\wt{h}_{\,\psi}
=\int_{\cZ}h_{\,\psi}\star 1
=\int_{\cZ}\,\left[\,
\bDelta_{\,p}^{\,\zeta \psi}\wedge \bF_{\,\psi}^{\,\zeta p}\,
+\bDelta_{\,q}^{\,\zeta \psi}\wedge \left(\,-\bF_{\,\psi}^{\,\zeta q}\,\right)
+\Gamma_{\,\zeta \psi}\left(\,\bDelta_{\,0}^{\,\zeta \psi}\,\right)\,\star 1
\,\right],
$$
where $\Delta_{\,0}^{\,\zeta \psi},\bDelta_{\,p}^{\,\zeta \psi},\bDelta_{\,q}^{\,\zeta \psi}$
are defined in \fr{adapted-mixed-form},  
$h_{\,\psi}\in\Gamma\Lambda_{\,\mbbH,\mbbV}^{\,0,0}\cK$, 
$\bF_{\,\psi}^{\,\zeta p}\in\Gamma\Lambda_{\,\mbbH,\mbbV}^{\,p,0}\cK$,
$\bF_{\,\psi}^{\,\zeta q}\in\Gamma\Lambda_{\,\mbbH,\mbbV}^{\,q,0}\cK$ 
are 
$$
h_{\,\psi}
=\star^{\,-1}\left[\,
\bDelta_{\,p}^{\,\zeta \psi}\wedge\,\bF_{\,\psi}^{\,\zeta p}\,
\,\right]
+\star^{\,-1}\left[\,
\bDelta_{\,q}^{\,\zeta \psi}\wedge\left(\,-\, \bF_{\,\psi}^{\,\zeta q}\,\right)
\,\right]+\Gamma_{\,\zeta \psi}\left(\,\bDelta_{\,0}^{\,\zeta \psi}\,\right),
$$
$$
\bF_{\,\psi}^{\,\zeta p}
:=(-1)^{\,r}\dr\left(\,\frac{\delta\wt{\psi}}{\delta \balpha_{\,q}}
\right)
=(-1)^{\,r}\,\dr\star\balpha_{\,q},\qquad 
\bF_{\,\psi}^{\,\zeta q}
:=\dr\left(\frac{\delta\wt{\psi}}{\delta\balpha_{\,p}}\right)
=\dr\star\balpha_{\,p},
$$
respectively, and $\Gamma_{\,\zeta \psi}$ is such that 
$$
\Gamma_{\,\zeta \psi}\left(\,\Delta_{\,0}^{\,\zeta \psi}\,\right)
=\left\{
\begin{array}{cc}
0&\mbox{for}\quad \Delta_{\,0}^{\,\zeta \psi}=0\\
\mbox{non-zero}&\mbox{for}\quad \Delta_{\,0}^{\,\zeta \psi}\neq 0.
\end{array}
\right.
$$
Also, define 
$$
\bbf_{\,\partial}
=\left.\frac{\delta\wt\psi}{\delta\balpha_{\,p}}\right|_{\partial\cZ},
\quad\mbox{and}\quad
\bfe_{\,\partial}
=(-1)^{\,p}\left.\frac{\delta\wt\psi}{\delta\balpha_{\,q}}\right|_{\partial\cZ}.
$$
Then, the restricted contact Hamiltonian vertical vector field 
$X_{\,\wt{h}_{\,\psi}}|_{\,\cA_{\,\psi}^{\,\cK}}$ onto 
the phase space for distributed-parameter port-Hamiltonian system 
(\,see Definition\,
\ref{definition-phase-space-distributed-parameter-port-Hamiltonian-system}\,)
gives distributed-parameter 
port-Hamiltonian systems given in  
Definition\,\ref{definition-distributed-parameter-port-Hamiltonian-system}. 
In addition, 
\fr{from-power-conserving-structure} is satisfied.
%%%%%%%%%%5
\end{Thm} 
%%%%%%%%%
\begin{Proof}
%%%%%%%%%%%%%%
One can follow the procedure given in the proof of 
Theorem\,\ref{theorem-n-3-dirac-contact}. 
Throughout this proof, $\star\star\balpha=(-1)^{p}\balpha$ and 
$\star^{\,-1}\balpha=(-1)^{\,p}\star\balpha$ 
for any $p$-form $\balpha$  (\,see Lemma\,\ref{2-d-Riemannian-Hodge}\,),
and and $\cA_{\,\zeta \psi}^{\cC}=\{\,\wt{h}_{\,\psi}=0\,\}$ are used.  
\\
(Proof for the case of $p=1,q=2$): 
% Write 
% $$
% \balpha_{\,p}
% =(\,\balpha_{\,p}\,)_{\,a}\,\bsigma^{\,a},\quad
% \balpha_{\,q}
% =(\,\balpha_{\,q}\,)_{\,0}\,\star 1,
% \quad\mbox{and}\quad 
% \bfe_{\,p}
% =(\,\bfe_{\,p}\,)_{\,a}\,\bsigma^{\,a},\quad
% \bfe_{\,q}
% =(\,\bfe_{\,q}\,)_{\,0},
% $$
% Let $x=\{\,x_{\,(p)}^{\,a},x_{\,(q)}\,\}$ and 
% $y=\{\,y_{\,a}^{\,(p)},y^{\,(q)}\,\}$ such that 
% \beqa
% x&=&\{\,(\,\balpha_{\,p}\,)^{\,1},
% (\,\balpha_{\,p}\,)^{\,2},
% (\,\balpha_{\,q}\,)^{\,0}
% \},
% \non\\
% &&x_{\,(p)}^{\,a}
% =\delta^{\,ab}(\,\balpha_{\,p}\,)_{\,b},\qquad 
% x_{\,(q)}
% =(\,\balpha_{\,q}\,)^{\,0}
% =(\,\balpha_{\,q}\,)_{\,0},
% \non\\ 
% y&=&
% \{\,(\,\bfe_{\,p}\,)_{\,1},(\,\bfe_{\,p}\,)_{\,2},
% (\,\bfe_{\,q}\,)_{\,0}\}. 
% \non\\
% &&
% y_{\,a}^{\,(p)}
% =(\,\bfe_{\,p}\,)_{\,a},\qquad 
% y_{\,a}^{\,(q)}
% =(\,\bfe_{\,q}\,)_{\,0}.
% \non
% \eeqa
Since 
$$
\star^{-1}\left[\,\left(\,
-\bF_{\,\psi}^{\,\zeta p}\,\right)\wedge\bDelta_{\,p}^{\,\zeta \psi}\,\right]
=g^{-1}\left(\,\bDelta_{\,p}^{\,\zeta \psi},-\,\star\,\bF_{\,\psi}^{\,\zeta p}\,\right),\quad\mbox{and}\quad
\star^{-1}\left[\,
\bDelta_{\,q}^{\,\zeta \psi}\wedge\left(-\bF_{\,\psi}^{\,\zeta q}\right)
\,\right]
=\bDelta_{\,q}^{\,\zeta \psi}\left(-\star\bF_{\,\psi}^{\,\zeta q}\right),
$$
the contact Hamiltonian density function $h_{\,\psi}$ is written as 
\beqa
h_{\,\psi}
&=&g^{-1}
\left(\,\bDelta_{\,p}^{\,\zeta \psi},-\,\star\,\bF_{\,\psi}^{\,\zeta p}\right)
+\bDelta_{\,q}^{\,\zeta \psi}\left(-\star\,\bF_{\,\psi}^{\,\zeta q}\right)
+\Gamma_{\,\zeta \psi}\left(\,\bDelta_{\,0}^{\,\zeta \psi}\,\right)
\non\\
&=&\delta^{\,ab}\left(\,\bDelta_{\,p}^{\,\zeta \psi}\right)_{\,a}
\left(\,-\,\star\,\bF_{\,\psi}^{\,\zeta p}\right)_{\,b}
+
\left(\,\bDelta_{\,q}^{\,\zeta \psi}\,\right)
\left(\,-\star\,\bF_{\,\psi}^{\,\zeta q}\,\right)
+\Gamma_{\,\zeta \psi}\left(\,\bDelta_{\,0}^{\,\zeta \psi}\,\right),
\non
\eeqa
where 
$$
\bDelta_{\,p}^{\,\zeta \psi}
=\left(\,\bDelta_{\,p}^{\,\zeta \psi}\right)_{\,a}\,\bsigma^{\,a}.
$$
In addition, one has from \fr{adapted-mixed-form} that
$$
\left(\,\bDelta_{\,p}^{\,\zeta \psi}\,\right)_{\,a}
=\left(\,\star\,\balpha_{\,p}\,\right)_{\,a}-(\,\bfe_{\,p}\,)_{\,a},\quad
\left(\,\bDelta_{\,q}^{\,\zeta \psi}\,\right)
=\left(\,\star\,\balpha_{\,q}\,\right)-(\,\bfe_{\,q}\,)
=(\,\balpha_{\,q}\,)_{\,0}-(\,\bfe_{\,q}\,)_{\,0}.
$$
The component expression of the restricted contact vertical vector field is 
obtained from \fr{coordinate-expression-contact-Hamiltonian-vertical-vector} 
as 
\beqa
\left.\dot{x}_{\,(p)}^{\,a}\right|_{\,\cA_{\,\zeta \psi}^{\cC}}
&=&-\,\left.\frac{\partial h_{\,\psi}}{\partial y_{\,a}^{\,(p)}}
\right|_{\,\cA_{\,\zeta \psi}^{\cC}}
=-\,\left.\delta^{\,ab}\left(\star\,\bF_{\,\psi}^{\,\zeta p}\right)_{\,b}
\right|_{\,\cA_{\,\zeta \psi}^{\cC}},
\non\\
\left.\dot{x}_{\,(q)}\right|_{\,\cA_{\,\zeta \psi}^{\cC}}
&=&-\,\left.\frac{\partial h_{\,\psi}}{\partial y^{\,(q)}}
\right|_{\,\cA_{\,\zeta \psi}^{\cC}}
=-\,\left.\star\,\bF_{\,\psi}^{\,\zeta q}
\right|_{\,\cA_{\,\zeta \psi}^{\cC}},
\non\\
\left.\dot{y}_{\,a}^{\,(p)}\right|_{\,\cA_{\,\zeta \psi}^{\cC}}
&=&\left.\left(\frac{\partial h_{\,\psi}}{\partial x_{\,(p)}^{\,a}}
+y_{\,a}^{\,(p)}\frac{\partial h_{\,\psi}}{\partial z}\right)
\right|_{\,\cA_{\,\zeta \psi}^{\cC}}
=-\,\left.\left(\star\,\bF_{\,\psi}^{\,\zeta p}\right)_{\,a}
\right|_{\,\cA_{\,\zeta \psi}^{\cC}},
\non\\
\left.\dot{y}^{\,(q)}\right|_{\,\cA_{\,\zeta \psi}^{\cC}}
&=&\left.\left(\frac{\partial h_{\,\psi}}{\partial x_{\,(q)}}
+y^{\,(q)}\frac{\partial h_{\,\psi}}{\partial z}\right)
\right|_{\,\cA_{\,\zeta \psi}^{\cC}}
=-\,\left.\star\,\bF_{\,\psi}^{\,\zeta q}
\right|_{\,\cA_{\,\zeta \psi}^{\cC}},
\non\\
\left.\dot{z}\right|_{\,\cA_{\,\zeta \psi}^{\cC}}
&=&\left.
\left(h_{\,\psi}-y_{\,a}^{\,(p)}\frac{\partial h}{\partial y_{\,a}^{\,(p)}}
-y_{\,a}^{\,(q)}\frac{\partial h}{\partial y_{\,a}^{\,(q)}}
\right)\right|_{\,\cA_{\,\zeta \psi}^{\cC}}
=-\,\left.\left[\,g^{\,-1}\left(\,\bfe_{\,p},\star\,\bF_{\,\psi}^{\,\zeta p}\right)
+\,\bfe_{\,q}\left(\,\star\,\bF_{\,\psi}^{\,\zeta q}\right)\,\right]
\right|_{\,\cA_{\,\zeta \psi}^{\cC}}.
\non
\eeqa
These are equivalent to write 
$$
-\,\frac{\partial\,\balpha_{\,p}}{\partial t}
=\bF_{\,\psi}^{\,\zeta p},\qquad 
-\,\frac{\partial\,\balpha_{\,q}}{\partial t}
=\bF_{\,\psi}^{\,\zeta q},
$$
$$
\frac{\partial\, \bfe_{\,p}}{\partial t}
=-\,\star\,\bF_{\,\psi}^{\,\zeta p}
=\frac{\partial\, }{\partial t}\,\left( \star\,\balpha_{\,p}\,\right),\qquad 
\frac{\partial\, \bfe_{\,q}}{\partial t}
=-\,\star\,\bF_{\,\psi}^{\,\zeta q}
=\frac{\partial\,}{\partial t}\left(\,\star\, \balpha_{\,q}\,\right)\quad
\mbox{on}\ \cA_{\,\zeta \psi}^{\cC},
$$
and 
\beqa
\dot{z}
&=&\dot{\psi}
=-\,\star\,\left(\,\bfe_{\,p}\wedge \bF_{\,\psi}^{\,\zeta p}
+\bfe_{\,q}\wedge \bF_{\,\psi}^{\,\zeta q}\,\right)
=-\,\star\,\left[\,-\,\frac{\delta\wt{\psi}}{\delta\balpha_{\,p}}\wedge
\dr\left(\frac{\delta\wt{\psi}}{\delta\balpha_{\,q}}
\right)+\frac{\delta\wt{\psi}}{\delta\balpha_{\,q}}\wedge
\dr\left(\frac{\delta\wt{\psi}}{\delta\balpha_{\,p}}\right)
\,\right]
\non\\
&=&-\,\star\,\dr\left(\,
\frac{\delta\wt{\psi}}{\delta\balpha_{\,p}}\wedge
\frac{\delta\wt{\psi}}{\delta\balpha_{\,q}}
\,\right)
\quad
\mbox{on}\ \cA_{\,\zeta \psi}^{\cC}.
\non
\eeqa
The last equation above yields the following 
$$
\frac{\dr\wt{\psi}}{\dr t}
=\int_{\cZ}\dot{\psi}\star 1
=-\,\int_{\cZ}\dr\,\left(\,
\frac{\delta\wt{\psi}}{\delta\balpha_{\,p}}\wedge
\frac{\delta\wt{\psi}}{\delta\balpha_{\,q}}
\,\right)
=-\,\int_{\partial\cZ}\left(\,
\frac{\delta\wt{\psi}}{\delta\balpha_{\,p}}\wedge
\frac{\delta\wt{\psi}}{\delta\balpha_{\,q}}
\,\right)
=\int_{\partial\cZ}\bbf_{\,\partial}\wedge\bfe_{\,\partial}
=-\,\int_{\partial\cZ}\bfe_{\,\partial}\wedge\bbf_{\,\partial},
$$
where 
$$
\bbf_{\,\partial}
=\left.\frac{\partial\wt{\psi}}{\delta\balpha_{\,p}}\right|_{\,\partial\cZ},
\qquad\mbox{and}\qquad 
\bfe_{\,\partial}
=-\left.\frac{\partial\wt{\psi}}{\delta\balpha_{\,q}}\right|_{\,\partial\cZ}
$$
have been used.
Thus, one obtains 
$$
\frac{\dr\wt{\psi}}{\dr t}
=-\int_{\cZ}\left(\,\bfe_{\,p}\wedge\bbf_{\,p}+\bfe_{\,q}\wedge\bbf_{\,q}\,
\right)
=\int_{\partial\cZ}\bfe_{\,\partial}\wedge\bbf_{\,\partial},
$$ 
from which one has \fr{from-power-conserving-structure}.
% \noindent
% (Proof for the case of $p=2,q=1$):  
% A way to prove is analogous to that for the case of $p=1,q=2$.  

So far the discussion above is for $\cA_{\,\zeta \psi}^{\cC}$ and that is 
valid for a covering $U_{\,i}$ containing $\zeta$. Taking into account this, 
one completes the proof. 
\qed
%%%%%%%%%%
\end{Proof}
%%%%%%%%%%%%%
%%%%%%%%%%%%
\begin{Remark}
%%%%%%%%%%%%%%
The dimension of the contact manifold $\cC_{\,\zeta}$ over the $n$-dimensional 
base space $\cZ$  with $\zeta\in\cZ$ is given by 
$\dim\,\cC_{\zeta}=2({}_{n}C_{\,p}+{}_{n}C_{\,q})+1$, with ${}_{n}C_{\,p}=n!/(p!(n-p)!)$.
%%%%%%%%%%%%
\end{Remark}
%%%%%%%%%%%%%%

%%%%%%%%%%%%%%%%%%%%%%%%
\subsection{One-dimensional Riemannian manifold}
%%%%%%%%%%%%%%%%%%%%%%% 
For the case of $n=1$, 
systems with and without quadratic energy functionals are concerned.
In many physical problems, systems on $1$-dimensional spatial manifolds 
are considered as toy models, and sometimes 
these models are exactly solvable.  
Thus it is expected that the theorem for the case of $n=1$ will be useful. 
For this case, $n=1$, the possible combination of $0\leq p,q\leq n$ that 
satisfy $p+q=n+1$ is 
\begin{itemize}
\item
$\{\,p=1,q=1\,\}$.
\end{itemize}
Taking into account this combination, one has the following theorem.
%%%%%%%%%%%%%
\begin{Thm}
\label{theorem-n-1-dirac-contact}
%%%%%%%%%%%
(Distributed-parameter port-Hamiltonian system as 
contact vertical vector field):
Let $(\cZ,g)$ be a $1$-dimensional Riemannian manifold, and $\psi$ an 
energy density function specified later. 
let $(\cK,\lambda_{\,\mbbV},\pi,\cZ)$ 
be a $5$-dimensional contact 
manifold over the base space $\cZ$ with 
$\lambda_{\,\mbbV}=\dr_{\mbbV} z-p_{\,a}\,\dr_{\,\mbbV} x^{\,a}$,   
$\bsigma\in\Gamma\Lambda_{\,\mbbH}^{1}\cK$ the orthonormal co-frame,  
$$
\balpha_{\,p}
=(\,\balpha_{\,p}\,)_{\,0}\,\star 1
\in\Gamma\Lambda_{\,\mbbH,\mbbV}^{1,0}\cK,\qquad
\balpha_{\,q}
=(\,\balpha_{\,q}\,)_{\,0}\,\star 1\
\in\Gamma\Lambda_{\,\mbbH,\mbbV}^{1,0}\cK,\quad\mbox{where}\quad 
\star 1
=\bsigma,
$$
$$
\bfe_{\,p}
=(\,\bfe_{\,p}\,)_{\,0}
\in\Gamma\Lambda_{\,\mbbH,\mbbV}^{0,0}\cK,\qquad
\bfe_{\,q}
=(\,\bfe_{\,q}\,)_{\,0}\
\in\Gamma\Lambda_{\,\mbbH,\mbbV}^{0,0}\cK
$$ 
mixed forms,   
$(x,y,z)$ canonical coordinates with $x=\{\,x_{\,(p)},x_{\,(q)}\,\}$, 
$y=\{\,y^{\,(p)},y^{\,(q)}\,\}$,  
% Let $x=\{\,x_{\,(p)}^{\,a},x_{\,(q)}\,\}$ and 
% $y=\{\,y_{\,a}^{\,(p)},y^{\,(q)}\,\}$ such that 
\beqa
x&=&\{\,\star\,\balpha_{\,p}\,, 
\star\,\balpha_{\,q}\,
\},\qquad
x_{\,(p)}
=\star\,\balpha_{\,p}
=(\,\balpha_{\,p}\,)_{\,0},\quad 
x_{\,(q)}
=\star\,\balpha_{\,q}
=(\,\balpha_{\,q}\,)_{\,0},
\non\\ 
y&=&
\{\,\bfe_{\,p}\,,\bfe_{\,q}\,\}. \qquad
y^{\,(p)}
=(\,\bfe_{\,p}\,)_{\,0},\qquad 
y^{\,(q)}
=(\,\bfe_{\,q}\,)_{\,0}.
\non\\
z&=&\cE,
\non
\eeqa
where 
$\cE$ is the value of the energy density such that $\cE=\psi$ on  
$\cA_{\,\zeta \psi}^{\,\cC}$ with $\psi$ depending only on $x$. 
In addition let $\wt{\psi}$ be a functional  as 
$$
\wt{\psi}
=\int_{\cZ}\psi(\,(\,\balpha_{\,p}\,)_{\,0},(\,\balpha_{\,q}\,)_{\,0}\,)\star 1,\quad
\mbox{so that}\quad
\bfe_{\,p}
=\frac{\delta\wt{\psi}}{\balpha_{\,p}},\quad\mbox{and}\quad
\bfe_{\,q}
=\frac{\delta\wt{\psi}}{\balpha_{\,q}}, 
$$
and the contact Hamiltonian functional as 
$$
\wt{h}_{\,\psi}
=\int_{\,\cZ}h_{\,\psi}\star 1
=\int_{\,\cZ}\left[\,
\bDelta_{\,p}^{\,\zeta}\,\left(-\,\bF_{\,\psi}^{\,\zeta p}\,\right)
+\bDelta_{\,q}^{\,\zeta}\,\left(-\,\bF_{\,\psi}^{\,\zeta q}\,\right)
+\Gamma_{\,\zeta\psi}
\left(\,
\bDelta_{\,0}^{\,\zeta \psi}
\,\right)\,\star 1
\,\right],
$$
where 
$\Delta_{\,0}^{\,\zeta \psi},\bDelta_{\,p}^{\,\zeta \psi},\bDelta_{\,q}^{\,\zeta \psi}$
are defined in \fr{adapted-mixed-form},  
$h_{\,\psi}\in\Gamma\Lambda_{\,\mbbH,\mbbV}^{\,0,0}\cK$,
$\bF_{\,\psi}^{\,\zeta p}\in\Gamma\Lambda_{\,\mbbH,\mbbV}^{\,1,0}\cK$,
$\bF_{\,\psi}^{\,\zeta q}\in\Gamma\Lambda_{\,\mbbH,\mbbV}^{\,1,0}\cK$ are 
$$
h_{\,\psi}
=\star^{-1}\left[\,
\bDelta_{\,p}^{\,\zeta}\,\left(-\,\bF_{\,\psi}^{\,\zeta p}\,\right)
+\bDelta_{\,q}^{\,\zeta}\,\left(-\,\bF_{\,\psi}^{\,\zeta q}\,\right)
\,\right]+\Gamma_{\,\zeta\psi}
\left(\,
\bDelta_{\,0}^{\,\zeta \psi}
\,\right),
$$
$$
\bF_{\,\psi}^{\,\zeta p}
:=(-1)^{\,r}\dr\left(\,\frac{\delta\wt{\psi}}{\delta \balpha_{\,q}}
\right)
%=(-1)^{\,r}\dr\star\balpha_{\,q}
,\qquad 
\bF_{\,\psi}^{\,\zeta q}
:=\dr\left(\frac{\delta\wt{\psi}}{\delta\balpha_{\,p}}\right),
%=\dr\star\balpha_{\,p},
$$
respectively, and $\Gamma_{\,\zeta \psi}$ is such that 
$$
\Gamma_{\,\zeta \psi}\left(\,\Delta_{\,0}^{\,\zeta \psi}\,\right)
=\left\{
\begin{array}{cc}
0&\mbox{for}\quad \Delta_{\,0}^{\,\zeta \psi}=0\\
\mbox{non-zero}&\mbox{for}\quad \Delta_{\,0}^{\,\zeta \psi}\neq 0.
\end{array}
\right.
$$
Also, define 
$$
\bbf_{\,\partial}
=\left.\frac{\delta\wt\psi}{\delta\balpha_{\,p}}\right|_{\partial\cZ},
\quad\mbox{and}\quad
\bfe_{\,\partial}
=(-1)^{\,p}\left.\frac{\delta\wt\psi}{\delta\balpha_{\,q}}\right|_{\partial\cZ}.
$$
Then, the restricted contact Hamiltonian vertical vector field 
$X_{\,\wt{h}_{\,\psi}}|_{\,\cA_{\,\psi}^{\,\cK}}$ 
onto 
the phase space for distributed-parameter port-Hamiltonian system 
(\,see Definition\,
\ref{definition-phase-space-distributed-parameter-port-Hamiltonian-system}\,)
gives distributed-parameter 
port-Hamiltonian systems given in  
Definition\,\ref{definition-distributed-parameter-port-Hamiltonian-system}. 
In addition, 
\fr{from-power-conserving-structure} is satisfied.
%%%%%%%%%%
\end{Thm} 
%%%%%%%%%
\begin{Proof}
%%%%%%%%%%%%%%
One can follow the procedure given in the proof of 
Theorem\,\ref{theorem-n-3-dirac-contact}.  
Throughout this proof, $\star\star\balpha=\balpha$ and 
$\star^{\,-1}\balpha=\star\,\balpha$ 
for any $p$-form $\balpha$  (\,see Lemma\,\ref{1-d-Riemannian-Hodge}\,), 
and  $\cA_{\,\zeta \psi}^{\cC}=\{\,\wt{h}_{\,\psi}=0\,\}$ are used. 
% Since $\bfe_{\,p}=\delta \wt{h}_{\,\psi}/\delta\balpha_{\,p}$ 
% and $\bfe_{\,q}=\delta\wt{h}_{\,\psi}/\delta\balpha_{\,q}$ are $0$-forms, 
% one can write them as 
% $e_{\,p}^{\,0},e_{\,q}^{\,0}\in\Gamma\Lambda_{\,\mbbH,\mbbV}^{0,0}\cK$ : 
% $$
% \bfe_{\,p}
% =e_{\,p}^{\,0}\,,\qquad 
% \bfe_{\,q}
% =e_{\,q}^{\,0}.
% $$ 
% Let $x=\{x_{\,(p)},x_{\,(q)}\}$ 
% and $y=\{y^{\,(p)},y^{\,(q)}\}$ 
% such that 
% $$
% x=\{\,(\,\alpha_{\,p}^{\,0}\,),
% (\,\alpha_{\,q}^{\,0}\,)
% \,\},
% \qquad 
% y=\{\,
% e_{\,p}^{\,0},
% e_{\,q}^{\,0}\,\}.
% $$
Since 
$$
\star^{-1}\left[\,\bDelta_{\,p}^{\,\zeta \psi}\left(\,
-\bF_{\,\psi}^{\,\zeta p}\,\right)\,\right]
=\bDelta_{\,p}^{\,\zeta \psi}\left(\,-\,\star\,\bF_{\,\psi}^{\,\zeta p}\,\right),\quad\mbox{and}\quad
\star^{-1}\left[\,\bDelta_{\,q}^{\,\zeta \psi}\left(\,
-\bF_{\,\psi}^{\,\zeta q}\,\right)\,\right]
=\bDelta_{\,q}^{\,\zeta \psi}\left(\,-\,\star\,\bF_{\,\psi}^{\,\zeta q}\,\right),
$$
the contact Hamiltonian density function $h_{\,\psi}$ is written as 
$$
h_{\,\psi}
=\bDelta_{\,p}\left(\,-\,\star\bF_{\,\psi}^{\,\zeta p} \,\right)
+\bDelta_{\,q}\left(\,-\,\star\bF_{\,\psi}^{\,\zeta q} \,\right)
+\Gamma_{\,\zeta \psi}\left(\,\Delta_{\,0}^{\,\zeta \psi}\,\right).
$$
The component expression of the restricted 
contact vertical vector field is obtained 
from \fr{coordinate-expression-contact-Hamiltonian-vertical-vector} as 
\beqa
\left.\dot{x}_{\,(p)}\right|_{\,\cA_{\,\zeta \psi}^{\cC}}
&=&-\,\left.\frac{\partial h_{\,\psi}}{\partial y^{\,(p)}}
\right|_{\,\cA_{\,\zeta \psi}^{\cC}}
=-\,\left.\star\,\bF_{\,\psi}^{\,\zeta p}
\right|_{\,\cA_{\,\zeta \psi}^{\cC}},
\non\\
\left.\dot{x}_{\,(q)}\right|_{\,\cA_{\,\zeta \psi}^{\cC}}
&=&-\,\left.\frac{\partial h_{\,\psi}}{\partial y^{\,(q)}}
\right|_{\,\cA_{\,\zeta \psi}^{\cC}}
=-\,\left.\star\,\bF_{\,\psi}^{\,\zeta q}
\right|_{\,\cA_{\,\zeta \psi}^{\cC}},
\non\\
\left.\dot{y}^{\,(p)}\right|_{\,\cA_{\,\zeta \psi}^{\cC}}
&=&\left.\left(\frac{\partial h_{\,\psi}}{\partial x_{\,(p)}}
+y^{\,(p)}\frac{\partial h_{\,\psi}}{\partial z}\right)
\right|_{\,\cA_{\,\zeta \psi}^{\cC}}
=-\,\left.\star\,\bF_{\,\psi}^{\,\zeta p}
\right|_{\,\cA_{\,\zeta \psi}^{\cC}},
\non\\
\left.\dot{y}^{\,(q)}\right|_{\,\cA_{\,\zeta \psi}^{\cC}}
&=&\left.\left(\frac{\partial h_{\,\psi}}{\partial x_{\,(q)}}
+y^{\,(q)}\frac{\partial h_{\,\psi}}{\partial z}\right)
\right|_{\,\cA_{\,\zeta \psi}^{\cC}}
=-\,\left.\star\,\bF_{\,\psi}^{\,\zeta q}
\right|_{\,\cA_{\,\zeta \psi}^{\cC}},
\non\\
\left.\dot{z}\right|_{\,\cA_{\,\zeta \psi}^{\cC}}
&=&\left.
\left(h_{\,\psi}-y^{\,(p)}\frac{\partial h_{\,\psi}}{\partial y^{\,(p)}}
-y^{\,(q)}\frac{\partial h_{\,\psi}}{\partial y^{\,(q)}}
\right)\right|_{\,\cA_{\,\zeta \psi}^{\cC}}
=\left.\left[\,\bfe_{\,p}\left(\,-\star\,\bF_{\,\psi}^{\,\zeta p}\,\right)
+\bfe_{\,q}\left(\,-\star\ \bF_{\,\psi}^{\,\zeta q}\,\right)\,\right]
\right|_{\,\cA_{\,\zeta \psi}^{\cC}}.
\non
\eeqa
These are equivalent to write 
$$
-\,\frac{\partial\,\balpha_{\,p}}{\partial t}
=\bF_{\,\psi}^{\,\zeta p},\qquad 
-\,\frac{\partial\,\balpha_{\,q}}{\partial t}
=\bF_{\,\psi}^{\,\zeta q},
$$
$$
\frac{\partial\, e_{\,p}^{\,0}}{\partial t}
=\frac{\partial\,}{\partial t}\frac{\delta \wt{h}_{\psi}}{\delta\balpha_{\,p}}
=-\,\star\,\bF_{\,\psi}^{\,\zeta p}
%=\frac{\partial\, }{\partial t}\,\left( \star\,\balpha_{\,p}\,\right)
,\qquad 
\frac{\partial\, e_{\,q}^{\,0}}{\partial t}
=-\,\star\,\bF_{\,\psi}^{\,\zeta q}
=\frac{\partial\,}{\partial t}\left(\,\star\, \balpha_{\,q}\,\right)\quad
\mbox{on}\ \cA_{\,\zeta \psi}^{\cC},
$$
and 
\beqa
\dot{z}
&=&\dot{\psi}
=-\star\,\left(\,\bfe_{\,p}\, \bF_{\,\psi}^{\,\zeta p}
+\bfe_{\,q}\, \bF_{\,\psi}^{\,\zeta q}\,\right)
=-\,\star\,\left[\,\frac{\delta\wt{\psi}}{\delta\balpha_{\,p}}%\wedge
\dr\left(\frac{\delta\wt{\psi}}{\delta\balpha_{\,q}}
\right)+\frac{\delta\wt{\psi}}{\delta\balpha_{\,q}}%\wedge
\dr\left(\frac{\delta\wt{\psi}}{\delta\balpha_{\,p}}\right)
\,\right]
\non\\
&=&-\,\star\,\dr\left(\,
\frac{\delta\wt{\psi}}{\delta\balpha_{\,p}}%\wedge
\frac{\delta\wt{\psi}}{\delta\balpha_{\,q}}
\,\right)
\quad
\mbox{on}\ \cA_{\,\zeta \psi}^{\cC}.
\non
\eeqa
The last equation above yields the following 
$$
\frac{\dr\wt{\psi}}{\dr t}
=\int_{\cZ}\dot{\psi}\star 1
=-\,\int_{\cZ}\dr\,\left(\,
\frac{\delta\wt{\psi}}{\delta\balpha_{\,p}}%\wedge
\frac{\delta\wt{\psi}}{\delta\balpha_{\,q}}
\,\right)
=-\,\int_{\partial\cZ}\left(\,
\frac{\delta\wt{\psi}}{\delta\balpha_{\,p}}%\wedge
\frac{\delta\wt{\psi}}{\delta\balpha_{\,q}}
\,\right)
=\int_{\partial\cZ}\bbf_{\,\partial}\,%\wedge
\bfe_{\,\partial}
=\int_{\partial\cZ}\bfe_{\,\partial}\bbf_{\,\partial},
$$
where 
$$
\bbf_{\,\partial}
=\left.\frac{\partial\wt{\psi}}{\delta\balpha_{\,p}}\right|_{\,\partial\cZ},
\qquad 
\bfe_{\,\partial}
=-\,\left.\frac{\partial\wt{\psi}}{\delta\balpha_{\,q}}\right|_{\,\partial\cZ},
$$
have been used.
Thus, 
one obtains 
$$
\frac{\dr\wt{\psi}}{\dr t}
=-\int_{\cZ}\left(\,\bfe_{\,p}\,\bbf_{\,p}+\bfe_{\,q}\,\bbf_{\,q}\,
\right)
=\int_{\partial\cZ}\bfe_{\,\partial}\,\bbf_{\,\partial},
$$ 
from which one has \fr{from-power-conserving-structure}.

So far the discussion above is for $\cA_{\,\zeta \psi}^{\cC}$ and that is 
valid for a covering $U_{\,i}$ containing $\zeta$. Taking into account this, 
one completes the proof. 
\qed
%%%%%%%%%%
\end{Proof}
%%%%%%%%%%%
%%%%%%%%%%%%
\begin{Remark}
%%%%%%%%%%%%%%
The dimension of the contact manifold $\cC_{\,\zeta}$ over the $n$-dimensional 
base space $\cZ$ with $\zeta\in\cZ$ is given by 
$\dim\,\cC_{\zeta}=2({}_{n}C_{\,p}+{}_{n}C_{\,q})+1$, 
with ${}_{n}C_{\,p}=n!/(p!(n-p)!)$.
%%%%%%%%%%%%
\end{Remark}
%%%%%%%%%%%%%%

In this section with $n=1,2$ and $3$, classes of 
distributed-parameter port-Hamiltonian systems can be written in terms of 
restricted contact Hamiltonian vertical vector fields. 
This contact geometric description gives the 
following points. Since contact Hamiltonian vertical vector fields 
are on the Legendre submanifolds, 
the relations between effort variables $\{\bfe_{\,p}, \bfe_{\,q}\}$    
and energy variables $\{\balpha_{\,p},\balpha_{\,q}\}$ are preserved. 
In addition,  
the values of energy are appropriately chosen 
along the contact Hamiltonian vertical vector fields on phase space.
%%%%%%%%%%%%%
\section{Information geometry for distributed-parameter port-Hamiltonian systems}
\label{section-information-geometry-distributed-parameter-port-Hamiltonian}
%%%%%%%%%%%
It has been shown in Ref.\,\cite{Goto2015} 
that a contact manifold and a strictly convex function induce  
a dually flat space that is used in information geometry.

When an energy density functional is strictly convex, 
one can introduce a dually flat space in  
a fiber space of a bundle for the distributed-parameter port-Hamiltonian 
systems.
First,  one introduces a metric tensor field as follows. 
%%%%%%%%%%%%%%%%%%
\begin{Def}
\label{definition-fiber-metric} 
%%%%%%%%%%%%%%% 
(Fiber metric tensor field for distributed-parameter 
port-Hamiltonian systems in Stokes-Dirac structure): 
Let $(\cK,\lambda_{\mbbV},\pi,\cB)$ be 
a $(2\wt{n}_{\,pq}+1)$-dimensional contact manifold over a base space with 
$\dim\cB=n$ and $\wt{n}_{\,pq}={}_{n}C_{p}+{}_{n}C_{\,q}$ with $p,q$ such that 
$0\leq p,q\leq n$ and $p+q=n+1$,  
$(x,y,z)$ the canonical coordinates for the fiber space 
such that $\lambda_{\mbbV}=\dr_{\,\mbbV} z-y_{\,a}\dr_{\mbbV} x^{\,a}$ with 
$x=\{\,x^{\,1},\ldots,x^{\,n}\,\}$ and $y=\{\,y_{\,1},\ldots,y_{\,n}\,\}$, and 
$\psi$ an energy density function 
(\,see Definition\, \ref{definition-energy-functional-energy-density-function}\,).
Then  
the metric tensor field 
$g^{\,\zeta}=g_{\,ab}^{\,\zeta}\dr_{\,\mbbV}\, x^{\,a}\otimes \dr_{\mbbV}\, x^{\,b}$ 
on $\cA_{\,\zeta\psi}^{\,\cC}(\subset\pi^{-1}(\zeta),\zeta\in\cB)$ with 
\beq
g_{\,ab}^{\,\zeta}
=\frac{\partial^{\,2}\,\psi}{\partial x^{\,a}\partial x^{\,b}},\quad 
a,b\in\{1,\ldots,\wt{n}_{\,pq}\,\}
\label{fiber-metric-tensor-field}
\eeq
is referred to as the fiber metric tensor field of $\cA_{\,\zeta \psi}^{\,\cC}$ 
for the contact manifold over the base space. 
%%%%%%%%%%%%%%%%%
\end{Def}
%%%%%%%%%%%%%%%%%
Given the energy functionals used in 
Section\,\ref{section-distributed-Hamiltonian-systems-Dirac}, 
the corresponding energy density functions and co-energy density functions  
are calculated as follows.
%%%%%%%%%%%%5
\begin{Proposition}
\label{proposition-total-Legendre-transform-quadratic-functional}
%%%%%%%%%%%%5
(Total Legendre transform of energy density function and co-energy density function):
Let $\wt{\psi}$ and $\wt{\varphi}$ be an energy functional 
and co-energy functionals given in  
\fr{quadratic-energy-functional-p-q} and 
\fr{quadratic-co-energy-functional-p-q}, respectively. 
In addition, let $\psi$ and $\varphi$ be 
the energy density function and the co-energy density function such 
that 
$$
\wt{\psi}
=\int_{\cZ}\psi\star1,\qquad\mbox{and}\qquad 
\wt{\varphi}
=\int_{\cZ}\varphi\star1.
$$
First, one has the following explicit expressions.
%%%%%%%%%%%%%%%
\begin{itemize}
%%%%%%%%%%%%%%%
\item
%%%%%%
For the case $n=3,\{p=1,q=3\}$, one has 
$$
\psi(x_{(p)},x_{(q)})
=\frac{1}{2}\left[\,g^{-1}\left(\,\balpha_{\,p},\balpha_{\,p}\,\right)
+\left(\,\star\,\balpha_{\,q}\,\right)^2
\,\right]
=\frac{1}{2}\left[\,\left(x_{(p)}^{\,1}\right)^{\,2}
+\left(x_{(p)}^{\,2}\right)^{\,2}+\left(x_{(p)}^{\,3}\right)^{\,2}
+\left(x_{(q)}\right)^{\,2}
\,\right],
$$
$$
\varphi(y^{(p)},y^{(q)})
=\frac{1}{2}\left[\,g^{-1}\left(\,\star\,\bfe_{\,p},\star\,\bfe_{\,p}\,\right)
+\left(\,\bfe_{\,q}\,\right)^2
\,\right]
=\frac{1}{2}\left[\,\left(y_{\,1}^{(p)}\right)^{\,2}
+\left(y_{\,2}^{(p)}\right)^{\,2}+\left(y_{\,3}^{(p)}\right)^{\,2}
+\left(y^{(q)}\right)^{\,2}
\,\right],
$$
where $x_{(p)}=\{\,x_{(p)}^{\,a}\,\}$, $y^{(p)}=\{\,y_{\,a}^{(p)}\,\}$ and 
$x_{(q)},y^{(p)}$ have been defined in Theorem\,\ref{theorem-n-3-dirac-contact}.
%%%%%%
\item
%%%%%
For the case $n=3,\{p=2,q=2\}$, one has 
\beqa
\psi(x_{(p)},x_{(q)})
&=&\frac{1}{2}\left[\,g^{-1}\left(\,\star\,\balpha_{\,p},\star\,\balpha_{\,p}\,\right)
+\,g^{-1}\left(\,\star\,\balpha_{\,q},\star\,\balpha_{\,q}
\,\right)
\,\right]
\non\\
&=&\frac{1}{2}\left[\,
 \left(x_{(p)}^{\,1}\right)^{\,2}
+\left(x_{(p)}^{\,2}\right)^{\,2}
+\left(x_{(p)}^{\,3}\right)^{\,3}
+\left(x_{(q)}^{\,1}\right)^{\,2}
+\left(x_{(q)}^{\,2}\right)^{\,2}
+\left(x_{(q)}^{\,3}\right)^{\,2}
\,\right],
\non\\
\varphi(y^{(p)},y^{(q)})
&=&\frac{1}{2}\left[\,g^{-1}\left(\,\bfe_{\,p},\bfe_{\,p}\,\right)
+\,g^{-1}\left(\,\bfe_{\,q},\bfe_{\,q}
\,\right)
\,\right]
\non\\
&=&\frac{1}{2}\left[\,
 \left(y_{\,1}^{(p)}\right)^{\,2}
+\left(y_{\,2}^{(p)}\right)^{\,2}
+\left(y_{\,3}^{(p)}\right)^{\,3}
+\left(y_{\,1}^{(q)}\right)^{\,2}
+\left(y_{\,2}^{(q)}\right)^{\,2}
+\left(y_{\,3}^{(q)}\right)^{\,2}
\,\right],
\non
\eeqa

where $x_{(p)}=\{\,x_{(p)}^{\,a}\,\}$, 
$y^{(p)}=\{\,y_{\,a}^{(p)}\,\}$ and 
$x_{(q)}=\{\,x_{(q)}^{\,a}\,\}$, $y^{(q)}=\{\,y_{\,a}^{(q)}\,\}$ 
have been defined in 
Theorem\,\ref{theorem-n-3-dirac-contact}.
%%%%%
\item
%%%%%%
For the case $n=2,\{p=1,q=2\}$, one has 
$$
\psi(x_{(p)},x_{(q)})
=\frac{1}{2}\left[\,g^{-1}\left(\,\star\,\balpha_{\,p},\star\,\balpha_{\,p}\,\right)
+\left(\,\star\,\balpha_{\,q}\,\right)^2
\,\right]
=\frac{1}{2}\left[\,
 \left(x_{(p)}^{\,1}\right)^{\,2}
+\left(x_{(p)}^{\,2}\right)^{\,2}
+\left(x_{(q)}\right)^{\,2}
\,\right],
$$
$$
\varphi(y^{(p)},y^{(q)})
=\frac{1}{2}\left[\,g^{-1}\left(\,\bfe_{\,p},\bfe_{\,p}\,\right)
+\left(\,\bfe_{\,q}\,\right)^2
\,\right]
=\frac{1}{2}\left[\,
 \left(y_{\,1}^{(p)}\right)^{\,2}
+\left(y_{\,2}^{(p)}\right)^{\,2}
%+\left(y_{\,3}^{(p)}\right)^{\,2}
+\left(y_{(q)}\right)^{\,2}
\,\right],
$$
where $x_{(p)}=\{\,x_{(p)}^{\,a}\,\}$, $y^{(p)}=\{\,y_{\,a}^{(p)}\,\}$ and 
$x_{(q)},y^{(q)}$ have been defined in 
Theorem\,\ref{theorem-n-2-dirac-contact}.
%%%%%
\item
%%%%%%
For the case $n=1,\{p=1,q=1\}$, one has 
$$
\psi(x_{(p)},x_{(q)})
=\frac{1}{2}\left[\,\left(\,\star\,\balpha_{\,p}\,\right)^{\,2}
+\left(\,\star\,\balpha_{\,q}\,\right)^{\,2}
\,\right]
=\frac{1}{2}\left[\,
\left(x_{(p)}\right)^{\,2}+\left(x_{(q)}\right)^{\,2}
\,\right],
$$
$$
\varphi(y^{(p)},y^{(q)})
=\frac{1}{2}\left[\,\left(\,\bfe_{\,p}\,\right)^{\,2}
+\left(\,\bfe_{\,q}\,\right)^{\,2}
\,\right]
=\frac{1}{2}\left[\,
\left(y^{(p)}\right)^{\,2}+\left(y^{(q)}\right)^{\,2}
\,\right],
$$
where $x_{(p)}$, $y^{(p)}$  and $x_{(q)}$, $y^{(q)}$ have been defined in 
Theorem\,\ref{theorem-n-1-dirac-contact}.
 %%%%%%%%%%%%%%
\end{itemize}
%%%%%%%%%%%%
Then all of the cases above, one has 
$$
\Leg\left[\,\psi\,\right](y^{(p)},y^{(q)})
=\varphi(y^{(p)},y^{(q)}).
$$
%%%%%%%%%%%%%%%%%
\end{Proposition}
%%%%%%%%%55
\begin{Proof}
One can prove these with straightforward calculations. 
\qed
\end{Proof}

With this Proposition, 
one can show the following. 
%%%%%%%%%%%%%%%%
\begin{Proposition}
%%%%%%%%%%%%
(Components of covariant fiber metric tensor field for distributed-port Hamiltonian systems): 
The inverse matrix of $\{g_{\,ab}^{\,\zeta}\}$ with $x=\{x_{(p)},x_{(q)}\}$ in
\fr{fiber-metric-tensor-field} is given as 
$$
g_{\,\zeta}^{\,ab}
=\frac{\partial^{\,2}\,\varphi}{\partial y_{\,a}\partial y_{\,b}},\quad 
a,b\in\{\,1,\ldots,\wt{n}_{\,pq}\,\}, 
$$
where $y=\{y^{(p)},y^{(q)}\}$ and $\wt{n}_{\,pq}={}_{n}C_{\,p}+{}_{n}C_{\,q}$.
%%%%%%
\end{Proposition}  
%%%%%%%
%%%%%%%%%%%%%
\begin{Proof}
%%%%%%%%%%%
A proof is similar to that found in Ref.\,\cite{AN}.
\qed
%%%%%%%%%%%
\end{Proof}
%%%%%%%%%%%

In the standard information geometry summarized in  
Section\,\ref{subsection-information-geometry},   
there are two coordinates referred to as dual coordinates
(\,see Definition\,\ref{definition-dual-coordinate}\,), and 
analogous coordinates exist for our formulation of systems on 
contact bundles discussed in 
Section\,\ref{section-distributed-Hamiltonian-systems-Dirac}. 
The following proposition shows 
that $x$ and $y$ are such analogous coordinates.  
%%%%%%%%%%%%%%%%%%%
\begin{Proposition}
%%%%%%%%%%%%%%%%%%  
(Dual connections for distributed-parameter port-Hamiltonian systems): 
With  
$x^{\,j}=\partial\,\varphi/\partial y_{\,j}$, one has
$$
g^{\,\zeta}
\left(\,\frac{\partial}{\partial x^{\,b}},\frac{\partial}{\partial y_{\,a}}\,
\right)
=\delta_{\,b}^{\,a},\qquad
a,b\in\{1,\ldots,\wt{n}_{\,pq}\}, 
$$
where $\{\partial/\partial x^{\,a}\},\{\partial/\partial y_{\,a}\}$ 
are vertical vectors fields, and $\wt{n}_{\,pq}={}_{n}C_{\,p}+{}_{n}C_{\,q}$.
%%%%%%%%%%%%
\end{Proposition}  
%%%%%%%%%%%%%
\begin{Proof}
%%%%%%%%%%%
It follows from 
$$
\frac{\partial x^{\,j}}{\partial y_{\,a}}
=\frac{\partial^2\,\varphi}{\partial y_{\,a}\partial y_{\,j}}
=g_{\,\zeta}^{\,aj}  
$$
that 
$$
g^{\,\zeta}
\left(\,\frac{\partial}{\partial x^{\,b}},\frac{\partial}{\partial y_{\,a}}\,
\right)
=g_{\,ij}^{\,\zeta}\delta_{\,b}^{\,i}\frac{\partial x^{\,j}}{\partial y_{\,a}}
=g_{\,ij}^{\,\zeta}\delta_{\,b}^{\,i}\,g_{\,\zeta}^{\,aj}
=\delta_{\,b}^{\,a}.
$$
\qed
%%%%%%%%%%%
\end{Proof}
%%%%%%%%%%%

In the standard information geometry, 
the dual connections are often discussed
( see Definition\,\ref{definition-dual-coordinate} ).  
Analogous connections can appear in the present geometry. 
%%%%%%%%%%%%%
\begin{Def}
%%%%%%%%%%%
(Dual connections on contact bundle): 
Let $(\cK,\lambda_{\,\mbbV},\pi,\cB)$ a contact bundle, 
$\cC_{\,\zeta}$ a contact manifold over a point of the base space $\zeta\in\cB$, 
$\cA_{\,\zeta\psi}^{\cC}$ a Legendre submanifold generated by a function $\psi$ 
over a base point $\zeta\in\cB$, $g^{\,\zeta}$ the metric tensor field on 
$\cA_{\,\zeta\psi}^{\cC}$ ( see Definition\, \ref{definition-fiber-metric} ), 
$\nabla^{\,\zeta}$ a connection 
on the Riemannian manifold $(\cA_{\,\zeta\psi}^{\cC},g^{\,\zeta})$, and 
$X_{\,\mbbV},Y_{\,\mbbV},Z_{\,\mbbV}$ vertical vector fields. 
If another connection $\nabla^{\,\zeta\,\prime}$ satisfies 
$$
X_{\,\mbbV}\left[\, g^{\,\zeta}(Y_{\,\mbbV},Z_{\,\mbbV})\,\right]
=g^{\,\zeta}\left(\nabla_{\,X_{\,\mbbV}}^{\,\zeta}Y_{\,\mbbV},Z_{\,\mbbV}\right)+
g^{\,\zeta}\left(Y_{\,\mbbV},\nabla_{\,X_{\,\mbbV}}^{\,\zeta\,\prime}Z_{\,\mbbV}\right).
$$
Then the two connections $\nabla^{\,\zeta}$ and $\nabla^{\,\zeta\,\prime}$
are referred to as dual connections on with respect to $g^{\,\zeta}$ on the 
contact bundle. 
%%%%%%%%%%5
\end{Def}
%%%%%%%%%%

A realization of connection components of dual connections have been known. 
In our present case of 
$\psi$ the following is a trivial identity since $\psi$ is a 
quadratic function. 
%%%%%%%%%%%%%%%%%%%%
\begin{Proposition}
%%%%%%%%%%%%%%%%
(Component expression of dually flat space on contact bundle): 
Defining 
$$
\Gamma_{\,abc}^{\,\zeta\,(\alpha)}
:=\frac{1-\alpha}{2}
\frac{\partial^{\,3}\,\psi}{\partial x^{\,a}\partial x^{\,b}\partial x^{\,c}},
\qquad \alpha\in\mbbR,
$$
one has 
$$
\frac{\partial}{\partial x^{\,a}}g_{\,bc}^{\,\zeta}
=\Gamma_{\,abc}^{\,\zeta\,(\alpha)}
+\Gamma_{\,acb}^{\,\zeta\,(-\alpha)},
\quad 
a,b\in\{1,\ldots,{}_{n}C_{\,p}+{}_{n}C_{\,q}\}.
$$
where $p$ and $q$ are given natural 
numbers such that $0\leq p,q\leq n$ and $p+q=n+1$.
%%%%%%%%%%%%%%%%%
\end{Proposition}
%%%%%%%%%%%%%%%%%% 
\begin{Proof}
%%%%%%%%%%%%%%
Substituting \fr{fiber-metric-tensor-field} 
into the left hand side of the equation above, one completes the proof.
\qed
%%%%%%%%%%%%
\end{Proof}
%%%%%%%%%%%5
\begin{Remark}
%%%%%%%%%%%%%
The dual connections $\nabla^{\,\zeta}$ and $\nabla^{\,\zeta\,\prime}$ with respect to
$g^{\,\zeta}$ are constructed such that 
$$
\nabla^{\,\zeta}_{\partial/\partial x^{\,a}}\frac{\partial}{\partial x^{\,b}}
=\Gamma_{\,ab}^{\,\zeta(\alpha)\, c}\frac{\partial}{\partial x^{\,c}},\qquad
\nabla^{\,\zeta\,\prime}_{\partial/\partial x^{\,a}}\frac{\partial}{\partial x^{\,b}}
=\Gamma_{\,ab}^{\,\zeta(-\alpha)\, c}\frac{\partial}{\partial x^{\,c}},
$$
where $\Gamma_{\,ab}^{\,\zeta\,(\alpha)\,c}$ and $\Gamma_{\,ab}^{\,\zeta\,(-\alpha)\,c}$ are 
such that 
$$
\Gamma_{\,abc}^{\,\zeta\,(\alpha)}
=g_{\,cj}^{\,\zeta}\Gamma_{\,ab}^{\,\zeta\,(\alpha)\,j},\qquad
\mbox{and}\qquad
\Gamma_{\,abc}^{\,\zeta\,(-\alpha)}
=g_{\,cj}^{\,\zeta}\Gamma_{\,ab}^{\,\zeta\,(-\alpha)\,j}.
$$
%%%%%%%%%%%%%
\end{Remark}
%%%%%%%%%%

With discussions above, one finds the following main theorem in this section.
%%%%%%%%%%%
\begin{Thm}
\label{theorem-Stokes-Dirac-dually-flat-space}
%%%%%%5%%%%%
(Distributed-parameter port-Hamiltonian systems induce dually flat space): 
The phase 
space for distributed-parameter port-Hamiltonian systems $\cA_{\,\psi}^{\,\cK}$
( see Definition\,\ref{definition-phase-space-distributed-parameter-port-Hamiltonian-system} ) and the quadratic energy functional 
\fr{quadratic-energy-functional-p-q}  
induce   
the quadruplet 
$(\cA_{\,\zeta\psi}^{\,\cC}, g^{\,\zeta},\nabla^{\,\zeta},\nabla^{\,\zeta\,\prime})$.
%%%%%%%%%
\end{Thm}
%%%%%%%%%

% On any Riemannian manifold $(\cM,g)$ with a connection $\nabla$, 
% one can find a unique dual connection $\nabla^{\,\prime}$. Then 
% the quadruplet $(\cM,g,\nabla,\nabla^{\,\prime})$ is referred to as a 
% dually flat space \cite{AN}. 

In accordance with a dually flat space in the standard 
information geometry ( see Definition\,\ref{definition-dually-flat-space} ), 
one can introduce 
such a space in the present geometry as follows. 
%%%%%%%%%%
\begin{Def}
%%%%%%%%%%%%
(Dually flat space in contact bundle): 
The quadruplet introduced in 
Theorem\,\ref{theorem-Stokes-Dirac-dually-flat-space}
is referred to as a dually flat space in a contact bundle.  
%%%%%%%%%%%% 
\end{Def}
%%%%%%%%%%%%
Analogous to 
Proposition\,\ref{proposition-inner-product-paring-dually-flat-space},
one has the following.
%%%%%%%%%%%%%%%%%%%%%
\begin{Proposition}
%\label{proposition-inner-product-paring-dually-flat-space}
%%%%%%%%%%%%%%%%%%%
(Inner products and pairings on a dually flat space in contact bundle):  
Let \\
$(\cA_{\,\zeta\psi}^{\,\cC}, g^{\,\zeta},\nabla^{\,\zeta},\nabla^{\,\zeta\,\prime})$
be an $\wt{n}$-dimensional 
dually flat space, $x=\{\,x^{\,1},\ldots,x^{\,\wt{n}}\,\}$ a set of 
$\nabla$-affine coordinates, $y=\{\,y_{\,1},\ldots,y_{\,\wt{n}}\,\}$ a set of 
$\nabla^{\,*}$-affine coordinates.
If the inner products 
$T_{\xi}\cA_{\,\zeta\psi}^{\,\cC}\times T_{\xi}\cA_{\,\zeta\psi}^{\,\cC}\to\mbbR, (\xi\in\cA_{\,\zeta\psi}^{\,\cC})$ 
between the bases
$\{\,\partial/\partial x^{\,1},\ldots,\partial/\partial x^{\,\wt{n}}\,\}$ and 
$\{\,\partial/\partial y_{\,1},\ldots,\partial/\partial y_{\,\wt{n}}\,\}$ are given as 
\beqa
g^{\,\zeta}\left(\,\frac{\partial}{\partial x^{\,a}},\frac{\partial}{\partial x^{\,b}}\,\right)
&=&g_{\,ab}^{\,\zeta},\qquad
g^{\,\zeta}\left(\,\frac{\partial}{\partial x^{\,a}},\frac{\partial}{\partial y_{\,b}}\,\right)
=\delta_{\,a}^{\,b},
\non\\
g^{\,\zeta}\left(\,\frac{\partial}{\partial y_{\,a}},\frac{\partial}{\partial x^{\,b}}\,\right)
&=&\delta_{\,b}^{\,a},\qquad
g^{\,\zeta}\left(\,\frac{\partial}{\partial y_{\,a}},\frac{\partial}{\partial y_{\,b}}\,\right)
=g_{\,\zeta}^{\,ab},
\non
\eeqa
( i.e., $x$ and $y$ are mutually dual with respect to $g^{\,\zeta}$ ), then 
one has 
the following pairings  
$T_{\xi}^{\,*}\cA_{\,\zeta\psi}^{\,\cC}\times T_{\xi}\cA_{\,\zeta\psi}^{\,\cC}\to\mbbR, (\xi\in\cM)$
\beqa
\dr x^{\,a}\left(\,\frac{\partial}{\partial x^{\,b}}\,\right)
&=&\delta^{\,ab},\qquad
\dr x^{\,a}\left(\,\frac{\partial}{\partial y_{\,b}}\,\right)
=g_{\,\zeta}^{\,ab},
\non\\
\dr y_{\,a}\left(\,\frac{\partial}{\partial x^{\,b}}\,\right)
&=&g_{\,ab}^{\,\zeta},\qquad
\dr y_{\,a}\left(\,\frac{\partial}{\partial y_{\,b}}\,\right)
=\delta_{\,ab}.
\non
\eeqa
%%%%%%%%%%%%%%%
\end{Proposition}
%%%%%%%%%%%%%%%%

The canonical divergence plays a role in information geometry 
( see Definition\,\ref{definition-canonical-divergence-information-geometry} ),
and 
that can be defined in the fiber space as follows. 
%%%%%%%%%%%
\begin{Def}
%%%%%%%%%%%
(Canonical divergence on phase space of distributed-parameter port-Hamiltonian systems): 
The function 
$\mbbD^{\,\zeta}:\cA_{\,\zeta\psi}^{\,\cC}\times\cA_{\,\zeta\psi}^{\,\cC}\to\mbbR,(\zeta\in\cZ)$ such that 
$$
\mbbD^{\,\zeta}\,(\,\xi\,\|\,\xi'\,)
:=\psi(\xi)+\varphi(\xi')-x^{\,a}|_{\,\xi}\,y_{\,a}|_{\,\xi'}.
$$
is referred to as canonical divergence 
on phase space of distributed-parameter  port-Hamiltonian systems.
%%%%%%%%%
\end{Def}
%%%%%%%%% 
The generalized Pythagorean theorem plays a role in the standard information 
geometry ( see Theorem\,\ref{theorem-Pythagorean} ), 
and an analogous theorem exists in the present geometry.
%%%%%%%%%%%%%
\begin{Thm}
%%%%%%%%%%%
(Generalized Pythagorean theorem for phase space of distributed-parameter 
port-Hamiltonian systems):  
Let  $\xi^{\,\prime},\xi^{\,\prime\prime}$ and 
$\xi^{\,\prime\prime\prime}$ be points of of  $\cA_{\,\zeta\psi}^{\,\cC}$ such that 
1. 1. $\xi^{\,\prime}$ and $\xi^{\,\prime\prime}$ are 
connected with the $\nabla^{\,\zeta\,\prime}$-geodesic curve $\gamma^{\,\zeta\,\prime}$ 
and 
2. $\xi^{\,\prime\prime}$ and $\xi^{\,\prime\prime\prime}$ are connected with 
the $\nabla^{\,\zeta}$-geodesic curve $\gamma^{\,\zeta}$. 
If at the intersection $\xi^{\,\prime\prime}$ the curves $\gamma^{\,\zeta}$ 
and $\gamma^{\,\zeta\,\prime}$ 
are orthogonal with respect to $g^{\,\zeta}$, 
then one has that
$$
\mbbD^{\,\zeta}\,(\,\xi^{\,\prime\prime\prime}\,\|\,\xi^{\,\prime}\,)
=\mbbD^{\,\zeta}\,(\,\xi^{\,\prime\prime\prime}\,\|\,
\xi^{\,\prime\prime}\,)
+\mbbD^{\,\zeta}\,(\,\xi^{\,\prime\prime}\,\|\,\xi^{\,\prime}\,).
$$
%%%%%%%%%
\end{Thm}
%%%%%%%%%%
%%%%%%%%%%%%%
\begin{Proof}
%%%%%%%%%%%
A proof is similar to that found in Ref.\,\cite{AN}.
\qed
%%%%%%%%%%%
\end{Proof}
%%%%%%%%%%%

%%%%%%%%%%%%%%%%%%%%%%%%%%%%%%%%%%%%%%%%%%%5
 \section{Concluding remarks}
\label{sec-summary}
%%%%%%%%%%%%%%%%%%%%%%%%%%%%%%%%%%%%%%%%
% This paper offers 
% a contact geometric description of distributed-parameter systems on 
% Riemannian manifolds with respect to    
% %Dirac structures and that defined on 
% Stokes-Dirac structures.
This paper provides a viewpoint that 
%implicit Hamiltonian systems with respect to Dirac structures 
%can be written by means of contact geometry, and 
%a viewpoint that 
distributed-parameter port-Hamiltonian systems with respect to 
Stokes-Dirac structures 
can be written in terms of bundles whose fiber spaces are 
contact manifolds and base spaces are Riemannian manifolds.  
% Then, it has been shown with 
% our approaches that 
% various mathematical statements on contact geometry can be applied 
%to systems with respect to Dirac structures and Stokes-Dirac structures.  
Also, it has been shown that one can introduce information geometry for 
a class of distributed parameter port-Hamiltonian systems.  
%Throughout this paper, Legendre duality has explicitly been stated.  
%In addition, some applications of the developed general theories to   
%science and foundation of engineering have also been given.     

There are some potential future works that follow from this paper. 
One is to study the case where energy functionals contains higher order terms, 
since this paper has only considered the case where 
energy functionals are quadratic except for $1$-dimensional 
Riemannian manifolds. In addition, the present approach should be extended 
such that distributed-parameter port-Hamiltonian systems with external 
sources can be described. 
%in terms of an extention of the present contact formulation.
%it should be mentioned that 
%some class of distributed-parameter 
%port-Hamiltonian systems cannot be described 
%in terms of our contact formulation. 
Such a class with external sources includes 
Maxwell's equations with some external sources. 
%Another one is to compare the present approach and the one in Ref.\cite{Noda} 
We believe 
that the elucidation of these remaining questions 
will develop the geometric theories in mathematical sciences and 
foundation of engineering.

% One is to compare this work with other existing works. Such other works 
% include the work on symplectic geometry adopted 
% to a dually flat space discussed in Ref.\,\cite{Noda2011}, 
% and relaxation dynamics on a statistical manifold in Ref.\cite{OW09}. 
% In addition, it is interesting to construct 
% the vector field associated with the Brayton-Moser equations 
% on a dually flat space, where the Brayton-Moser equations describe 
% a class of electric circuit models\cite{Eberard2006}. 
% Another important future work is to import mathematical 
% findings in contact geometry and topology to 
% science and engineering. 
% We believe 
% that the elucidation of these remaining questions 
% will develop the geometric theories in mathematical sciences and 
% foundation of engineering.

%%%%%%%%%%%
\section*{Acknowledgments }
%%%%%%%%%%%
The author would like to thank Ken Umeno (Kyoto University)   
for supporting my work, 
 and thank %T. Wada, M. Koga, 
Yosuke Nakata (Shinshu University) %, D. Tarama, 
% and an anonymous referee  
for giving various comments on this work. 
% , 
% and thank M. Koga, T. Wada, and an anonymous referee 
% for giving various comments on this paper. 

%%%%%%%%%%%%%%%%%%%%%%%%%%%

%%%%%%%%%%%%%%%%%%%%%%
%%%%%%%%%%%%%%%
\end{document}